\documentclass[longauth]{aa}
\usepackage{graphicx}
\usepackage{caption}
\usepackage{txfonts}
\usepackage{booktabs}
\usepackage{color}
\usepackage[dvipsnames]{xcolor}
\usepackage{natbib,twoopt}
\usepackage{lscape}
\usepackage{caption}
\usepackage{subcaption}
\usepackage[symbol]{footmisc}
\usepackage{tablefootnote}
\usepackage[bookmarks=true, pdfnewwindow=true, colorlinks=true, allcolors=blue, breaklinks=true, final=true]{hyperref}
\usepackage{orcidlink}
\bibpunct{(}{)}{;}{a}{}{,}
\newcommand{\massrate}{$M_{\odot}$\,yr$^{-1}$}

\newcommand{\hii}{H\textsc{ii}}
\newcommand{\msun}{$ M_\odot$}

\newcommand{\kms}{km\,s$^{-1}$}
\newcommand{\jybeam}{Jy\,beam$^{-1}$}
\newcommand{\mjybeam}{mJy\,beam$^{-1}$}
\newcommand{\ujybeam}{$\mu$Jy\,beam$^{-1}$}

\newcommand{\parcsec}{\mbox{$.\!\arcsec$}}

\newcommand{\cc}{\mbox{cm$^{-3}$}}

\newcommand{\meth}{\mbox{CH$_3$OH}}

\newcommand{\water}{\mbox{H$_2$O}}

\newcommand{\ctw}{the 20~\kms{} cloud}
\newcommand{\cfi}{the 50~\kms{} cloud}
\newcommand{\gzp}{G0.253+0.016}
\newcommand{\sgc}{Sgr~C}

\newcommand{\zenodo}{\href{https://doi.org/10.5281/zenodo.15068333}{10.5281/zenodo.15068333}}
\begin{document}

\title{Dual-band Unified Exploration of three CMZ Clouds (DUET).}

\subtitle{Cloud-wide census of continuum sources showing low spectral indices}

\author{Fengwei Xu \inst{\ref{kiaa}, \ref{pkudoa}, \ref{unikoeln}} \orcidlink{0000-0001-5950-1932} \and
Xing Lu\inst{\ref{shao}}\thanks{xinglu@shao.ac.cn} \orcidlink{0000-0003-2619-9305} \and
Ke Wang\inst{\ref{kiaa}}\thanks{kwang.astro@pku.edu.cn} \orcidlink{0000-0002-7237-3856} \and
Hauyu Baobab Liu\inst{\ref{nsysu}, \ref{ntnu}} \orcidlink{0000-0003-2300-2626} \and
Adam Ginsburg\inst{\ref{UF}} \orcidlink{0000-0001-6431-9633} \and
Tie Liu\inst{\ref{shao}} \orcidlink{0000-0002-5286-2564} \and
Qizhou Zhang\inst{\ref{cfa}} \orcidlink{0000-0003-2384-6589} \and
Nazar Budaiev \inst{\ref{UF}} \orcidlink{0000-0002-0533-8575} \and
Xindi Tang \inst{\ref{xao}} \orcidlink{0000-0002-4154-4309} \and
Peter Schilke\inst{\ref{unikoeln}} \orcidlink{0000-0003-2141-5689} \and
Suinan Zhang \inst{\ref{shao}} \orcidlink{0000-0002-8389-6695} \and
Sihan Jiao \inst{\ref{naoc}} \and
Wenyu Jiao \inst{\ref{shao}} \orcidlink{0000-0001-9822-7817} \and
Siqi Zheng\inst{\ref{shao}, \ref{unikoeln}} \orcidlink{0000-0001-9047-846X} \and
Beth Jones\inst{\ref{unikoeln}} \orcidlink{0000-0002-0675-0078} \and
J. M. Diederik Kruijssen\inst{\ref{cool}} \orcidlink{0000-0002-8804-0212} \and
Cara Battersby\inst{\ref{uconn}} \orcidlink{0000-0002-6073-9320} \and
Daniel L. Walker\inst{\ref{ukarc}} \orcidlink{0000-0001-7330-8856} \and
Elisabeth A.C. Mills\inst{\ref{ukans}} \orcidlink{0000-0001-8782-1992} \and
Jens Kauffmann\inst{\ref{mit}} \orcidlink{0000-0002-5094-6393} \and
Steven N. Longmore\inst{\ref{liver}, \ref{cool}} \orcidlink{0000-0001-6353-0170} \and
Thushara G.S. Pillai\inst{\ref{mit}} \orcidlink{0000-0003-2133-4862}
}

\institute{
\label{kiaa} Kavli Institute for Astronomy and Astrophysics, Peking University, Beijing 100871, People's Republic of China
\and
\label{pkudoa} Department of Astronomy, School of Physics, Peking University, Beijing, 100871, People's Republic of China
\and
\label{unikoeln} I. Physikalisches Institut, Universit{\"a}t zu K{\"o}ln, Z{\"u}lpicher Stra{\ss}e 77, 50937 Cologne, Germany 
\and
\label{shao} Shanghai Astronomical Observatory, Chinese Academy of Sciences, 80 Nandan Road, Shanghai 200030, People's Republic of China
\and
\label{nsysu} Department of Physics, National Sun Yat-Sen University, No. 70, Lien-Hai Road, Kaohsiung City 80424, Taiwan, R.O.C.
\and
\label{ntnu} Center of Astronomy and Gravitation, National Taiwan Normal University, Taipei 116, Taiwan
\and
\label{UF} Department of Astronomy, University of Florida, PO Box 112055, Florida, USA \and
\label{cfa} Center for Astrophysics $|$ Harvard \& Smithsonian, 60 Garden Street, Cambridge, MA 02138, USA \and
\label{xao} Xinjiang Astronomical Observatory, Chinese Academy of Sciences, Urumqi 830011, People’s Republic of China \and
\label{naoc} National Astronomical Observatories, Chinese Academy of Sciences, 20A Datun Road, Chaoyang District, Beijing 100012, China \and
\label{cool} Cosmic Origins Of Life (COOL) Research DAO, \href{https://coolresearch.io}{https://coolresearch.io} \and
\label{uconn} Department of Physics, University of Connecticut, 196A Auditorium Road, Unit 3046, Storrs, CT 06269, USA \and
\label{ukarc} UK ALMA Regional Centre Node, Jodrell Bank Centre for Astrophysics, The University of Manchester, Manchester M13 9PL, UK \and
\label{ukans} Department of Physics and Astronomy, University of Kansas, 1251 Wescoe Hall Drive, Lawrence, KS 66045, USA \and
\label{mit} Haystack Observatory, Massachusetts Institute of Technology, 99 Milsstone Road, Westford, MA 01886, USA \and
\label{liver} Astrophysics Research Institute, Liverpool John Moores University, IC2, Liverpool Science Park, 146 Brownlow Hill, Liverpool L3 5RF, UK
}

\date{accepted: Mar. 20th, 2025}

\titlerunning{Cloud-wide continuum sources with low spectral indices in CMZ}
\authorrunning{Fengwei Xu et al.}

\abstract
{The Milky Way's central molecular zone (CMZ) has been measured to form stars ten times less efficiently than in the Galactic disk, based on emission from high-mass stars. However, the CMZ's low-mass ($\lesssim$2~\msun) protostellar population, which accounts for most of the initial stellar mass budget and star formation rate (SFR), is poorly constrained observationally due to limited sensitivity and resolution.}
{We aim to perform a cloud-wide census of the protostellar population in three massive CMZ clouds.}
{We present the Dual-band Unified Exploration of three CMZ Clouds (DUET) survey, targeting the \ctw, \sgc, and the dust ridge cloud “e” using the Atacama Large Millimeter/submillimeter Array (ALMA) at 1.3 and 3~mm. The mosaicked observations achieve a comparable resolution of 0\parcsec2--0\parcsec3 ($\sim$2000~au) and a sky coverage of 8.3--10.4~arcmin$^2$, respectively.}
{We report 563 continuum sources at 1.3~mm and 330 at 3~mm, respectively, and a dual-band catalog with 450 continuum sources. These sources are marginally resolved at a resolution of 2000~au. We find a universal deviation ($>70$\% of the source sample) from commonly used dust modified blackbody (MBB) models, characterized by either low spectral indices or low brightness temperatures.}
{Three possible explanations are discussed for the deviation. (1) Optically thick class 0/I young stellar objects (YSOs) with a very small beam filling factor can lead to lower brightness temperatures than what MBB models predict. (2) Large dust grains with millimeter or centimeter in size have more significant self-scattering, and frequency-dependent albedo could therefore cause lower spectral indices. (3) Free-free emission over 30~$\mu$Jy can severely contaminate dust emission and cause low spectral indices for milli-Jansky sources, although the number of massive protostars (embedded UCH{\sc ii} regions) needed is infeasibly high for the normal stellar initial mass function. A reliable measurement of the SFR at low protostellar masses will require future work to distinguish between these possible explanations.}


\maketitle

\section{Introduction}\label{sec:intro}

The central molecular zone (CMZ), the inner $\sim500$~pc of the Milky Way, contains more than $10^7$~\msun{} of dense molecular gas \citep{Morris1996, Ferriere2007}. It is a unique star-forming environment, characterized by extreme conditions similar to the ones under which most of the cosmic star formation may have occurred \citep{Mezger1996, Kruijssen2013}. These conditions include strong magnetic fields, intense turbulence, elevated cosmic ray flux, and gravitational shear, along with feedback from the supermassive black hole Sgr~A$^\star$ \citep{Genzel1987, Kruijssen2014, Mills2017Review, Bryant2021Review, Henshaw2023Review}. Several molecular clouds with masses of over $10^4$~\msun{} and densities of $\gtrsim$10$^4$~\cc{} are found within the CMZ \citep{Bally1987, Bally1988, Lis1994, Tsuboi1999, Kendrew2013, Walker2015, Kauffmann2017a, Kauffmann2017b, Lu2019SFR, Lu2020Jeans, Battersby2020}. Despite the presence of a large reservoir of molecular gas at volume densities high enough to generate widespread star formation in the solar neighborhood \citep[e.g.,][]{Lada2010}, the CMZ is notoriously inefficient at forming stars, with a star formation rate (SFR) of $0.07^{+0.08}_{-0.02}$~\massrate\ inferred from massive ($>8$~\msun) stellar emission \citep[see review Table.~1 in][]{Henshaw2023Review} which is about ten times lower than expected based on dense gas star formation relations \citep[e.g.,][]{Longmore2013, Barnes2017}. 

The measurements of the SFR in the CMZ initially relied on free-free emission from massive stars at an extremely low angular resolution \citep{Murray2010}. Follow-up interferometric observations have provided direct evidence of high-mass ($>8$~\msun) star formation. For instance, observations with the Submillimeter Array (SMA) have resolved several hundred dense cores at 0.2~pc resolution \citep{Liu2013CMZ, Kauffmann2017a, Lu2015, Lu2017, Lu2019SFR, Hatchfield2020}. The Very Large Array (VLA) surveys found class \textsc{ii} \meth{} masers, \water{} masers, and ultracompact \hii{} (UC\hii{}) regions associated with these dense cores of $10^2$--$10^3$~\msun{} \citep{Lu2019Census, Lu2019SFR}. 

However, the low-mass ($<2$~\msun) protostellar population in the CMZ remains unclear, limited by resolution and sensitivity. Theories suggest that both high-mass and low-mass protostars form in a common accretion reservoir. Therefore, low-mass protostars should be found around high-mass ones \citep[e.g.,][]{Bonnell2001, Smith2009}. Recent high-resolution and high-sensitivity Atacama Large Millimeter/submillimeter Array (ALMA) observations toward star-forming regions in the Galactic disk have detected faint low-mass cores surrounding massive ones, revealing crowded mini-clusters that were previously identified as massive cores \citep[e.g.,][]{Wang2014, Zhang2015, Pillai2019, Xu2023SDC335, Xu2024Protocluster}. These low-mass members are crucial to measuring the SFR. For a normal stellar initial mass function (IMF, e.g.,\ \citealt{Bastian2010}), they represent the dominant source of both the stellar mass and the SFR. In particular, a top-heavy IMF has been suggested for young massive clusters in the CMZ \citep{Figer1999, Wang2006, Hosek2019}, and studying low-mass star formation in CMZ protoclusters will be crucial to place these claims in the context of the incipient stellar population. 

\section{DUET: dual-band mosaic of three CMZ clouds} \label{sec:duet}

\begin{figure*}[!ht]
\centering
\includegraphics[width=\linewidth]{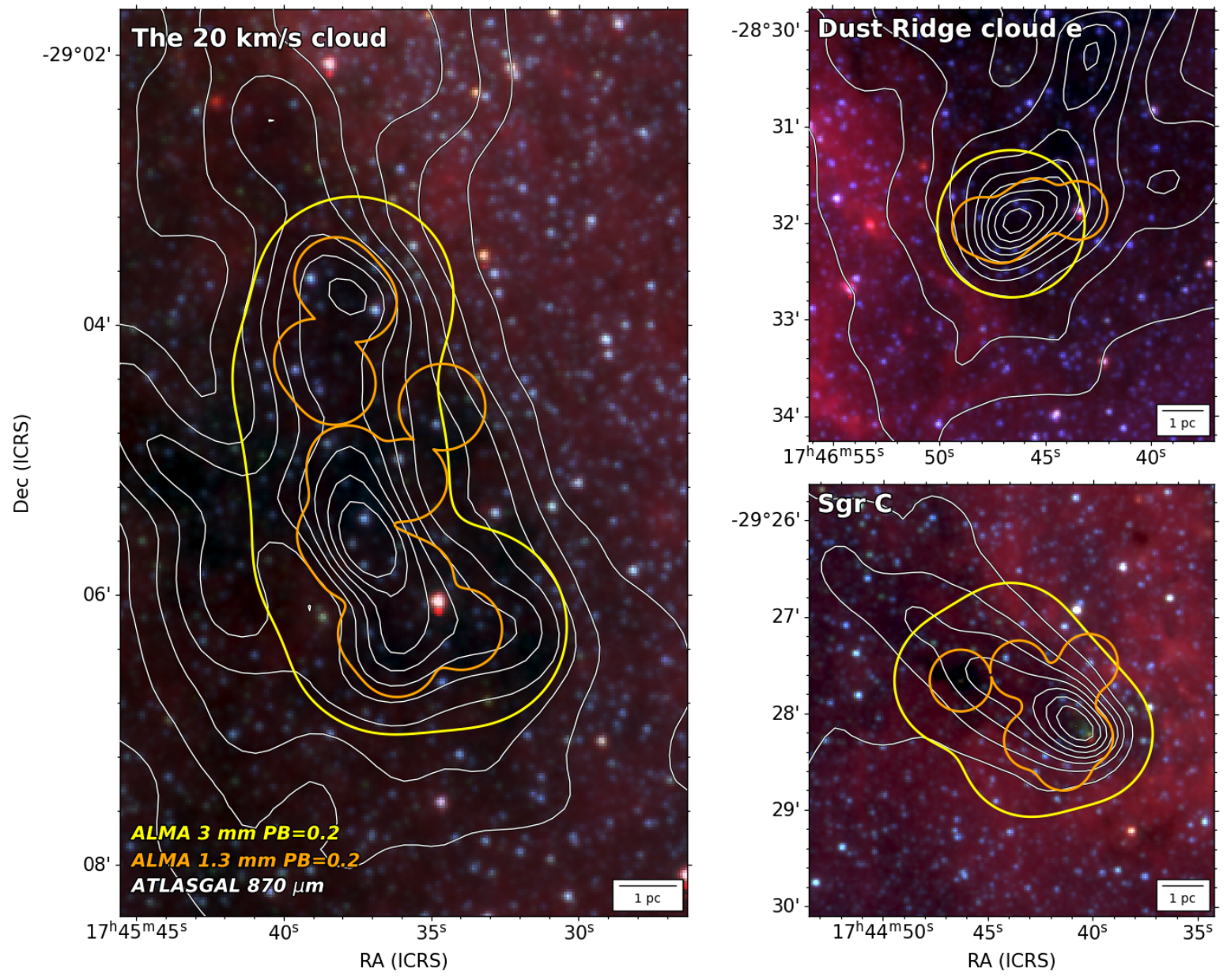}
\caption{Overview of the three CMZ clouds, the 20~\kms, Sgr~C, and the dust ridge cloud~e by the dual-band ALMA observations. The background pseudo-color maps show the composite \texttt{Spitzer} 3.6, 4.5 and 8.0~$\mu$m emission. The white contours show the ATLASGAL 870~$\mu$m emission, with contour levels increasing from 3 by a step of 1~\jybeam, up to 12~\jybeam. The dashed yellow and orange loops outline the primary beam response of 0.2 for ALMA 3~mm and 1.3~mm observations, respectively. The scale bar of 1~pc is shown on the bottom right.
\label{fig:infrared}}
\end{figure*}

\begin{table*}[!th]
\centering
\caption{Basic information on the three CMZ molecular clouds \label{tab:cloud}}
\begin{tabular}{cccccccccc}
\hline
\hline \noalign{\smallskip}
CMZ Cloud & R.A. & Dec & $l$ & $b$ & $V_{\rm lsr}$ & $R_{\rm eff}$ & $T_{\rm dust}$ & Mass & $n_{\rm H_2}$ \\
 & (hh:mm:ss) & (dd:mm:ss) & (deg) & (deg) & (km~s$^{-1}$) & (pc) & (K) & ($10^3$ \msun) & ($10^{4}$ cm$^{-3}$) \\
\hline \noalign{\smallskip}
20~\kms & 17:45:37.4 & $-$29:05:20 & 359.87 & $-$0.08 & 12.5 & 5.1 & 15.8 & 33.9 & 0.9 \\
Sgr~C & 17:44:44.1 & $-$29:27:46 & 359.45 & $-$0.11 & $-$52.6 & 1.7 & 20.6 & 2.5 & 1.8 \\
cloud~e & 17:46:45.5 & $-$28:31:34 & 0.48 & 0.00 & 31.1 & 3.6 & 16.4 & 14.5 & 1.1 \\
\hline \\
\end{tabular}
\tablefoot{
The cloud effective radius, total gas mass, and H$_2$ volume density are retrieved from \citet{Kauffmann2017a}. The averaged dust temperature is the mean value over a column density threshold of $9\times10^{22}$ cm$^{-2}$ in the fore/background-subtracted column density map of \citet{Battersby2024}. 
}
\end{table*}

To shed light on the low-mass population in the Milky Way CMZ, we have conducted the Dual-band Unified Exploration of three CMZ Clouds (DUET) survey at both 1.3~mm and 3~mm by ALMA. DUET serves as part of the Coordinated Observations of Nebulae in the CMZ Exploring Gas Recycling and Transformation (CONCERT) campaign. 

Our sample selection took advantage of the previous SMA survey of all major molecular clouds in the CMZ outsider of Sgr~B2 \citep{Kauffmann2017a, Kauffmann2017b}. We selected three massive molecular clouds namely \ctw, \sgc, and cloud~e, which present abundant fragmentation in the 1.3~mm continuum emission \citep{Lu2020Jeans}. Compared to \gzp{} and \cfi{} \citep{Rathborne2015, Walker2021}, the three DUET clouds have been found to contain a larger number of dense cores. 

In the Galactic context, \ctw{} (M$-$0.13$-$0.08) is projected to within 10~pc of Sgr~A$^\star$. It has been suggested that it feeds gas and dust to the 2-pc circum-nuclear disk \citep{Ho1991}, and it shows a moderate level of star formation itself \citep{Kauffmann2017a, Lu2019SFR, Lu2021Outflow}. \sgc{} is the most massive and luminous star-forming region on the western side of the CMZ, and it has been suggested that it is a connection point to a stream of gas and dust linking the CMZ and the nuclear stellar disk to the Galactic bar \citep{Henshaw2023Review}. Star formation in Sgr~C has been found to be active, with an overall efficiency comparable to ones in Galactic disk clouds \citep{Kauffmann2017a, Lu2019SFR}. The dust ridge cloud~“e” is one of the massive clouds ($>10^5$~\msun{}) along the dust ridge extending to the eastern side of Sgr~A$^\star$ \citep{Immer2012}. It is largely devoid of signatures of ongoing star formation, except for two dense cores associated with masers and a UC\hii{} region candidate \citep{Lu2019SFR, Lu2019Census}. Therefore, the three clouds constitute a small yet representative sample in terms of their locations in the CMZ and their levels of star formation. Fig.~\ref{fig:infrared} shows an overview of the three clouds, with their properties including the cloud radii, masses, and H$_2$ volume densities listed in Table~\ref{tab:cloud}. 

\begin{table*}[!ht]
\centering
\caption{Summary of dual-band observations and continuum images \label{tab:dualband_obs}}
\begin{tabular}{cccccccc}
\hline
\hline \noalign{\smallskip}
CMZ Cloud & \multicolumn{2}{c}{On-source time (LB/SB)} & \multicolumn{2}{c}{Resolution} & MRS & Noise & Dynamic Range \\
\cline{4-5}
 & \multicolumn{2}{c}{(mins per pointing)} & (\arcsec$\times$\arcsec) & (au$\times$au) & (\arcsec) & (\ujybeam) & \\
\hline \noalign{\smallskip}
\multicolumn{8}{c}{ALMA band 3} \\
\hline \noalign{\smallskip}
20 \kms & \multicolumn{2}{c}{35.1 / 15.6} & $0.32\times0.23$ & $2600\times1900$ & 14.4 & 12 & 257 \\
Sgr C & \multicolumn{2}{c}{38.6 / 17.2} & $0.34\times0.23$ & $2800\times1900$ & 13.6 & 11 & 340 \\
cloud~e & \multicolumn{2}{c}{71.4 / 8.6} & $0.30\times0.20$ & $2400\times1600$ & 9.7 & 10 & 191 \\
\hline \noalign{\smallskip}
\multicolumn{8}{c}{ALMA band 6} \\
\hline \noalign{\smallskip}
20 \kms & \multicolumn{2}{c}{5.6 / 1.8} & $0.25\times0.17$ & $2000\times1400$ & 6.4 & 45 & 459  \\
Sgr C & \multicolumn{2}{c}{5.3 / 1.8} & $0.25\times0.17$ & $2000\times1400$ & 6.4 & 45 & 1006 \\
cloud~e & \multicolumn{2}{c}{5.4 / 1.8} & $0.25\times0.17$ & $2000\times1400$ & 6.4 & 43 & 676 \\
\hline \\
\end{tabular}
\tablefoot{
On-source integration time (per pointing) is listed for two configurations: longer baseline (LB) and shorter baseline (SB), respectively. For band 3, they are C7 and C4; for band 6, C5 and C3. Resolution: Continuum image resolution. MRS: Maximum recoverable scale. Noise: Rms measured from the images without primary beam correction. Dynamic range: Image dynamic range, defined as the peak value over noise rms.
}
\end{table*}

DUET primary motivation lies in penetrating the ambiguous nature of millimeter emission in these clouds. Previous ALMA single-band 1.3~mm continuum observations have revealed hundreds of compact sources on 2000-au scales, with flux densities as low as $\lesssim$1~mJy \citep{Lu2020Jeans}. Assuming that the 1.3~mm emission arises purely from optically thin dust at a uniform temperature of 25~K, these compact continuum sources correspond to a population of low-mass dense cores with masses between 0.6 and 2~\msun. However, there are several \texttt{caveats} in using a single-band millimeter continuum observation to characterize the dense cores. First, dust emission is not always optically thin, particularly at shorter wavelengths, potentially leading to an underestimation of the derived masses. Second, these sources may consist of a mix of thermal and non-thermal emissions, particularly considering the widespread non-thermal radio continuum emission associated with strong magnetic fields or cosmic rays in the CMZ \citep{Yusef-Zadeh2013, Heywood2022, Bally2024}. Third, thermal free-free emission from ionized gas can contaminate the millimeter continuum \citep[e.g.,][]{Ginsburg2018, Lu2019SFR}. Last but not least, previous observations are mostly single pointings, which could miss the population outside fields. The dual-band mosaics by DUET allow us to address these problems, providing an unprecedented insights into the nature of these cloud-wide compact sources. 

In the following, we introduce the ALMA observations and data reduction in Sect.~\ref{sec:obs}. We present the source extraction processes, the derived source catalogs, and the quality assessment of the catalogs in Sect.~\ref{sec:source}. We then discuss the intrinsic source size in Sect.~\ref{sec:size}. Leveraging the dual-band catalog, we find a universal ($>$\,70\% in the source sample) deviation from commonly used dust modified blackbody (MBB) emission models, and discuss three possible interpretations -- class 0/I young stellar objects (YSOs) with a very small source size, the existence of large dust grains, and free-free emission contamination -- in Sect.~\ref{sec:si}. Throughout this paper, we adopt a uniform heliocentric distance to these three clouds of 8.277~kpc \citep{Gravity2022}. 

\section{Observations and data reduction}  \label{sec:obs}

\subsection{ALMA 3~mm} \label{subsec:obs_band3}

The ALMA 3~mm observations were taken in the C43-4 and C43-7 configurations of the 12-m array (project code: 2018.1.01420.S; PI: X.\ Lu). The C43-4 observations were carried out in November 2018, and the C43-7 observations were carried out in August 2019, and August and July 2021. A brief summary of the observational logs can be found in Table~\ref{tab:obslogs}. 

The correlators were configured with four spectral windows to cover several important spectral lines. The frequency coverage and an example of observed spectra are presented in Fig.~\ref{fig:spw}. Two 0.9375-GHz spectral windows with a spectral resolution of 0.488~MHz ($\sim$1.67~\kms) were put at central rest frequencies of 86.71~GHz and 88.29~GHz, respectively. Two 1.875-GHz spectral windows with a spectral resolution of 0.976~MHz ($\sim$2.95~\kms) were centered at the rest frequencies of 98.10~GHz and 100.00~GHz, respectively. They were set to cover dense gas tracers including H$^{13}$CN (1--0), H$^{13}$CO$^+$ (1--0), HN$^{13}$C (1--0), HCN (1--0), OCS (8--7), and CS (2--1), shock tracers including SiO (2--1), SO $^3\Sigma$ v=0 3(2)--2(1), CH$_3$OH 2(1,1)--1(1,0)A, and HNCO 4(0,4)--3(0,3), and an ionized gas tracer, H40$\alpha$. We mark several strong lines in Fig.~\ref{fig:spw}, but note that there are plenty of unlabeled complex organic molecule (COM) lines. Analyses of COM lines will be presented in forthcoming papers of the series. A total bandwidth of $\sim$5.6~GHz for the continuum was achieved, although when we imaged the continuum a portion of the bandwidth was flagged to remove spectral lines. 

All the data were calibrated using the standard ALMA pipeline \citep{Hunter2023} implemented in the Common Astronomy Software Applications package \citep[CASA;][]{CASA2022}. Due to the three different epochs of observations, three different CASA versions were used, including 5.4.0-68, 6.1.1-15, and 6.2.1-7. Channels with spectral line emission in the visibility data were visually identified with the \texttt{plotms} task in CASA, and were flagged to keep only line-free channels. For the 20 km/s cloud, 6.3\% of the bandwidth is flagged. This is 4.1\% for Sgr~C and 9.2\% for cloud e. The calibrated and flagged visibility data of the C43-4 and C43-7 configurations were then merged using the \texttt{concat} task. 

\begin{figure*}[!ht]
\centering
\includegraphics[width=0.9\linewidth]{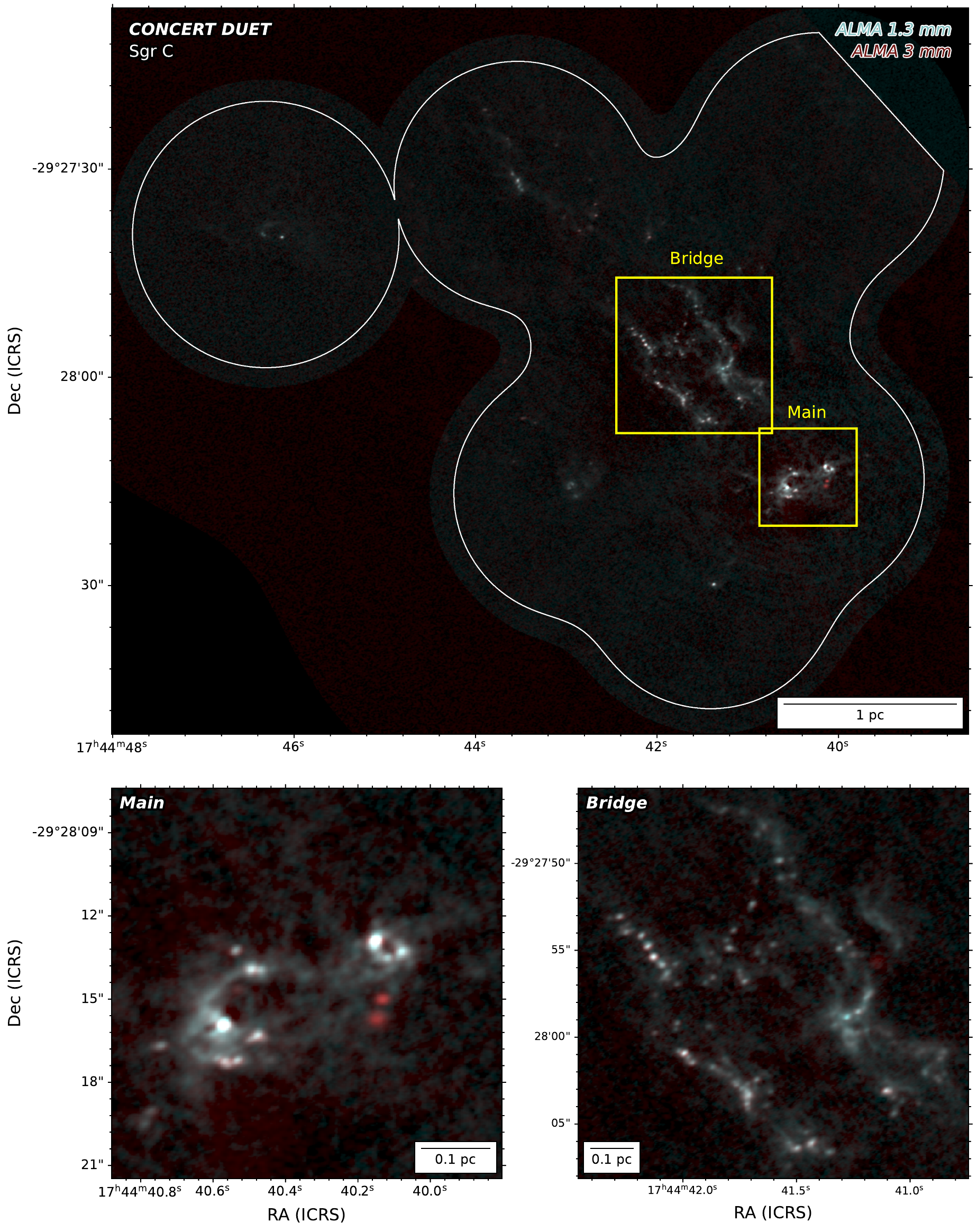}
\caption{Composite two-color map of the ALMA 1.3~mm (cyan) and 3~mm (red) continuum emission in \sgc, with two zoom-in panels of subregions ``Main'' and ``Bridge''. The white loop marks the overlapped area of where primary beam responses are both $>0.2$ for 1.3 and 3~mm bands. An interactive zoom-in webpage of all the three clouds can be found in \href{https://xfengwei.github.io/magnifier/index.html}{https://xfengwei.github.io/magnifier/index.html}.}
\label{fig:dualband_sgrc}
\end{figure*}

Continuum images were reconstructed from the line-free channels, using the multi-term, multifrequency synthesis method (\texttt{mtmfs}) with nterm=2 in the \texttt{tclean} task in CASA 6.5.2-26. Briggs weighting with a robust parameter of 0.5 was used. The multi-scale deconvolution algorithm was adopted with multi-scale parameters of [0, 5, 15, 50, 150] for a pixel size of 0\farcs{04}. The reference frequency of the continuum map produced by \texttt{mtmfs} is 93.6~GHz. For \ctw{} and \sgc{}, where multiple pointings were observed, we produced the mosaicked images. The resulting synthesized beam sizes are on average 0\farcs{32}$\times$0\farcs{22} (equivalent to 2600$\times$1800~au). Due to the lack of short baselines, the data are not sensitive to structures larger than 14\arcsec~($\sim$0.56~pc). This value is larger than that of 1.3~mm by a factor of $2.3$, which potentially bias the spectral index analyses. So, when discussing spectral index in Sect.~\ref{si:low}, we re-cleaned the continuum image by constraining the same $uv$-range at two bands. The 3~mm continuum image is published in \zenodo, and its emission is shown in a red color in Fig.~\ref{fig:dualband_sgrc}, compared to the cyan color showing the 1.3~mm emission. The continuum noise root mean square (rms) in Col.~4 of Table~\ref{tab:dualband_obs} was measured on the image without primary beam corrections with uniform thermal noise by iteratively clipping until convergence at $\pm3\sigma$ around its median \citep{Stetson1987, SExtractor}. 

\subsection{ALMA 1.3~mm}
\label{subsec:obs_band6}

The ALMA 1.3~mm continuum and spectral line data have been reported in \citet{Lu2020Jeans,Lu2021Outflow}. Details of the observation setups and data calibration can be found therein. The ALMA project code is 2016.1.00243.S. We highlight relevant information as follows. We have used the standard CASA pipeline to calibrate the data, and used the \texttt{tclean} task in CASA to image the visibility data after combining data from two array configurations C40-5 and C40-3. The continuum reference frequency is 226.0~GHz. The properties of the band 6 continuum images are summarized in Sect.~\ref{tab:dualband_obs}. The pixel size is 0\farcs{04}, the same as that of the band~3 images. The resulting synthesized beam is 0\farcs{25}$\times$0\farcs{17} (equivalent to 2000~au $\times$ 1400~au). The maximum recoverable scale is 6\farcs{4} ($\sim$0.26~pc). The measured continuum noise rms is 43--45~\ujybeam. The 1.3~mm continuum image is published in \zenodo, and its emission is shown in a cyan color in Fig.~\ref{fig:dualband_sgrc}, compared to the red color showing the 3~mm emission. 

\section{Continuum source catalogs} \label{sec:source}

\subsection{Source extraction strategy} \label{source:extraction}

The continuum images contain multiscale structures ranging from beam scales (0\parcsec2--0\parcsec3) to maximum recoverable scales (7\arcsec--14\arcsec), and exhibit a high intensity dynamic range up to 1000 (c.f.\ Col.~7 in Table~\ref{tab:dualband_obs}). The getsf (v231026) extraction algorithm \citep{Men2021getsf} was adopted to identify and characterize sources, as it can identify sources from multi-scale and also from multiband data. We adopted the getsf definition of sources, which are roundish and significantly stronger than local surrounding fluctuations (of background and noise). The getsf algorithm is recommended for interferometric data, as it effectively handles various backgrounds from filamentary structures to negative bowls originated from imperfect uv-sampling and can separate blended sources in crowded fields like protoclusters \citep{Xu2023SDC335,Cheng2024}. 

Our getsf procedure was designed to take full advantages of the dual-band observations. Details of the settings and the main workflow can be found in Appendix~\ref{app:getsf}. Here we only provide a brief overview. The inputs are the 1.3~mm and 3~mm continuum images at the original resolutions to avoid potential loss of detection due to the smoothing process. The images are without primary beam corrections for a more uniform noise distribution. The extraction mask includes regions where the primary beam response is greater than 0.3 (\texttt{pb}$\geq$0.3), to exclude ambiguous detections on the edges. The minimum source size is set as the original resolution, and the maximum source size was set as five times the 1.3~mm beam size, 1.2$\arcsec$ (approximately 0.05~pc or 10,000~au at the CMZ distance). The choice of maximum source size can not only effectively exclude those extended sources that are not core-like structures, but also preserve the completeness of the core-like source catalog. The main getsf procedure was divided into three rounds: preparing flattened components across the two bands; performing single-band extractions for monochromatic catalogs; and rerunning a dual-band detection to obtain a combined catalog. 

\subsection{Catalog description} \label{source:cat}

Our procedure (see Sect.~\ref{source:extraction}) produced two types of source catalogs: mono-band and dual-band. The mono-band source catalogs (\texttt{mbcat}) provide detections and measurements at individual bands, while the dual-band source catalogs (\texttt{dbcat}) combine sources\footnote{getsf accumulates the clean images (background removed) over dual bands by a natural weight, as is described by Eq.~(29) in \citet{Men2021getsf}.} decomposed from the two bands for detection and measurement. In the following, we introduce the mono-band catalogs in Sect.~\ref{cat:mono} and the dual-band catalogs in Sect.\ref{cat:dual}, and discuss their caveats and complementarity in Sect.~\ref{cat:caveats}. The source catalogs are publicly available at \href{https://doi.org/10.5281/zenodo.15063109}{10.5281/zenodo.15063109}.

\subsubsection{Mono-band source catalogs} \label{cat:mono}

The \texttt{mbcat} catalogs were obtained in the second round (see Appendix~\ref{app:getsf}), where the sources were identified and measured from either 1.3~mm or 3~mm band. A total of 563 sources (including 81 in cloud~e, 224 in Sgr~C, and 258 in \ctw) at 1.3~mm and 330 sources (including 39 in cloud~e, 128 in Sgr~C and 163 in \ctw) at 3~mm were identified. The sources have passed the getsf recommended filtering criteria: 1) goodness values (GOOD) greater than 2 and significance values (Sig) greater than 5; 2) peak and total fluxes higher than twice the local rms noise; and 3) an elliptical aspect ratio $\leq$2 to be roundish \citep{Men2021getsf}. The \texttt{mbcat} sources overlaid on the corresponding continuum images are shared in \zenodo. 

The 1.3~mm and 3~mm \texttt{mbcat}s are presented in Tables~\ref{tab:mbcat01} and \ref{tab:mbcat02}, respectively. The significance value (Sig) is equivalent to the signal-to-noise ratio (S/N) of a source on the scale where it is best detected by getsf. The sources are sorted by the Sig value in the descending order, as indicated by the ID in the table. Throughout the DUET series, we use the nomenclature of ``\{cloud name\}-\{band\}\{\#ID\}'', where ``cloud name'' refers to either of cloude, SgrC, and 20kms. For example, SgrC-1mm\#1 refers to the source with ID of 1 in the 1.3~mm \texttt{mbcat} in the Sgr~C cloud, and 20kms-3mm\#2 refers to the source with ID of 2 in the 3~mm \texttt{mbcat} in \ctw. Following the convention from \citet{Motte2018NA}, Sig greater than 7 corresponds to a robust detection. There are 437 and 278 robust detections at 1.3~mm and 3~mm in \texttt{mbcat}s, corresponding to about 77\% and 84\% of the monochromatic detections, respectively. To obtain the intrinsic source sizes, we deconvolved the synthesized beams from the observed sizes by the analytical solution introduced in Appendix~\ref{app:deconv}. The deconvolved FWHM sizes were converted into physical scales at the distance of 8.277~kpc. The peak intensity and total flux were both primary-beam-corrected; the total flux of source is aperture-corrected by the radio beam shape rather than directly summarizing the flux per pixel that belongs to the source. 

\begin{table*}[!ht]
\centering
\caption{1.3~mm mono-band source catalog \label{tab:mbcat01}}
\begin{tabular}{ccccccccc}
\hline
\hline \noalign{\smallskip}
ID & R.A. & Dec. & Sig & $\theta_{\rm maj}\times\theta_{\rm min}$ & PA & Size & $I^{\rm peak}_\nu$ & $S^{\rm tot}_\nu$ \\
 & (J2000) & (J2000) &  & ($\arcsec\times\arcsec$) & (deg) & (au) & (\mjybeam) & (mJy) \\
\hline \noalign{\smallskip}
\multicolumn{9}{c}{20 \kms} \\ 
\hline \noalign{\smallskip}
1 & 17:45:37.47 & -29:03:49.94 & 449.3 & 0.36$\times$0.29 & 60.46 & 2000 & 14.12(0.61) & 47.61(0.90) \\
2 & 17:45:36.33 & -29:05:49.71 & 298.0 & 0.28$\times$0.22 & 130.1 & 1100 & 8.25(0.12) & 18.49(0.16) \\
\hline \noalign{\smallskip}
\multicolumn{9}{c}{Sgr C} \\
\hline \noalign{\smallskip}
1 & 17:44:40.57 & -29:28:15.89 & 861.1 & 0.29$\times$0.26 & 157.1 & 1400 & 39.59(0.77) & 108.67(1.10) \\
2 & 17:44:40.15 & -29:28:12.85 & 645.5 & 0.32$\times$0.21 & 158.4 & 1300 & 33.65(0.37) & 98.13(0.59) \\
\hline \noalign{\smallskip}
\multicolumn{9}{c}{cloud~e} \\
\hline \noalign{\smallskip}
1 & 17:46:47.06 & -28:32:07.14 & 611.9 & 0.32$\times$0.22 & 105.2 & 1300 & 19.25(0.25) & 52.80(0.40) \\
2 & 17:46:48.23 & -28:32:02.05 & 139.9 & 0.28$\times$0.21 & 132.5 & 1000 & 4.13(0.05) & 8.60(0.06) \\
\hline \\
\end{tabular}
\tablefoot{
The 1.3~mm mono-band source catalog is presented. Only the top two rows of each cloud are shown and the complete version can be found in \zenodo. ID: Source index ordered by decreasing significance values (Sig) as defined in the getsf algorithm. R.A. and Dec.: Right ascension and declination. $\theta_{\rm maj}\times\theta_{\rm min}$: measured major and minor FWHM angular sizes. PA: position angle. Size: Intrinsic source size, beam deconvolved. $I^{\rm peak}_\nu$: Flux over beam solid angle at the brightest pixel. $S^{\rm tot}_\nu$: Integrated flux over source solid angle.
}
\end{table*}

\begin{table*}[!ht]
\centering
\caption{3~mm mono-band source catalog \label{tab:mbcat02}}
\begin{tabular}{ccccccccc}
\hline
\hline \noalign{\smallskip}
ID & R.A. & Dec. & Sig & $\theta_{\rm maj}\times\theta_{\rm min}$ & PA & Size & $I^{\rm peak}_\nu$ & $S^{\rm tot}_\nu$ \\
 & (J2000) & (J2000) &  & ($\arcsec\times\arcsec$) & (deg) & (au) & (\mjybeam) & (mJy) \\
\hline \noalign{\smallskip}
\multicolumn{9}{c}{20 \kms} \\ 
\hline \noalign{\smallskip}
1 & 17:45:37.47 & -29:03:49.95 & 250.7 & 0.40$\times$0.31 & 107.3 & 1800 & 2.37(0.05) & 5.86(0.09) \\
2 & 17:45:37.74 & -29:03:46.84 & 153.4 & 0.35$\times$0.31 & 74.43 & 1300 & 1.46(0.03) & 2.89(0.03) \\
\hline \noalign{\smallskip}
\multicolumn{9}{c}{Sgr C} \\
\hline \noalign{\smallskip}
1 & 17:44:40.15 & -29:28:12.86 & 323.1 & 0.35$\times$0.30 & 147.9 & 1100 & 2.85(0.07) & 5.95(0.08) \\
2 & 17:44:40.57 & -29:28:15.90 & 304.6 & 0.36$\times$0.31 & 78.34 & 1300 & 2.70(0.08) & 5.60(0.10) \\
\hline \noalign{\smallskip}
\multicolumn{9}{c}{cloud~e} \\
\hline \noalign{\smallskip}
1 & 17:46:47.06 & -28:32:07.14 & 168.6 & 0.43$\times$0.38 & 131.9 & 2500 & 1.15(0.03) & 3.98(0.05) \\
2 & 17:46:46.21 & -28:31:55.13 & 68.67 & 0.34$\times$0.30 & 131.1 & 1400 & 0.54(0.01) & 1.28(0.02) \\
\hline \\
\end{tabular}
\tablefoot{
The 3~mm mono-band source catalog is presented. Only the top two rows of each cloud are shown and the complete version can be found in \zenodo. ID: Source index ordered by decreasing significance values (Sig) as defined in the getsf algorithm. R.A. and Dec.: Right ascension and declination. $\theta_{\rm maj}\times\theta_{\rm min}$: Measured major and minor FWHM angular sizes. PA: Position angle. Size: Intrinsic source size, beam deconvolved. $I^{\rm peak}_\nu$: Flux over beam solid angle at the brightest pixel. $S^{\rm tot}_\nu$: Integrated flux over source solid angle.
}
\end{table*}

We did not perform any further filtering criteria based on their intensity profiles, to maintain \texttt{mbcat}'s completeness for future studies at individual bands. For example, some diffuse and extended sources can be analogous to nearby prestellar cores with sizes of $\lesssim$10,000~au \citep[see Table~4 of][and references therein]{Xu2024Scarcity}. The intensity profile of these prestellar cores should be much shallower than typically observed in protostellar cores because of a shallower density profile and a positive temperature gradient (colder in the inner part). Due to the missing flux effect, prestellar cores may appear even fainter than expected. 

\subsubsection{Dual-band source catalogs} \label{cat:dual}

\begin{figure*}[!ht]
\centering
\includegraphics[width=0.9\linewidth]{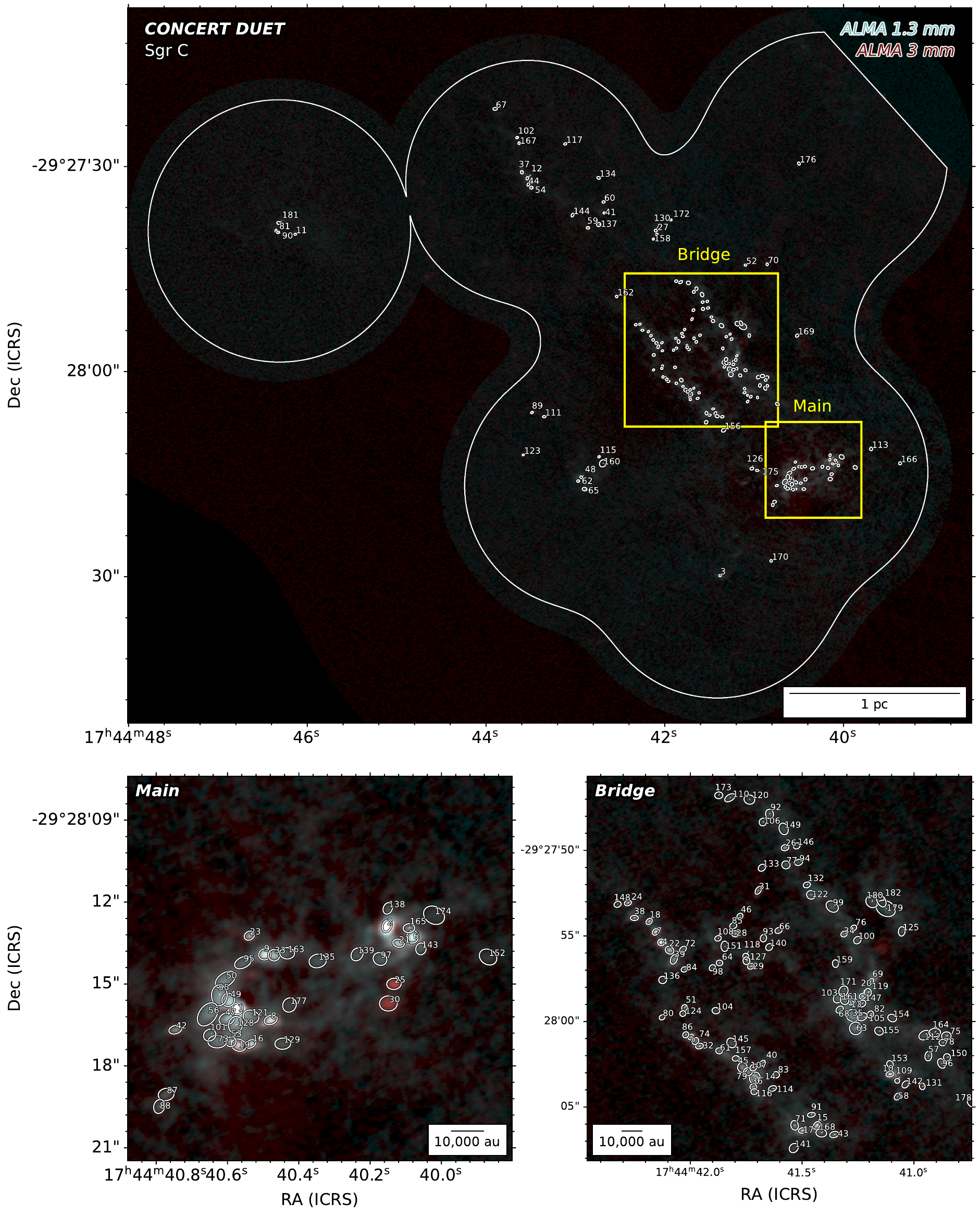}
\caption{Dual-band sources (white ellipses) overlaid on the two-color composite map (cyan: 1.3~mm; red: 3~mm) for Sgr~C. The ``Bridge'' and ``Main'' subregions indicated by yellow boxes are further zoomed in. An interactive zoom-in webpage can be found in \href{https://xfengwei.github.io/magnifier/index.html}{https://xfengwei.github.io/magnifier/index.html}.}
\label{fig:dbcat_SgrC}
\end{figure*}

The \texttt{dbcat} catalogs were obtained in the third round of getsf procedure (see Appendix~\ref{app:getsf}), where sources were identified and measured with the combination of both 1.3~mm and 3~mm continuum emissions. Since the FoVs of the two bands only partially overlap, the initial \texttt{dbcat}s include sources outside the \texttt{pb=0.3} mask of one of the bands. We removed such sources to obtain the cleanest source sample within the overlapping FoVs. Finally, a total of 450 sources (including 62 in cloud~e, 182 in Sgr~C, and 206 in \ctw) were included in the \texttt{dbcat}; they are listed in Table~\ref{tab:dbcat}. The complete version of \texttt{dbcat} is shared in \zenodo. Throughout the paper, we use the nomenclature of ``\{cloud name\}-\{db\}\#ID''. For example, SgrC-db\#1 refers to the source with ID of 1 in the \texttt{dbcat} in the Sgr~C cloud. As an example, the spatial distribution of the identified sources in the Sgr~C cloud is shown in Fig.~\ref{fig:dbcat_SgrC}. The complete version for all the three clouds are shared both in an interactive mode (\href{https://xfengwei.github.io/magnifier/index.html}{https://xfengwei.github.io/magnifier/index.html}) and in \zenodo. 

The \texttt{dbcat} sources have undergone a similar filtering criteria as the \texttt{mbcat} ones, but the goodness and significance are combined from both 1.3~mm and 3~mm. Therefore, some \texttt{dbcat} sources could have a small Sig value in one band because of a significant detection in the other band. For example, SgrC-db\#30 is marginally detected at 1.3~mm but filtered by Sig and goodness criteria, while it has S/N$>$65 at 3~mm. Another example is SgrC-db\#106, which has a S/N$>$17 detection at 1.3~mm but marginal detection at 3~mm. In the complete version of Table~\ref{tab:dbcat}, the monochromatic Sig values at 1.3~mm and 3~mm are also given. 

The \texttt{dbcat} can also resolve crowded sources thanks to the higher angular resolution at 1.3~mm. We refer to the cloud~e-Main as an example. We identified cloude-1mm\#1, 15, 21, 23, 28, 29, 55, and 61 from a clustered region at 1.3~mm, but only cloude-3mm\#1, 16, and 22 at 3~mm due to its coarser resolution. However, all eight sources were successfully separated and identified in the \texttt{dbcat}.

Source sizes and fluxes at each band were measured within individual monochromatic footprints, so the sizes can differ between the two bands. Two data columns for 1.3~mm and 3~mm, respectively, were include in the catalog, each of which contains the same parameters as the \texttt{mbcat}s. The intrinsic source sizes of both bands were obtained by deconvolving the synthesized beams from the observed sizes (see Appendix~\ref{app:deconv}). In Table~\ref{tab:dbcat} listed are the major and minor FWHM sizes, position angle, intrinsic source size, peak intensity, and integrated flux over the source.

\subsubsection{Completeness and robustness of the catalogs} \label{cat:caveats}

As is shown by solid orange and yellow lines in Fig.~\ref{fig:infrared}, the FoVs at 1.3~mm and 3~mm only partially overlap. The \texttt{mbcat}s contain all the detections within the FoVs at the individual bands, while the \texttt{dbcat} only includes sources appearing in the overlapped FoVs of the two bands. In this sense, the \texttt{mbcat} ensures the completeness of the sources within the FoV at each individual band. 

Meanwhile, \texttt{dbcat} considers both bands simultaneously for detection, raising the robustness of source detection. For example, the mass sensitivity of optically thin dust at 20~K at 1.3~mm is about four times better than 3~mm, so a faint dusty source can have a robust detection $>7\sigma$ at 1.3~mm but less than $2\sigma$ at 3~mm. In this case, the 1.3~mm \texttt{mbcat} will include this source, but the \texttt{dbcat} will not. Therefore, the 1.3~mm \texttt{mbcat} catalog is more complete than the \texttt{dbcat} at the low-flux end. In turn, sources in the \texttt{dbcat} are more robust, because they are verified in both bands. This is particularly important for interferometric images, where there can be sidelobes or artifacts from cleaning. Compared to the \texttt{mbcat}, the \texttt{dbcat} combines images to filter out false detections, resulting in a smaller but more robust source catalog. 

\begin{figure*}[!t]
\centering
\includegraphics[width=0.96\linewidth]{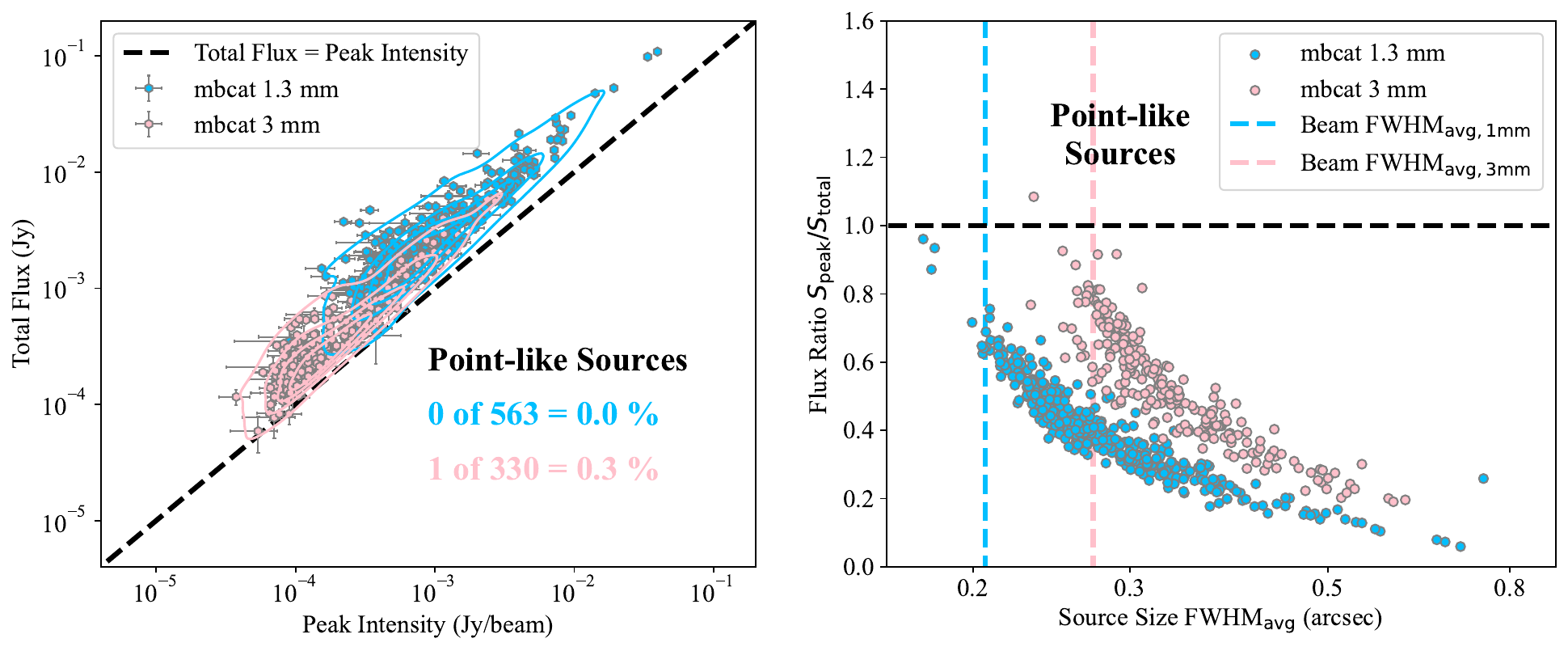}
\includegraphics[width=0.96\linewidth]{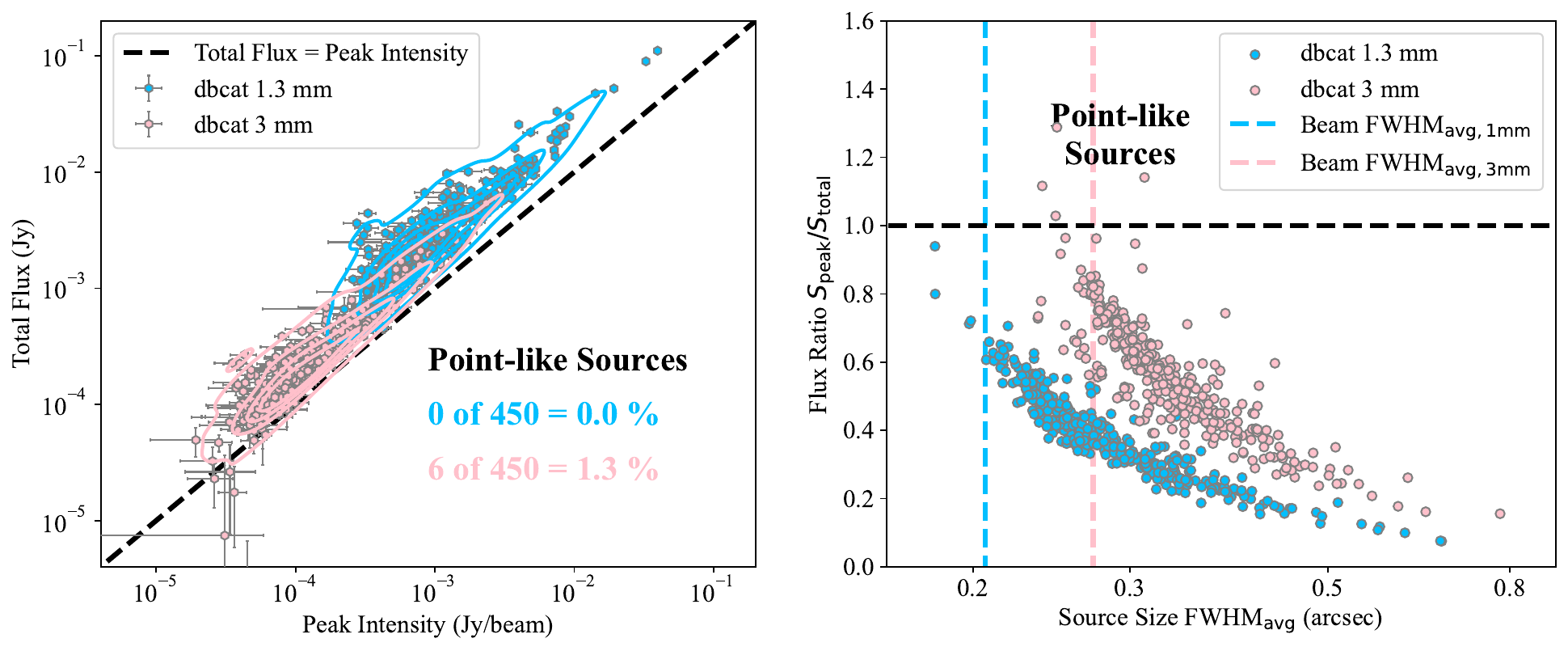}
\caption{\texttt{Left}: Measured total flux versus peak flux shown by KDE (contours). \texttt{Right}: Ratio of the flux per beam at the brightest pixel to the total integrated flux ($S_{\rm peak}/S_{\rm total}$) versus source size FWHM$_{\rm avg}$. The blue and pink data points show the 1.3~mm and 3~mm sources respectively. The dashed black lines mark where peak intensities equal total fluxes, i.e., true point-like sources. The dashed blue and pink lines mark the beam size of 1.3~mm and 3~mm observations, respectively. The top and bottom panels show the sources from \texttt{mbcat} and \texttt{dbcat}, respectively. \label{fig:flux}}
\end{figure*}

However, it should be noted that if a source is marginally detected and filtered out in one band but has a robust detection in another, the \texttt{dbcat} will contain it. It usually happens in two cases. The first case is also for optically thin dust core mentioned above, but the difference happens when a 1.3~mm detection reaches as much as 20$\sigma$ but 3~mm is still below 5$\sigma$. As such, the combined \texttt{dbcat} effect is that the significant 1.3~mm detection confirms the marginal detection at 3~mm, otherwise these sources could have been missed at 3~mm \texttt{mbcat}. As is shown in the two left panels in Fig.~\ref{fig:flux}, the \texttt{dbcat} sources have detections with fluxes down to 0.01~mJy at 3~mm, about one order of magnitude fainter compared to \texttt{mbcat}. The second case is that when a source evolves into an UC\hii{} region, the ionized gas produces strong free-free emission which flattens the spectral energy distribution at the millimeter wavelength. In this case, the 3~mm emission can be strongly enhanced, while the dust emission at 1.3~mm can be even weaker due to dust depletion by sputtering or sublimation and therefore becomes even more undetectable \citep{Hoare1991,Inoue2002,Draine2011}. For example, in Sgr~C-Main, there are two significant detections SgrC-3mm\#12 (Sig$\sim$140) and SgrC-3mm\#17 (Sig$\sim$101), and are either marginally detected as SgrC-1mm\#211 (Sig$\sim$5) or not detected at all. But in the \texttt{dbcat}, both of them are detected as SgrC-db\#25 (combined Sig$\sim$83) and SgrC-db\#30 (combined Sig$\sim$65), respectively. To summarize, \texttt{mbcat}s are optimized for monochromatic completeness, while \texttt{dbcat} has improved robustness by dual-band cross check.

\section{Intrinsic source size} \label{sec:size}

Recent interferometric observations on the order of thousands of astronomical units have detected a large number of beam-like sources in distant and massive protoclusters \citep[e.g.,][]{Liu2019G33, Motte2022, Xu2024Protocluster}. After deconvolution, their intrinsic sizes are generally smaller than the beam. A question arises whether they are objects with a uniform surface density across $\sim$1000 au or compact disks surrounded by less dense envelopes. Our DUET source catalog provides a large number of beam-like sources in a uniform spatial resolution. In Sect.~\ref{size:deconvolve}, we estimate the intrinsic source size and orientation after deconvolution, demonstrating that these sources are mostly unresolved. Furthermore, in Sect.~\ref{size:flux}, we use flux measurements to argue that the sources cannot be, or at least are not solely, point-like. Instead, they have finite source sizes but are still smaller than the beam. 

\subsection{Source deconvolution} \label{size:deconvolve}

\begin{figure*}[!ht]
\centering
\includegraphics[width=0.49\linewidth]{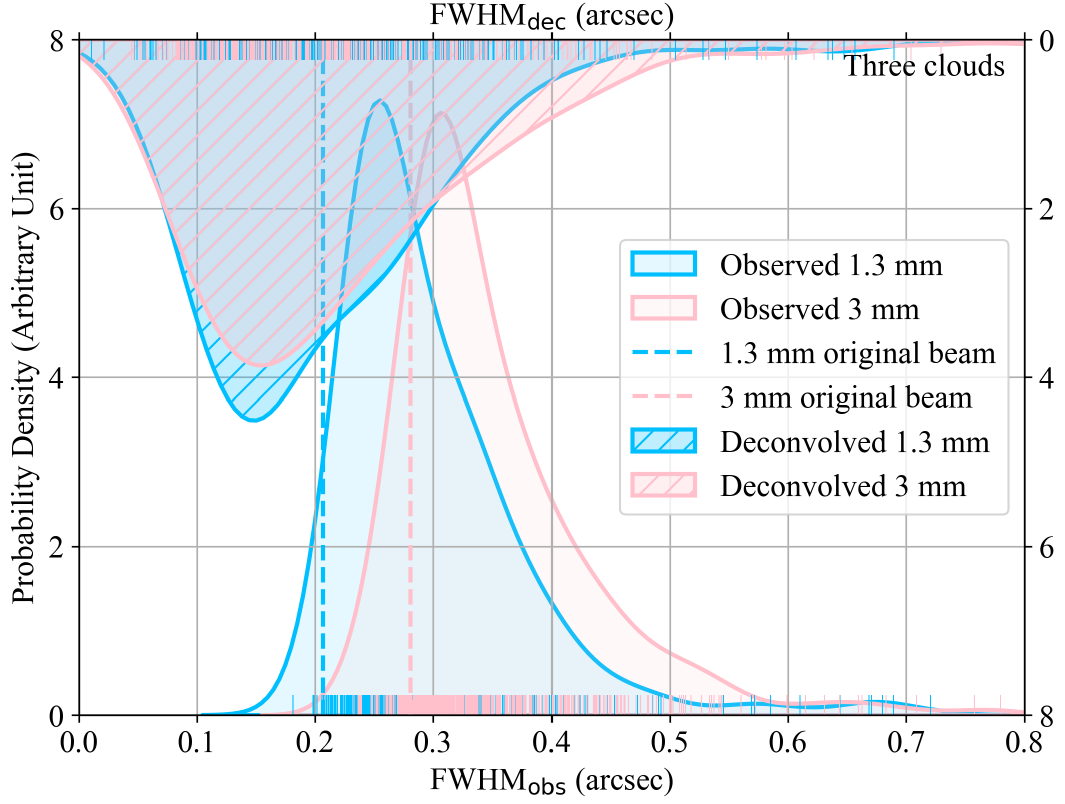}
\includegraphics[width=0.49\linewidth]{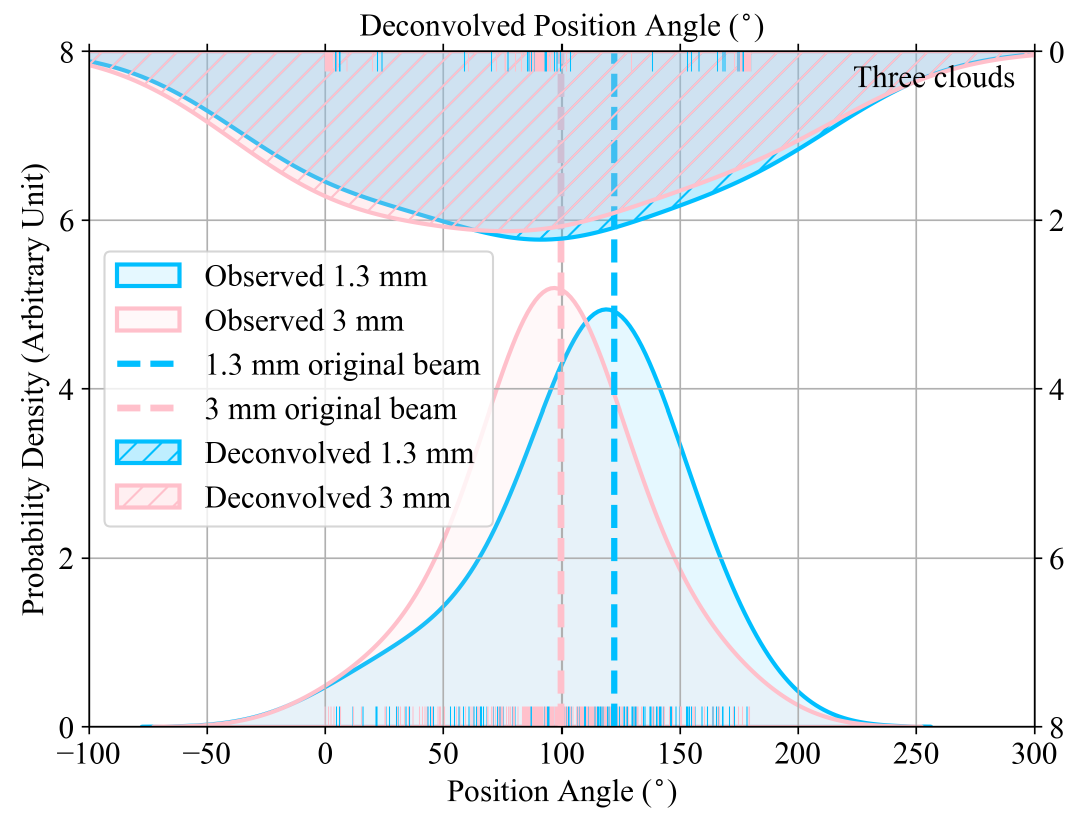}
\caption{\texttt{Left}: Probability distribution of the geometric mean source size FWHM directly measured from observations (FWHM$_{\rm obs}$) and deconvolved from observations (FWHM$_{\rm dec}$), shown by KDE. FWHM$_{\rm obs}$ and FWHM$_{\rm dec}$ are presented at the bottom and the top, respectively. \texttt{Right}: Probability distributions of measured and deconvolved position angle are shown by KDE at the bottom and the top, respectively. The blue and pink colors represent the 1.3~mm and 3~mm, respectively. The vertical dashed lines represent the beam sizes and position angles. Both panels show the \texttt{dbcat} measurements from all the three clouds.}
\label{fig:deconvolve}
\end{figure*}

The measured source size is the result of the intrinsic source size convolved with the beam size. Using the analytical method introduced in Appendix~\ref{app:deconv}, we can directly obtain the intrinsic major and minor axes as well as the position angle of the sources. We take \texttt{dbcat} as an example. As is shown in the left panel of Fig.~\ref{fig:deconvolve}, most measured source sizes are larger than the beam size for both bands. After beam deconvolution, 77\% sources have intrinsic sizes smaller than the beam size; these are referred to as marginally resolved sources. The distributions of deconvolved sizes of the 3~mm and 1.3~mm sources are more consistent between the two bands than the distributions of measured sizes, which indicates that the deconvolution successfully recovers the intrinsic source size. In the right panel, the observed source position angles show a clear concentration at the beam position angle, while the deconvolved position angles have a more scattered distribution. We interpret this as the convolution effects: the intrinsic source sizes should be smaller than or comparable to the beam size, so the beam-convolved position angles have a consistent beam-like direction. Once deconvolved from the beam, the tendency diminishes naturally. 

\subsection{Flux measurement} \label{size:flux}

Flux measurement serves as an additional metric to constrain the source size. If a source is truly point-like, the brightest pixel should contain all the flux from the source. Conversely, if a source has a spatial extent, the brightest pixel contains only a portion of the flux. In Fig.~\ref{fig:flux} left, we present the total flux versus peak intensity. The ratio of total flux to peak intensity is compared to the FWHM$_{\rm avg}$ of the source size (defined as the geometric average of the observed major and minor FWHM axes) in the right panels. The upper and lower panels show measurements in the two catalogs, \texttt{mbcat} and \texttt{dbcat}, respectively. In the left panels, the peak intensity and total flux are tightly correlated, indicating that these sources are mostly concentrated without extended structures. In the right panels, there is a clear decreasing global trend in the total-flux-to-peak-intensity ratios with increasing observed source sizes, as expected for extended sources that have higher total flux values than peak intensity values. There are some cases when peak intensity value is higher than total flux value, which appears intuitively wrong: this happens when there is a noisy background or the source is near an interferometric negative bowl, causing the total flux to be even lower than the peak intensity if these negative pixels are included.

In the \texttt{mbcat}s, all the observed source sizes at 1.3~mm are larger than the beam size, whereas only one at 3~mm is point-like. In the \texttt{dbcat}, more 3~mm sources that are faint and blended can be identified with the help of 1.3~mm, including seven point-like sources at the 3~mm low flux end. Comparing between the top left and bottom left panels in Fig.~\ref{fig:flux}, we find that the measured source fluxes in \texttt{dbcat}, especially at 3~mm, show a higher flux dynamic range than in \texttt{mbcat}. The limited flux dynamic range of the 3~mm sources is a direct result of the image sensitivity. However, when combined with 1.3~mm images, the detection sensitivity for sources at 3~mm can be improved from 0.1~mJy ($\sim$10$\sigma$) down to 0.01~mJy ($\sim$1$\sigma$). As is discussed in Sect.~\ref{cat:caveats}, these faint 3~mm sources would probably be missed in monochromatic detection due to their low S/Ns, which highlights the advantage of the \texttt{dbcat} compared to the \texttt{mbcat}s. 

\section{Spectral indices} \label{sec:si}

The spectral index is defined as
\begin{equation} \label{eq:SI}
\alpha \equiv \frac{\log(I_1/I_2)}{\log(\nu_1/\nu_2)},
\end{equation}
where $I_i$ is the monochromatic intensity or brightness at the frequency $\nu_i$. In this work, we are discussing the inter-band spectral index where its subscripts ($i = 1, 2$) represent 1.3~mm and 3~mm, respectively. By solving the equation of radiation transfer without dust scattering yields
\begin{equation} \label{eq:graybody}
    I_\nu = B (\nu,T) \left(1 - e^{-\tau_\nu}\right) = B (\nu,T) \left(1 - e^{-\kappa_\nu \Sigma}\right).
\end{equation}
where $B (\nu, T)$ is the monochromatic blackbody emission at the temperature $T$,
\begin{equation} \label{eq:blackbody}
    B(\nu,T) = \frac{2h\nu^3}{c^2}\left(e^{h\nu/kT} - 1\right)^{-1},
\end{equation}
and $\tau_\nu$ is the monochromatic dust optical depth and can be expressed by the product of the monochromatic dust absorption opacity, $\kappa_{\nu}$, and the dust surface density, $\Sigma$. 

We discuss $\alpha$ in different $\tau_\nu$ regime. In the optically thin case, Eq.~(\ref{eq:alphathin}) gives $\alpha=3.8$ assuming a dust emissivity spectral index $\beta=1.8$\footnote{The value is fitted from dust-ice mental models as described in \citet{Ossenkopf1994}}. However, recent ALMA observations of high-mass protostellar cores have shown surface densities $\Sigma\gtrsim50$~g~cm$^{-2}$ within $\lesssim2000$~au scale \citep[e.g.,][]{Motte2018AR, Traficante2023, Xu2024Protocluster, Xu2024QUARKS}, which makes dust optically thick. We adopt the opacity model of protostellar dust grain with ice mantle (see Fig.~\ref{fig:opacity}) in the $10^6$--$10^8$~\cc{} density regime \citep{Ossenkopf1994} and obtain $\kappa_{\rm 1.3~mm}=0.9 \pm 0.1$~cm$^2$/g and $\kappa_{\rm 3~mm}=0.18 \pm 0.02$~cm$^2$/g. Assuming a gas-to-dust ratio of $R_{\rm gd}=100$, the dust optical depths are then obtained as $\tau_{\rm 1.3~mm}=0.5$ and $\tau_{\rm 3~mm}=0.1$. As $\kappa_{\nu}\Sigma$ increases, $I_{\nu}$ converges toward $B(\nu, T)$, and the spectral index goes to 2.0 according to Eq.~(\ref{eq:alphablack}). 

\subsection{Low spectral indices versus low brightnesses} \label{si:low}

\begin{figure*}[!ht]
\centering
\includegraphics[width=0.8\linewidth]{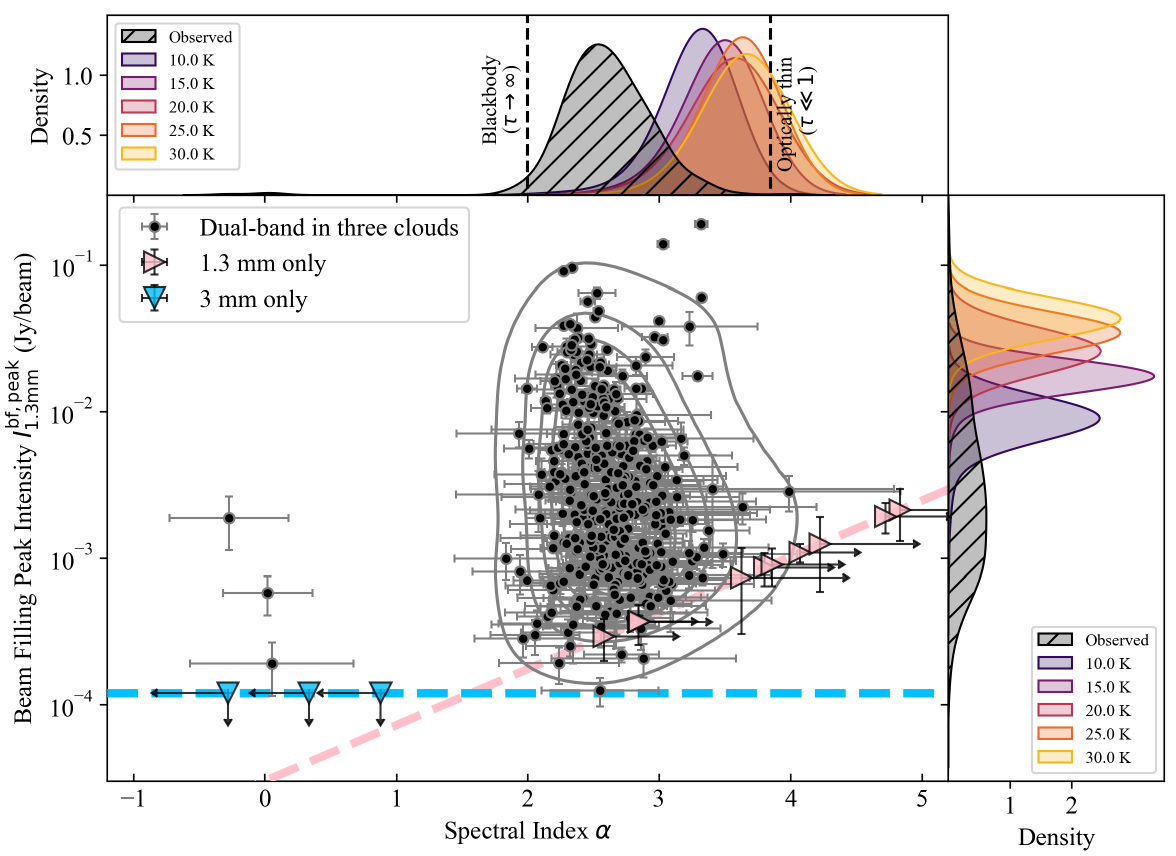}
\caption{Beam-filling peak intensity at 1.3~mm ($I^{\rm bf, peak}_{\rm 1.3~mm}$) versus dual-band spectral index ($\alpha$). The black circles with error bars show the dual-band detections in the common-beam images. Their KDE is outlined by gray contours. The dashed blue and pink lines show the $3\sigma$ detection limits at 1.3 and 3~mm, respectively. The triangles in corresponding colors show those only detected in individual bands. \texttt{Top:} KDE of observed spectral index shown in gray shade. Two special cases including optically thin ($\alpha\simeq3.8$) and blackbody ($\alpha=2$) emission are marked by vertical dashed black lines. The colored KDEs represent predicted $\alpha$ values from modified blackbody (MBB) emission. \texttt{Right:} Observed $I^{\rm bf, peak}_{\rm 1.3~mm}$ is shown in gray shade. The predicted $I^{\rm bf, peak}_{\rm 1.3~mm}$ from the MBB emission is given in colored KDEs.}
\label{fig:lowSI}
\end{figure*}

To remove the effect of different $uv$ samplings between 1.3~mm and 3~mm ALMA observations, we re-cleaned the 1.3 and 3~mm continuum images to achieve a common beam (see more details in Appendix~\ref{app:cbdual}). We reran the getsf to obtain common-beam versions of \texttt{dbcat}. The common-beam \texttt{dbcat} is specifically used for inter-band spectral index calculations. We note that in our published \texttt{dbcat} (Table~\ref{tab:dbcat}), the source extractions are still from the images with original beams, because common-beam \texttt{dbcat} will miss detections of blended sources due to its coarser resolution. 

The spectral indices were calculated from common-beam \texttt{dbcat} using Eq.~(\ref{eq:SI}). We define beam-filling peak intensity, $I^{\rm bf, peak}_{\nu}$, which represents the peak intensity when the beam filling factor is unit. For sources with deconvolved sizes smaller than the beam size, a beam filling factor of $\theta_{\rm deconv}^2 / (\theta_{\rm deconv}^2 + \theta_{\rm beam}^2)$ is divided from the measured peak intensities to obtain $I^{\rm bf, peak}_{\nu}$. The correction is convenient for comparison in later discussion. 

In Fig.~\ref{fig:lowSI}, the \texttt{dbcat} sources with reliable spectral indices are displayed in the $I^{\rm bf, peak}_{1.3~mm}$-$\alpha$ plane. The mean and median values of $\alpha$ are $2.6 \pm 0.5$ and $2.6 \pm 0.2$, respectively. As a comparison, we calculated how MBB (i.e., Eq.~(\ref{eq:graybody})) behaves on the $\alpha$-$I$ plane with a given temperature (see deduction in Appendix~\ref{app:isothermal}), which is further referred to as the isothermal track in form of Eq.~(\ref{eq:inverse}). To demonstrate the deviation from MBB emission, $\alpha$ is calculated for the given $I^{\rm bf, peak}$ of observed sources following isothermal track with a temperature from 10 to 30~K and $\beta = 1.8$. The distribution of these $\alpha$ values are shown by color-coded KDEs in the top panel of Fig.~\ref{fig:lowSI}. Significantly, the observed $\alpha$ distribution deviates from the MBB emission with the commonly assumed $\beta=1.8$ and with the assumed beam filling factor of unity. Such deviations account for more than 70\% of the detections. Since 
we observed lower $\alpha$ than MBB expects, we call it `low spectral index'' case. 

We now interpret the deviation in another way. In the right panel of Fig.~\ref{fig:lowSI}, the observed $I^{\rm bf, peak}$ spans a wide range from 0.1 to 100~mJy but peaks at 1~mJy. Following the MBB isothermal track with given low spectral indices by observation, the expected $I^{\rm bf, peak}$ is given in color-coded KDEs. These distributions, which peak at $\gtrsim 10$~mJy, cannot account for the observed values which are significantly lower. We refer to this as the ``low brightness temperature'' case. The two cases discussed above guides us to propose different scenarios and hypotheses in the following sections.

\subsection{Possible explanations} \label{si:explain}

One hypothesis to explain the observed low spectral indices and low brightnesses is that the flux densities are dominated by spatially compact and optically thick objects that are severely beam diluted. For example, the flux densities may be dominated by embedded class 0/I YSOs. In Sect.~\ref{explain:yso}, we explore what properties of such compact sources may be in order to explain the observed spectral indices and brightnesses. On the other hand, the presence of $\gtrsim0.1$~mm sized dust grains can also lead to attenuated $I_{\nu}$ and lower spectral indices, due to the high dust scattering opacity associated with larger grains. In case flux densities are not necessarily dominated by embedded class 0/I YSOs in some sources, in Sect.~\ref{explain:dust} we explore the extent of grain growth needed to explain the observed spectral indices and brightnesses. Finally, sources associated with free-free emission typically exhibit lower spectral indices, as is discussed in Sect.~\ref{explain:ff}. 

We note that the three hypotheses are not mutually exclusive. For example, with the presence of large dust grains, embedded class 0/I YSOs with larger angular scales (i.e., not severely beam diluted) can be consistent with the observations, as the attenuation due to both dust scattering and beam dilution helps explain the observed low brightnesses. 

\subsubsection{Beam diluted class 0/I YSOs} \label{explain:yso}

The hypothesis is motivated by recent ALMA long-baseline observations of CMZ clouds, revealing a large population of hundred-au scale compact sources \citep{Budaiev2024, Zhang2025disk}. Combining with the low brightness (Sect.~\ref{si:low}) and the small sizes discussed in Sect.~\ref{size:deconvolve}, our hypothesis is that these low-$\alpha$ sources could be unresolved class 0/I YSOs, of which the brightness temperature is significantly beam-diluted when being observed with a large beam size. 

\begin{figure}[!ht]
\centering
\includegraphics[width=0.9\linewidth]{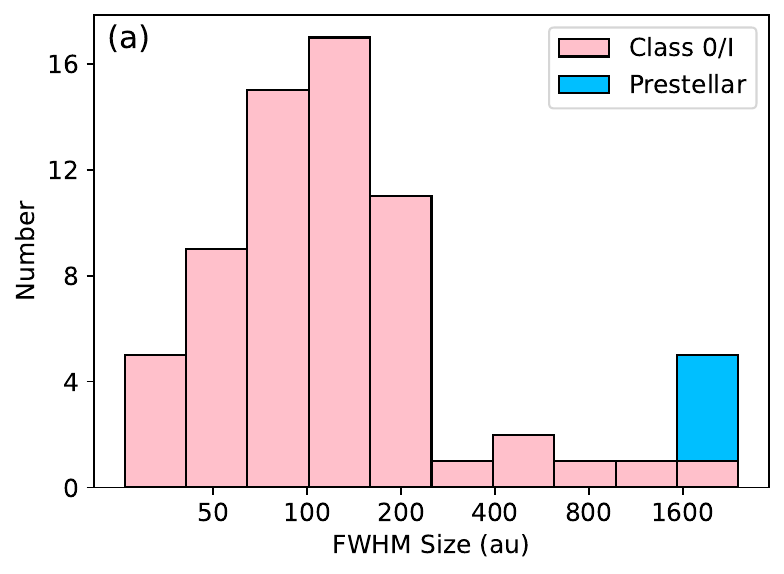}
\includegraphics[width=0.9\linewidth]{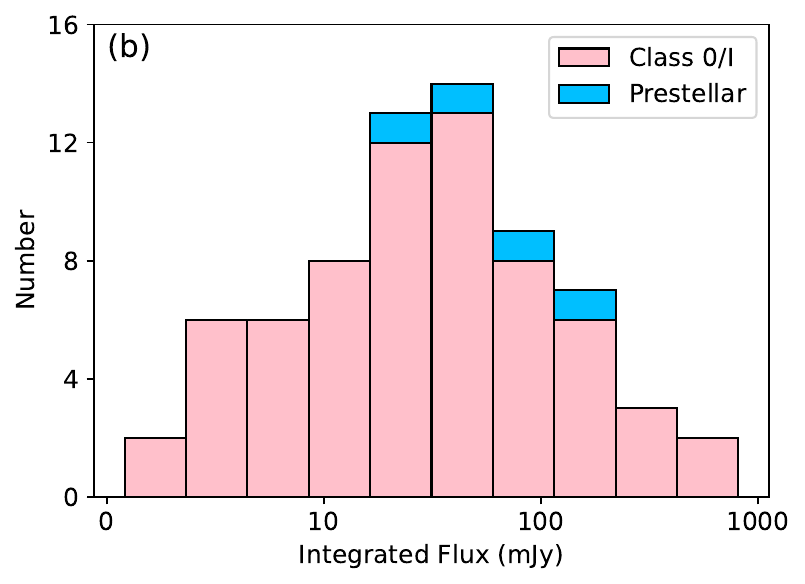}
\caption{Distributions of (a) the size FWHM and (b) the integrated flux of the ALMASOP sources. The pink and blue histograms show the cases of class 0/I YSOs and prestellar sources, respectively.}
\label{fig:almasop-cont}
\end{figure}

\begin{figure}[!ht]
\centering 
\includegraphics[width=\linewidth]{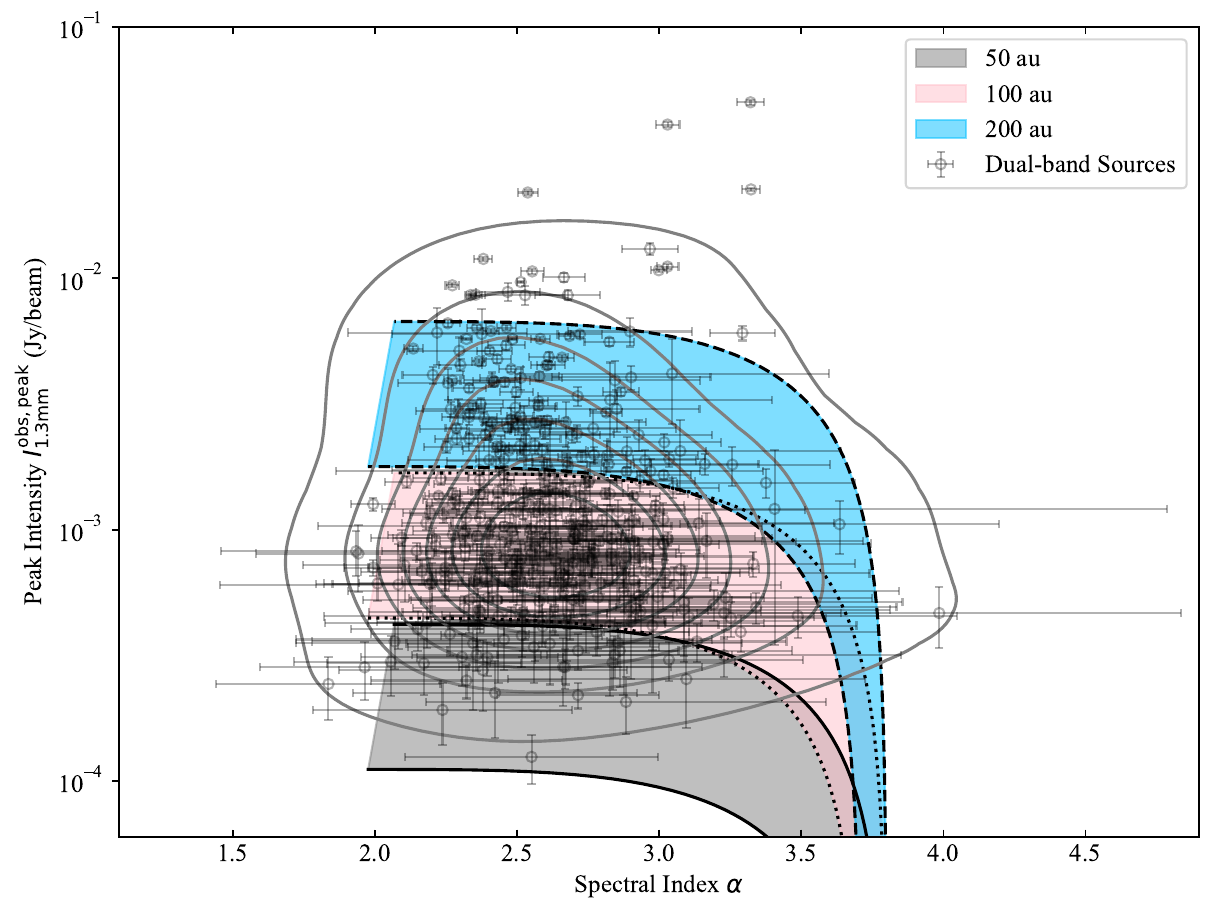}
\caption{Observed peak flux, $I^{\rm obs, peak}_{\rm 1.3~mm}$, versus dual-band spectral index, $\alpha$. The black circles with error bars show the dual-band detections, of which the KDE is shown in gray contours. The self-consistent dust model with micron-sized ($a_{\rm max}=0.01$~mm) dust grains \citep{Liu2019SelfScatter} are shown with shaded regions that cover a temperature range of 6--48~K. The blue, pink, and gray colors represent beam diluted class 0/I YSOs with FWHM sizes of 50, 100, and 200~au, respectively. 
\label{fig:thickYSOs}}
\end{figure}

To verify the hypothesis, we retrieve the 1.3~mm continuum sources from the ALMA Survey of Orion Planck Galactic Cold Clumps \citep[ALMASOP;][]{Dutta2020} with 140~au resolution. A total of 66 sources have been classified as class 0/I YSOs and four are prestellar core candidates. As is indicated in Fig.~\ref{fig:almasop-cont}~(a), the sizes of class 0/I YSOs are between 50 and 200~au (16\% and 85\% percentile for $1\sigma$), while the prestellar cores are usually as large as $>1000$~au, which is consistent with other well-resolved prestellar cores \citep[e.g.,][]{Caselli2022}. We adopted the dust models of 10-micron-sized grains (see description in Sect.~\ref{explain:dust}) and obtained their isothermal tracks with an optically thick diameter of 50, 100, and 200~au. The isothermal tracks with temperature range of 30--100~K are shown in shaded regions, shown with gray, pink, and blue colors in Fig.~\ref{fig:thickYSOs}. These models can cover the observed data points very well. In this case, a large fraction of the observed flux densities can be contributed by the embedded, optically thick low-mass class 0/I YSOs that are much smaller than the sizes of our synthesized beams. 

We further perform simulated DUET 1.3-mm observations of the ALMASOP sources by using the \texttt{simalma} task in CASA. As is shown in Fig.~\ref{fig:almasop-cont}~(b), the integrated fluxes of the ALMASOP class 0/I objects are primarily between 5~mJy and 120~mJy (16\% and 85\% percentile). So we choose four representative ALMASOP targets, G210.97$-$19.33S2, G208.68$-$19.20N3, G208.68$-$19.20N1, and G208.68$-$19.20N2, for simulated observations. The first three are identified as class 0/I YSO candidates, with integrated flux covering from 10 to 100 and to 800~mJy. The last one G208.68$-$19.20N2 is classified as a prestellar core with a 150-au extremely dense nucleus \citep{Hirano2024}. 

\begin{figure*}
\centering
\includegraphics[width=1.0\linewidth]{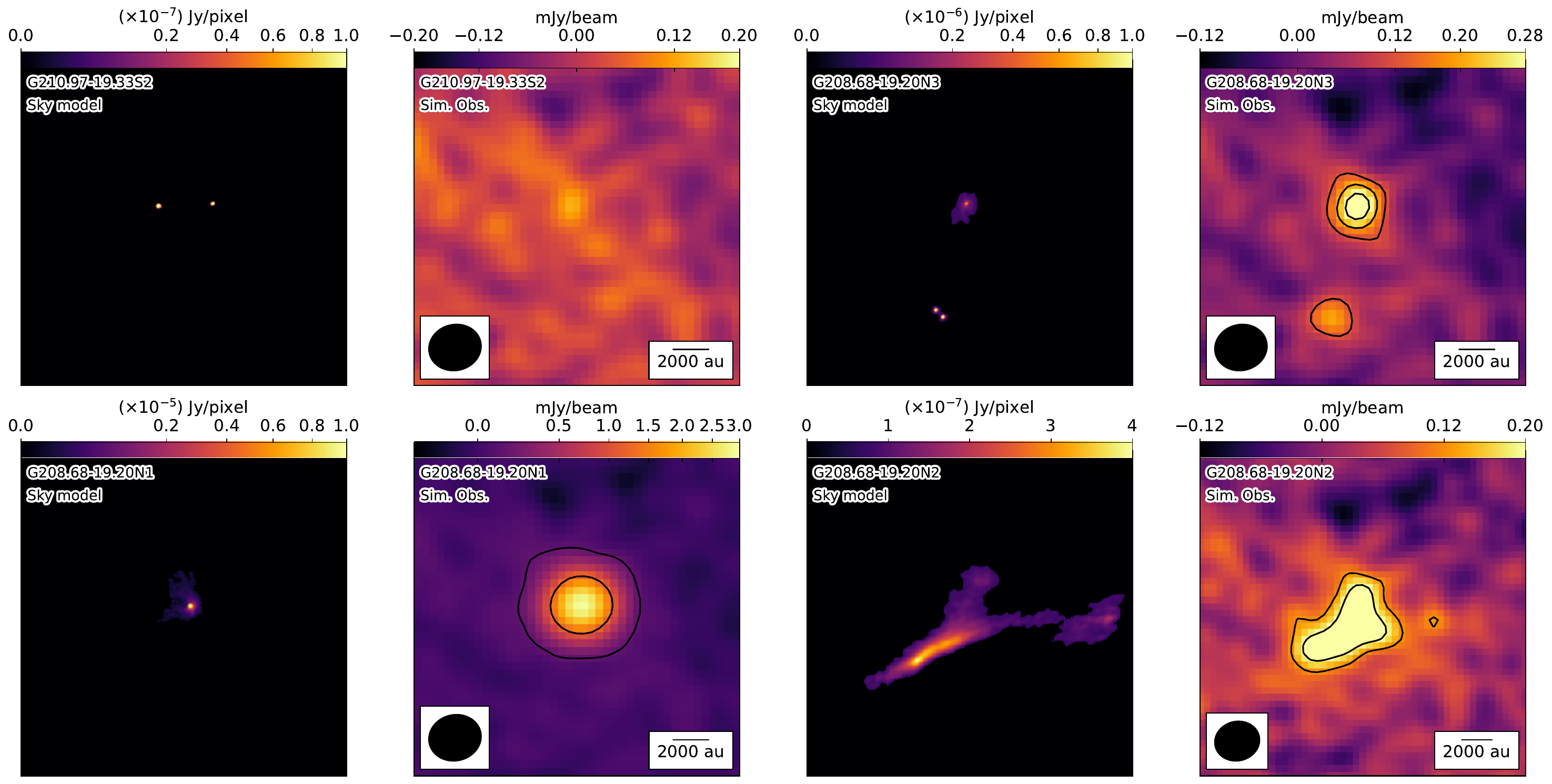}
\caption{Sky models and simulated observations (Sim. Obs.) by \texttt{simalma} are shown. The beam shape and scale bar of 2000~au are marked. No contour levels are shown for G210.97$-$19.33S2, because no pixels brighter than $3\sigma$ are found. The contour levels for G208.68$-$19.20N3 Sim.\ Obs.\ are $3\sigma$, $5\sigma$, and $7\sigma$ ($1\sigma=$0.04~\mjybeam). The contour levels for G208.68$-$19.20N1 Sim.\ Obs.\ are $3\sigma$ and $30\sigma$. The contour levels for G208.68$-$19.20N2 Sim.\ Obs.\ are $3\sigma$ and $5\sigma$. }
\label{fig:simalma}
\end{figure*}

Data from all three arrays combined (C-5, C-2, and ACA) are used to achieve the highest angular resolution as well as to recover diffuse emission and large-scale structures especially for the prestellar core G208.68$-$19.20N2. Details of the ALMASOP observations, array configuration combination, and imaging can be found in \citet{Dutta2020}. We extract the emission above $3\sigma$ for each image as sky models and put them at the same distance as our studied CMZ clouds, which are shown in Cols.~1 and 3 of Fig.~\ref{fig:simalma}. We set simulated observations with the same array configurations (C40-5 and C40-3 in Cycle 4) and the same integration time (270 and 90 seconds) as the actual DUET band 6 observations. The simulated visibility data are cleaned, and the achieved rms noises of cleaned images are 40~\ujybeam, consistent with our DUET images. 

G210.97$-$19.33S2 contains two point-like sources that are identified as class I YSOs, each with an integrated flux of 5--10~mJy. During protostellar evolution, their envelopes have dissipated, resulting in lower 1.3~mm fluxes compared to class 0 objects. These sources represent the low-flux end of the ALMASOP sample (cf.\ Fig.~\ref{fig:almasop-cont} (b)). When being placed at the CMZ, such objects would be challenging to be detected with the DUET observations. A brighter example is the southern binary system in G208.68$-$19.20N3. The system could be marginally detected at the $3\sigma$ level, as shown in the upper panel of Col.~4. So in our DUET observation, such a binary system could make up a marginally detected peak intensity of 0.2~\mjybeam{} in the CMZ. A more extended object is located in the northern part of G208.68$-$19.20N3, appearing as a point-like source with a pseudo-disk or an envelope. In the simulated observation, the flux of the extended structure is covered inside the beam, significantly increasing the peak intensity and yielding a $>7\sigma$ detection. A more extreme case is G208.68$-$19.20N1, but likely harbors a larger protostellar and disk mass. Its simulated images show a peak intensity of 3~\mjybeam, which is consistent with the majority of the actual \texttt{dbcat} sources in Fig.~\ref{fig:lowSI}. Thus, if the low-mass protostellar sources observed in the CMZ clouds are similar to these examples, they are likely at the class 0/I stage, where envelopes have not yet fully dissipated or retain significant remnants. 

Our simulated observation of G208.68$-$19.20N2 also proves the possibility of those sub-milli-Jansky sources to be prestellar cores. In the Col.~4 bottom panel, G208.68$-$19.20N2 is detected above $5\sigma$, with a peak intensity reaching 0.27~\mjybeam, when placed at the distance of the CMZ. Given that G208.68$-$19.20N2 is an extreme case—the brightest and densest prestellar core in the ALMASOP sample—prestellar cores could plausibly account for detections in the range of 0.1 to 0.3~\mjybeam. Examples might include SgrC-1mm\#208 in the SgrC ``Bridge'' and SgrC-1mm\#211 in the SgrC ``Main'' region. However, the possibility that even brighter sources ($>0.3$~\mjybeam) could also be prestellar cores remains to be verified, as the free-fall time in such cases might be even shorter, making these objects exceptionally rare. 

There are more supports from recent higher resolution observations towards the CMZ clouds. For example, there are 199 bright and compact millimeter sources with a median size of 370~au in three massive star-forming clumps in the Sgr~C and \ctw{}, which are suggested to be candidates of protostellar envelopes or disks \citep{Zhang2025disk}. With a bit lower spatial resolution of 500--700~au, \citet{Budaiev2024} have also identified more than 200 sources with similar sizes in the Sgr~B2 region, which are argued to be class 0/I YSO candidates with actively accreting protostars and rotation-supported envelopes or disks.

\subsubsection{Large dust grains} \label{explain:dust}

We discuss the hypothesis of millimeter- and/or centimeter-sized (mm/cm-sized) large dust grains at the scale of thousands of au. Large grains generally have lower millimeter absorption opacity spectral indices $\beta_{\rm mm}$ and therefore lower $\alpha_{\rm mm}$. Theoretically, \citet{Draine2006} shows that for typical dust in the interstellar medium, $\beta\simeq1$ at 1.3~mm if the power-law size distribution of the dust grains ($\mathrm{d}n/\mathrm{d}a \propto a^{-p}$ with $p\simeq3.5$) extends to sizes of $a_{\rm max}\gtrsim3$~mm. Recently, there are multiband observations in protostellar cores in Orion attribute the millimeter shallow spectral indices to larger grain sizes \citep{Schnee2014, Nozari2025}. 

We apply a self-consistent dust grain model which considers both absorption and scattering \citep{Liu2019SelfScatter}. In the model, the effective scattering opacity is $\kappa_{\nu}^{\rm sca,eff}$, which excludes the cross-section in the forward scattering. So, the total opacity $\kappa_\nu^{\rm ext}$ writes
\begin{equation} \label{eq:totkappa}
    \kappa_\nu^{\rm ext} = \kappa_\nu^{\rm abs} + \kappa_\nu^{\rm sca,eff}.
\end{equation}
The relative importance between scattering and absorption is characterized by dust albedo $\gamma \equiv \kappa_\nu^{\rm sca,eff}/\kappa_\nu^{\rm ext}$. In the $\gamma\ll 1$ regime, it reduces to the case of the usual MBB emission in Eq.~(\ref{eq:thindust}). For $\gamma \gtrsim 1$, the effect of dust scattering starts to become noticeable in the dust spectra at optical thick regime \citep{Birnstiel2018}. For most of the probable dust compositions (except for graphitic carbonaceous dust), this occurs when $2\pi a_{\rm max}\gtrsim\lambda$. First, dust scattering leads to attenuation of the observed intensity. Second, it changes the color (i.e., spectral indices), which depends on how the relative importance between $\kappa_\nu^{\rm abs}$ and $\kappa_\nu^{\rm sca,eff}$ varies with frequency. Spectral indices are lowered (as compared to the dust with zero scattering opacity) in the frequency range where albedo is increasing with frequency, which was referred to as anomalous reddening (\citealt{Liu2019SelfScatter}); conversely, spectral indices become larger in the frequency range where albedo is decreasing with frequency.

To demonstrate how the hypothesis explain the observations, we generated isothermal track based on the dust grain model, of which the procedure is explained as follows. We first produced a grid of dust spectra based on various dust temperature, dust column density, and the maximum grain size $a_{\rm max}$. The lowest and highest temperatures are 6~K and 48~K, which correspond to the lower limit of typical interstellar medium (ISM) temperature and Rayleigh-Jeans limit, respectively. We then adopted the dust opacity tables provided in \citet{Birnstiel2018} which assume that dust grains are spherical and compact and are composed of water ice, refractory organics, troilite, and astrophysical silicates. For the grain size, we assumed a power-law grain size distribution with $p\simeq3.5$ between the minimum grain size ($a_{\rm min}=$\,0.1~$\mu$m) and various $a_{\rm max}$. The size-averaged opacity is not sensitive to the assumption of $a_{\rm min}$. Last, surface densities start from $10^{-5}$ to 30~g~cm$^{-2}$ to generate isothermal tracks from optically thin through thick cases, with a given $a_{\rm max}$ and temperature. The Mie theory and the Henyey-Greenstein scattering approximation were considered.

\begin{figure}[!ht]
\centering
\includegraphics[width=\linewidth]{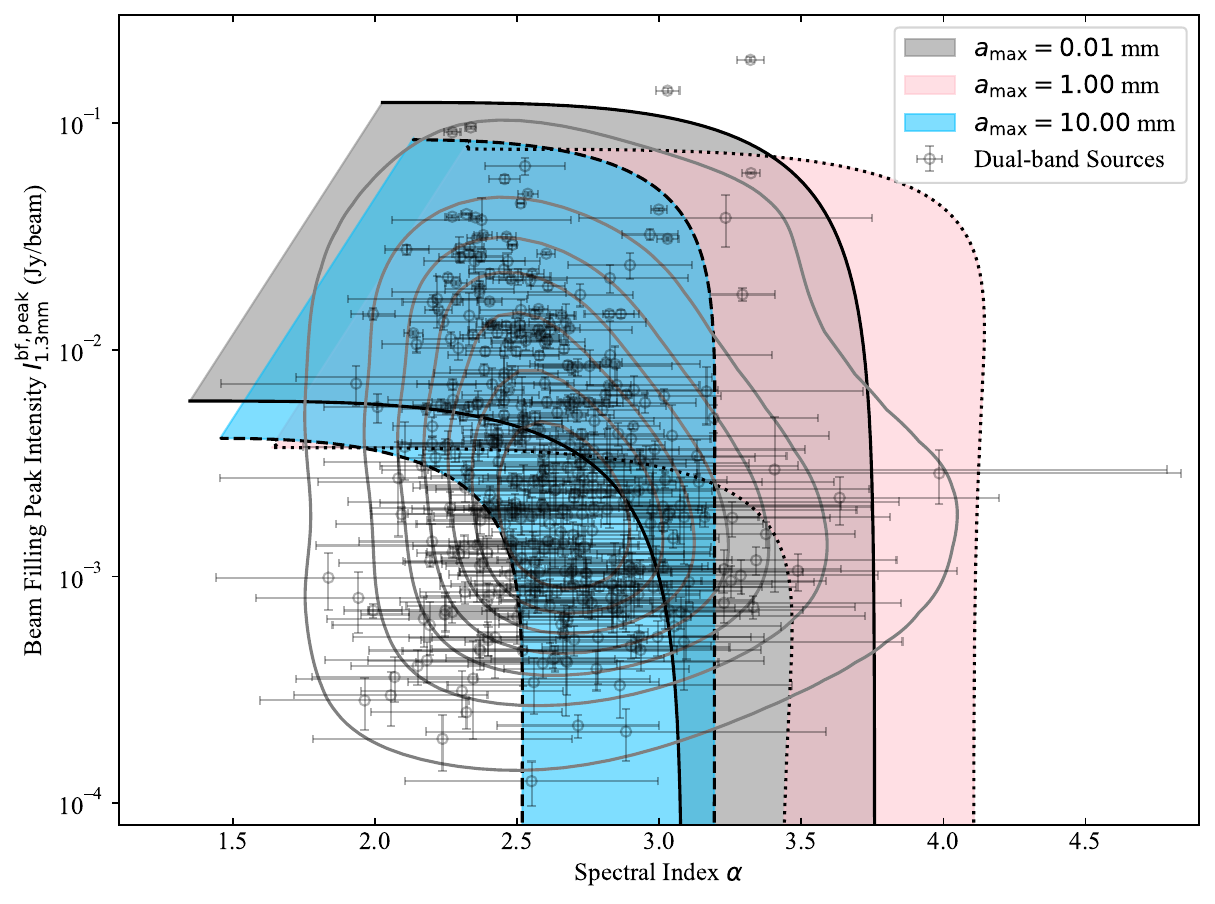}
\caption{Beam filling peak intensity, $I^{\rm bf, peak}_{\rm 1.3~mm}$, versus dual-band spectral index, $\alpha$. The black circles with error bars show the dual-band detections, of which the KDE is shown in gray contours. The self-consistent dust models \citep{Liu2019SelfScatter} are shown with shaded regions that cover the temperature range of 6--48~K. The gray, pink, and blue colors represent cases for $a_{\rm max}=$\,0.01, 1, and 10~mm, respectively. The tracks start with the optically thin case from the bottom where $I^{\rm bf, peak}_{\rm 1.3~mm}$ increases linearly with surface density and $\alpha$ keeps invariant. As it becomes optically thick, $I^{\rm bf, peak}_{\rm 1.3~mm}$ becomes more saturated, and $\alpha$ decreases down to $\simeq2$ as in the case of blackbody. 
\label{fig:largegrains}}
\end{figure}

In Fig.~\ref{fig:largegrains}, the cases for grain sizes $a_{\rm max}=$\,0.01, 1, and 10~mm are shown with gray, pink, and blue shades, respectively. The 6~K and 48~K isothermal tracks are the two extreme temperature cases to form the boundaries of the color-coded regions. Isothermal tracks start with the optically thin case from the bottom where $I^{\rm bf, peak}_{\rm 1.3~mm}$ increases linearly with surface density and $\alpha$ keeps invariant. As it becomes optically thick, $I^{\rm bf, peak}_{\rm 1.3~mm}$ is saturated, and $\alpha$ decreases down to $\simeq2$.

For micron-sized ($a_{\rm max}=0.01$~mm) dust grains, absorption dominates over scattering at millimeter bands. As is shown by the gray region in Fig.~\ref{fig:largegrains}, the spectral index is close to the case of Eq.~(\ref{eq:alphathin}), so the highest possible spectral index is $\sim$3.8. As is discussed in Sect.~\ref{si:low}, with $a_{\rm max}\sim0.01$~mm, a large amount of sources with low spectral indices $\alpha < 3$ cannot be explained. 

For mm-sized ($a_{\rm max}=1$~mm) dust grains, in the optically thin regime, the left and right boundaries of spectral indices shift toward higher values as compared to the case of $a_{\rm max}=0.01$~mm. This is because albedo decreases with frequency at 100--200~GHz when $a_{\rm max}=1$~mm (see Figures 2 and 3 of \citealt{Liu2019SelfScatter}), which leads to higher spectral indices in the regime of $\kappa_\nu^{\rm ext}\Sigma_{\rm dust}\gtrsim0.1$. As is shown by the shaded pink region, the 1~mm sized grain scenario can cover the ``high $\alpha$ bump'' that includes detections with $\alpha \simeq 4$. The high-$\alpha$ bump feature actually serves as an observational hint for the existence of intermediate-sized grains in astrophysical environments. In the optically thick regime, the intensity attenuation is noticeable as compared with the case of $a_{\rm max}=0.01$~mm, which better explains the data points with low $\alpha$ and low $S_{\rm 1.3~mm}^{\rm bf,peak}$.

For cm-sized ($a_{\rm max}=10$~mm) dust grains, in the optically thin regime, the left and right boundaries of spectral indices shift toward lower values as compared to the case of $a_{\rm max}=0.01$~mm, mainly due to the lower $\beta$ values. The intensity attenuation by dust scattering is also noticeable. Both features help explain the sources with low $\alpha$ and low $S_{\rm 1.3~mm}^{\rm bf,peak}$.

Therefore, a range of maximum grain sizes of 0.01--10~mm may explain the $\alpha$ and $I^{\rm bf,peak}_{\rm 1.3~mm}$ values in our observations. We note that there are two additional outliers with high peak intensities in Fig.~\ref{fig:largegrains} which need higher dust temperatures of 60 and 100~K, respectively. These two sources are two massive hot cores in Sgr~C which contain hot molecular species and are reported to be $>100$~K in gas temperature by fitting CH$_3$CN lines \citep{Zhang2025}. Combining the JWST data with other infrared and millimeter data, \citet{Crowe2024JWST} use a YSO radiative transfer model \citep{Zhang2018SED} to estimate the envelope dust temperature to be $\sim100$~K and stellar mass of $\sim$20~$M_{\odot}$. 

The origin of mm/cm-sized grains is uncertain. A typical envelope density of $n_{\rm H} \sim 10^{7-8}$~cm$^{-3}$ is insufficient to form grains of this size. Only at densities of $n_{\rm H} \gtrsim 10^{10}$~cm$^{-3}$ would the formation of such large grains be feasible \citep{Hirashita2009,Navarro-Almaida2024}. According to coagulation models, mm-sized grains observed in protostellar envelopes cannot form locally within the envelopes themselves \citep{Wong2016Origin,Silsbee2022}. In contrast to the inefficient dust growth in the envelope, dust can quickly grow to the order of mm in disks, with a typical timescale of $10^3$ years, thanks to higher densities in disks even in class 0/I YSOs. Numerical calculations have shown that large grains decoupled from gas can be ejected from disks through outflows due to the centrifugal force, enriching dust in envelopes \citep{Wong2016Origin,Tsukamoto2021}. The timescale is much shorter than the lifetime of YSOs, allowing large grains to survive in envelopes. 

In the context of the three CMZ clouds, ubiquitous molecular outflows have been identified to be associated with both low and high-mass protostellar cores \citep{Lu2021Outflow}. This fact supports the idea of outflow-driven large grain enhancement as discussed above. Additionally, the high density environment in the CMZ could further enhance this effects by increasing the drag force on the dust grain ($F_{\rm drag} \propto n_{\rm H_2}$) with certain outflow rates, opening angles, and outflow velocities \citep{Adachi1976Drag,McKee1987GrainDestruction}. 

\subsubsection{Free-free emission} \label{explain:ff}

Massive protostars can ionize the surrounding gas, which produces bremsstrahlung (or free-free) emission by electronic braking. Free-free emission spectrum shows a weak frequency dependence $\propto \nu^{-0.1}$ in the optically thin regime at short wavelengths \citep{Kurtz2005HCHII}, which can flatten the millimeter spectral index by MMB emission. 

To explore the free-free emission contamination, we use the archived JVLA 1.3~cm observations with representative frequency of 23~GHz \citep{Lu2019SFR} with $1\sigma$ rms noise of 0.05--0.15~\mjybeam{} as shown by yellow contours in Fig.~\ref{fig:jvla}. There exists extended filamentary 1.3~cm emission, which is suggested to have non-thermal origins \citep{Clark1976, Ho1985, Lu2003NT, Meng2019, Bally2024}. For example, the one at the northeast corner of \ctw{} has a spectral index of $-$0.4 \citep{Ho1985, Padovani2019}, which cannot be ascribed to free-free emission. In our analyses, two elongated structures (the aforementioned one in \ctw{} and one in cloud~e) are excluded, because they have non-thermal origins. 

At 23~GHz, the thermal free-free emission is optically thin even for the most compact HC\hii{} regions ($>1\times10^9$; turn-over frequency $\sim 15$~GHz) reported before \citep[e.g.,][]{yang2019, yang2023}. 
A spectral index of $-0.1$ is assumed for the free-free emission in order to extrapolate from 1.3~cm to millimeter wavelength in the optically thin regime. We discuss the free-free contamination of all the associated sources in Appendix~\ref{app:ff}. 

However, these cases can only explain the low spectral indices of a limited number of sources that are associated with detectable radio emission. Below the $\sim$0.1~mJy sensitivity at 1.3~cm, there could still be some undetected UCH{\sc ii} regions. Below we quantitatively estimate how much the centimeter continuum contamination would be in order to significantly reduce the spectral index. Considering an optically thin source with a spectral index of 3.8, let the flux at 1.3~mm from pure dust emission be $F_1$, the expected flux at 3~mm would then be $F_2 =(\nu_2/\nu_1)^{3.8}F_1 \simeq F_1/28.5$. Taking the free-free emission flux $F_{\rm ff}$ at both 1.3 and 3~mm into account \footnote{Free-free emission has $<10$\% flux difference at 1.3 and 3~mm due to the weak frequency dependence ($\nu^{-0.1}$).}, the differentiation of the spectral index $\Delta\alpha$ can be rewritten from Eq.~(\ref{eq:SI}) as,
\begin{equation}\label{eq:alphadif}
\begin{aligned}
\Delta \alpha &= \frac{\Delta\log\left[(F_1 + F_{\rm ff})/(F_2 + F_{\rm ff})\right]}{\log(\nu_1/\nu_2)} \\
&= \frac{\Delta\log\left[(28.5+x)/(1+x)\right]}{\log(\nu_1/\nu_2)} = \frac{-31.2 \Delta x}{(28.5+x)(1+x)}\vert_{x\to0} \\
&= -1.1 \Delta x,
\end{aligned}
\end{equation}
where $x \equiv F_{\rm ff}/F_2$. For a dusty source with a flux of 1~mJy at 1.3~mm, $F_2$ is about 0.035~mJy. Therefore, any free-free emission approaching this value could significantly reduce the spectral index. This is well below the sensitivity of the current JVLA data. More sensitive centimeter observations will be helpful to further test this hypothesis.

Given that more than 70\% of the sources exhibit low spectral indices, it is difficult to attribute them entirely to contamination by free-free emission. First, such a large number of massive protostars within a single cluster would conflict with expectations from the IMF, which predicts a decreasing population at the high-mass end. Second, low-mass protostars are unlikely able to contribute to the needed free-free emission either. In nearby molecular clouds \citep[e.g., Corona Australis, Taurus, Ophiuchus, and Perseus;][]{Liu2014,Dzib2015,Coutens2019,Tychoniec2018}, free-free emission from low-mass class 0/I objects typically exhibits flux densities of only a few milli-Jansky or less at centimeter wavelengths. When scaled to the Galactic center, these emissions would become fainter by approximately three orders of magnitude, therefore contributing less than what could result in significant contamination at millimeter bands. To firmly exclude the possibility, the sensitivity needs to be at least (sub-)$\mu$Jy. Indeed, a recent deep JVLA survey of the high-mass star-forming region G14.225$-$0.50 ($d\sim$ 2 kpc) indicates that free-free emission from low-mass class 0/I YSOs in distant star-forming regions is challenging to detect with current facilities (\citealt{Diaz2024}).

\section{Conclusions} \label{sec:conclude}

To characterize the dense core population and star formation activities in the CMZ of the Galaxy, we have carried out the Dual-band Unified Exploration of three CMZ Clouds (DUET) survey of three representative clouds, \ctw, \sgc, and the dust ridge cloud~e. DUET adopts cloud-wide mosaicked observations with ALMA, achieving a comparable resolution of 0\parcsec2--0\parcsec3 (about 2000~au at the distance of the CMZ) and a sky coverage of 8.3--10.4~arcmin$^2$ at the 1.3~mm and 3~mm bands, respectively. In this first paper of the series, we provide an overview of the survey, release the continuum images, and report mono-band and dual-band source catalogs. Our results and discussions are summarized as follows.

Catalog description: Using the getsf source extraction algorithm, we obtain mono-band catalogs (\texttt{mbcat}) of 563 sources at 1.3~mm, including 81 in cloud~e, 224 in Sgr~C, and 258 in \ctw{}, and 330 sources at 3~mm, including 39 in cloud~e, 128 in Sgr~C, and 163 in \ctw{}. Combining the two bands, we obtain dual-band catalogs (\texttt{dbcat}) of 450 sources, including 62 in cloud~e, 182 in \sgc, and 206 in \ctw. The two types of catalogs have their own advantages: \texttt{mbcat} is optimized for monochromatic completeness, while \texttt{dbcat} improves robustness and enables detections of faint and blended sources through a dual-band cross-check. 

Intrinsic source size and internal structure: By comparing the source sizes and position angles before and after deconvolution of the synthesized beam, we find that $\sim77$\% of the sources have intrinsic source sizes smaller than the 2000-au beam. On the other hand, we find that almost all the sources have total fluxes higher than peak intensities, suggesting that they have finite sizes rather than being true point sources. In this context, we refer to them as marginally resolved sources. 

Low spectral indices and possible explanations: \texttt{dbcat} sources enable studies of the inter-band (3~mm--1.3~mm) spectral index ($\alpha$). To avoid the systematic bias from different $uv$ ranges during the ALMA observations, we selected the same $uv$ ranges and re-cleaned the continuum images for both bands. With the same resolution and maximum recoverable scale, we find that $>70$\% of the \texttt{dbcat} sources deviate from MBB, being characterized either as the ``low spectral indices'' problem or as the ``low brightness temperature'' problem. To interpret the problems, we raise three hypotheses and discuss their possibilities.

\begin{enumerate}
\item Class 0/I YSOs. Assuming optically thick disk sizes of 50, 100, and 200~au, the isothermal tracks were calculated based on the same dust model. The theoretical tracks can explain the observations. The hypothesis is further supported by recent long-baseline ALMA observations towards the CMZ clouds that have resolved candidates of envelopes or disks. The hypothesis is also supported by the fact that there are always a large number of beam-like protostellar objects regardless of the spatial resolutions of observations, because they have hardly been resolved.
\item Existence of millimeter or centimeter-sized large grains. With a self-consistent dust radiation transfer model, we calculated isothermal tracks with various grain sizes from 10~$\mu$m to 10~mm. We find that centimeter-sized grains can explain most of sources with low $\alpha$ values, because non-negligible scattering flattens the frequency dependence of effective opacity and reduces $\alpha$ at millimeter band. More importantly, there is a ``high-$\alpha$ bump'' for which a set of detections with $\alpha \simeq 4$ can also be explained. 
\item Contamination of free-free emission. Cross-matching with previous K-band JVLA observations, the low spectral indices of a small amount of sources can be corrected by subtracting free-free emission contamination. Our calculation shows that even if free-free emission is only 30~$\mu$Jy in the K-band, it can still reduce $\alpha$ by a magnitude of 1 for a milli-Jansky source in our sample. However, the number of massive protostars (embedded UCH{\sc ii} regions) needed is infeasible for the IMF of star clusters (even considering a top-heavy IMF), whereas the free-free emission from low-mass YSOs cannot be constrained by current facilities. 
\end{enumerate}
Future work aimed at quantifying the SFR in the CMZ based on the complete protostellar mass spectrum will need to resolve the ambiguity between these possible explanations.

\section*{Data availability}

The 1.3~mm and 3~mm continuum fits files, monochromatic and dual-band continuum source catalogs, as well as the supplementary figures are published at \zenodo. The interactive zoom-in images can be accessed from the webpage \href{https://xfengwei.github.io/magnifier/index.html}{https://xfengwei.github.io/magnifier/index.html}.

\begin{acknowledgements}
We thank the anonymous referee for improving this work greatly. 
This work is supported by the National Key R\&D Program of China (No.\ 2022YFA1603101), the Strategic Priority Research Program of the Chinese Academy of Sciences (CAS) Grant No.\ XDB0800300.
F.W.X. acknowledges the funding from the European Union’s Horizon 2020 research and innovation programme under grant agreement No 101004719 (ORP). F.W.X. would like to thank the China Scholarship Council (CSC) for its support. 
X.L. acknowledges support from the National Natural Science Foundation of China (NSFC) through grant Nos.\ 12273090 and 12322305, the Natural Science Foundation of Shanghai (No.\ 23ZR1482100), and the CAS ``Light of West China'' Program No.\ xbzg-zdsys-202212. 
We acknowledge support from China-Chile Joint Research Fund (CCJRF No. 2211) and the Tianchi Talent Program of Xinjiang Uygur Autonomous Region. CCJRF is provided by Chinese Academy of Sciences South America Center for Astronomy (CASSACA) and established by National Astronomical Observatories, Chinese Academy of Sciences (NAOC) and Chilean Astronomy Society (SOCHIAS) to support China-Chile collaborations in astronomy. 
H.B.L. is supported by the National Science and Technology Council (NSTC) of Taiwan (Grant Nos. 111-2112-M-110-022-MY3, 113-2112-M-110-022-MY3).
Q. Zhang acknowledges the support from the National Science Foundation under Award No. 2206512. 
COOL Research DAO \citep{cool_whitepaper} is a Decentralized Autonomous Organization supporting research in astrophysics aimed at uncovering our cosmic origins.

We thank Alexander Men'shchikov for helpful discussion on the usage of getsf. This research made use of Montage. It is funded by the National Science Foundation under Grant Number ACI-1440620, and was previously funded by the National Aeronautics and Space Administration's Earth Science Technology Office, Computation Technologies Project, under Cooperative Agreement Number NCC5-626 between NASA and the California Institute of Technology.
\end{acknowledgements}

\bibliographystyle{aa}
\bibliography{duet}

\begin{thebibliography}{102}
\expandafter\ifx\csname natexlab\endcsname\relax\def\natexlab#1{#1}\fi

\bibitem[{{Adachi} {et~al.}(1976){Adachi}, {Hayashi}, \&
  {Nakazawa}}]{Adachi1976Drag}
{Adachi}, I., {Hayashi}, C., \& {Nakazawa}, K. 1976, Progress of Theoretical
  Physics, 56, 1756

\bibitem[{{Bally} {et~al.}(2024){Bally}, {Crowe}, {Fedriani}, {Ginsburg},
  {Sch{\"o}del}, {Andersen}, {Tan}, {Li}, {Nogueras-Lara}, {Cheng}, {Law},
  {Wang}, {Zhang}, \& {Zhang}}]{Bally2024}
{Bally}, J., {Crowe}, S., {Fedriani}, R., {et~al.} 2024, arXiv e-prints,
  arXiv:2412.10983

\bibitem[{{Bally} {et~al.}(1987){Bally}, {Stark}, {Wilson}, \&
  {Henkel}}]{Bally1987}
{Bally}, J., {Stark}, A.~A., {Wilson}, R.~W., \& {Henkel}, C. 1987, \apjs, 65,
  13

\bibitem[{{Bally} {et~al.}(1988){Bally}, {Stark}, {Wilson}, \&
  {Henkel}}]{Bally1988}
{Bally}, J., {Stark}, A.~A., {Wilson}, R.~W., \& {Henkel}, C. 1988, \apj, 324,
  223

\bibitem[{{Barnes} {et~al.}(2017){Barnes}, {Longmore}, {Battersby}, {Bally},
  {Kruijssen}, {Henshaw}, \& {Walker}}]{Barnes2017}
{Barnes}, A.~T., {Longmore}, S.~N., {Battersby}, C., {et~al.} 2017, \mnras,
  469, 2263

\bibitem[{{Bastian} {et~al.}(2010){Bastian}, {Covey}, \& {Meyer}}]{Bastian2010}
{Bastian}, N., {Covey}, K.~R., \& {Meyer}, M.~R. 2010, \araa, 48, 339

\bibitem[{{Battersby} {et~al.}(2020){Battersby}, {Keto}, {Walker}, {Barnes},
  {Callanan}, {Ginsburg}, {Hatchfield}, {Henshaw}, {Kauffmann}, {Kruijssen},
  {Longmore}, {Lu}, {Mills}, {Pillai}, {Zhang}, {Bally}, {Butterfield},
  {Contreras}, {Ho}, {Ott}, {Patel}, \& {Tolls}}]{Battersby2020}
{Battersby}, C., {Keto}, E., {Walker}, D., {et~al.} 2020, \apjs, 249, 35

\bibitem[{{Battersby} {et~al.}(2024){Battersby}, {Walker}, {Barnes},
  {Ginsburg}, {Lipman}, {Alboslani}, {Hatchfield}, {Bally}, {Glover},
  {Henshaw}, {Immer}, {Klessen}, {Longmore}, {Mills}, {Molinari}, {Smith},
  {Sormani}, {Tress}, \& {Zhang}}]{Battersby2024}
{Battersby}, C., {Walker}, D.~L., {Barnes}, A., {et~al.} 2024, arXiv e-prints,
  arXiv:2410.17334

\bibitem[{{Bertin} \& {Arnouts}(1996)}]{SExtractor}
{Bertin}, E. \& {Arnouts}, S. 1996, \aaps, 117, 393

\bibitem[{{Birnstiel} {et~al.}(2018){Birnstiel}, {Dullemond}, {Zhu}, {Andrews},
  {Bai}, {Wilner}, {Carpenter}, {Huang}, {Isella}, {Benisty}, {P{\'e}rez}, \&
  {Zhang}}]{Birnstiel2018}
{Birnstiel}, T., {Dullemond}, C.~P., {Zhu}, Z., {et~al.} 2018, \apjl, 869, L45

\bibitem[{{Bonnell} {et~al.}(2001){Bonnell}, {Bate}, {Clarke}, \&
  {Pringle}}]{Bonnell2001}
{Bonnell}, I.~A., {Bate}, M.~R., {Clarke}, C.~J., \& {Pringle}, J.~E. 2001,
  \mnras, 323, 785

\bibitem[{{Bryant} \& {Krabbe}(2021)}]{Bryant2021Review}
{Bryant}, A. \& {Krabbe}, A. 2021, \nar, 93, 101630

\bibitem[{{Budaiev} {et~al.}(2024){Budaiev}, {Ginsburg}, {Jeff}, {Goddi},
  {Meng}, {S{\'a}nchez-Monge}, {Schilke}, {Schmiedeke}, \& {Yoo}}]{Budaiev2024}
{Budaiev}, N., {Ginsburg}, A., {Jeff}, D., {et~al.} 2024, \apj, 961, 4

\bibitem[{{CASA Team} {et~al.}(2022){CASA Team}, {Bean}, {Bhatnagar}, {Castro},
  {Donovan Meyer}, {Emonts}, {Garcia}, {Garwood}, {Golap}, {Gonzalez Villalba},
  {Harris}, {Hayashi}, {Hoskins}, {Hsieh}, {Jagannathan}, {Kawasaki},
  {Keimpema}, {Kettenis}, {Lopez}, {Marvil}, {Masters}, {McNichols},
  {Mehringer}, {Miel}, {Moellenbrock}, {Montesino}, {Nakazato}, {Ott}, {Petry},
  {Pokorny}, {Raba}, {Rau}, {Schiebel}, {Schweighart}, {Sekhar}, {Shimada},
  {Small}, {Steeb}, {Sugimoto}, {Suoranta}, {Tsutsumi}, {van Bemmel},
  {Verkouter}, {Wells}, {Xiong}, {Szomoru}, {Griffith}, {Glendenning}, \&
  {Kern}}]{CASA2022}
{CASA Team}, {Bean}, B., {Bhatnagar}, S., {et~al.} 2022, \pasp, 134, 114501

\bibitem[{{Caselli} {et~al.}(2022){Caselli}, {Pineda}, {Sipil{\"a}}, {Zhao},
  {Redaelli}, {Spezzano}, {Maureira}, {Alves}, {Bizzocchi}, {Bourke},
  {Chac{\'o}n-Tanarro}, {Friesen}, {Galli}, {Harju}, {Jim{\'e}nez-Serra},
  {Keto}, {Li}, {Padovani}, {Schmiedeke}, {Tafalla}, \& {Vastel}}]{Caselli2022}
{Caselli}, P., {Pineda}, J.~E., {Sipil{\"a}}, O., {et~al.} 2022, \apj, 929, 13

\bibitem[{{Cheng} {et~al.}(2024){Cheng}, {Lu}, {Sanhueza}, {Liu}, {Zhang},
  {Galv{\'a}n-Madrid}, {Wang}, {Nakamura}, {Liu}, {Feng}, {Li}, {Jiao},
  {Tanaka}, {Liu}, {Li}, {Luo}, {Gu}, {Lin}, \& {Guzm{\'a}n}}]{Cheng2024}
{Cheng}, Y., {Lu}, X., {Sanhueza}, P., {et~al.} 2024, \apj, 967, 56

\bibitem[{{Chevance} {et~al.}(2025){Chevance}, {Kruijssen}, \&
  {Longmore}}]{cool_whitepaper}
{Chevance}, M., {Kruijssen}, J.~M.~D., \& {Longmore}, S.~N. 2025, arXiv
  e-prints, arXiv:2501.13160

\bibitem[{{Clark} \& {Caswell}(1976)}]{Clark1976}
{Clark}, D.~H. \& {Caswell}, J.~L. 1976, \mnras, 174, 267

\bibitem[{{Coutens} {et~al.}(2019){Coutens}, {Liu}, {Jim{\'e}nez-Serra},
  {Bourke}, {Forbrich}, {Hoare}, {Loinard}, {Testi}, {Audard}, {Caselli},
  {Chac{\'o}n-Tanarro}, {Codella}, {Di Francesco}, {Fontani}, {Hogerheijde},
  {Johansen}, {Johnstone}, {Maddison}, {Pani{\'c}}, {P{\'e}rez}, {Podio},
  {Punanova}, {Rawlings}, {Semenov}, {Tazzari}, {Tobin}, {van der Wiel}, {van
  Langevelde}, {Vlemmings}, {Walsh}, \& {Wilner}}]{Coutens2019}
{Coutens}, A., {Liu}, H.~B., {Jim{\'e}nez-Serra}, I., {et~al.} 2019, \aap, 631,
  A58

\bibitem[{{Crowe} {et~al.}(2024){Crowe}, {Fedriani}, {Tan}, {Kinman}, {Zhang},
  {Andersen}, {Bravo Ferres}, {Nogueras-Lara}, {Sch{\"o}del}, {Bally},
  {Ginsburg}, {Cheng}, {Yang}, {Kendrew}, {Law}, {Armstrong}, \&
  {Li}}]{Crowe2024JWST}
{Crowe}, S., {Fedriani}, R., {Tan}, J.~C., {et~al.} 2024, arXiv e-prints,
  arXiv:2410.09253

\bibitem[{{D{\'\i}az-M{\'a}rquez} {et~al.}(2024){D{\'\i}az-M{\'a}rquez},
  {Grau}, {Busquet}, {Girart}, {S{\'a}nchez-Monge}, {Palau}, {Povich},
  {A{\~n}ez-L{\'o}pez}, {Liu}, {Zhang}, \& {Estalella}}]{Diaz2024}
{D{\'\i}az-M{\'a}rquez}, E., {Grau}, R., {Busquet}, G., {et~al.} 2024, \aap,
  682, A180

\bibitem[{{Draine}(2006)}]{Draine2006}
{Draine}, B.~T. 2006, \apj, 636, 1114

\bibitem[{{Draine}(2011)}]{Draine2011}
{Draine}, B.~T. 2011, {Physics of the Interstellar and Intergalactic Medium}
  (Princeton University Press, 2011. ISBN: 978-0-691-12214-4)

\bibitem[{{Dutta} {et~al.}(2020){Dutta}, {Lee}, {Liu}, {Hirano}, {Liu},
  {Tatematsu}, {Kim}, {Shang}, {Sahu}, {Kim}, {Moraghan}, {Jhan}, {Hsu},
  {Evans}, {Johnstone}, {Ward-Thompson}, {Kuan}, {Lee}, {Lee}, {Traficante},
  {Juvela}, {Vastel}, {Zhang}, {Sanhueza}, {Soam}, {Kwon}, {Bronfman}, {Eden},
  {Goldsmith}, {He}, {Wu}, {Pelkonen}, {Qin}, {Li}, \& {Li}}]{Dutta2020}
{Dutta}, S., {Lee}, C.-F., {Liu}, T., {et~al.} 2020, \apjs, 251, 20

\bibitem[{{Dzib} {et~al.}(2015){Dzib}, {Loinard}, {Rodr{\'\i}guez},
  {Mioduszewski}, {Ortiz-Le{\'o}n}, {Kounkel}, {Pech}, {Rivera}, {Torres},
  {Boden}, {Hartmann}, {Evans}, {Brice{\~n}o}, \& {Tobin}}]{Dzib2015}
{Dzib}, S.~A., {Loinard}, L., {Rodr{\'\i}guez}, L.~F., {et~al.} 2015, \apj,
  801, 91

\bibitem[{{Ferri{\`e}re} {et~al.}(2007){Ferri{\`e}re}, {Gillard}, \&
  {Jean}}]{Ferriere2007}
{Ferri{\`e}re}, K., {Gillard}, W., \& {Jean}, P. 2007, \aap, 467, 611

\bibitem[{{Figer} {et~al.}(1999){Figer}, {Kim}, {Morris}, {Serabyn}, {Rich}, \&
  {McLean}}]{Figer1999}
{Figer}, D.~F., {Kim}, S.~S., {Morris}, M., {et~al.} 1999, \apj, 525, 750

\bibitem[{{Genzel} \& {Townes}(1987)}]{Genzel1987}
{Genzel}, R. \& {Townes}, C.~H. 1987, \araa, 25, 377

\bibitem[{{Ginsburg} {et~al.}(2018){Ginsburg}, {Bally}, {Barnes}, {Bastian},
  {Battersby}, {Beuther}, {Brogan}, {Contreras}, {Corby}, {Darling}, {De Pree},
  {Galv{\'a}n-Madrid}, {Garay}, {Henshaw}, {Hunter}, {Kruijssen}, {Longmore},
  {Lu}, {Meng}, {Mills}, {Ott}, {Pineda}, {S{\'a}nchez-Monge}, {Schilke},
  {Schmiedeke}, {Walker}, \& {Wilner}}]{Ginsburg2018}
{Ginsburg}, A., {Bally}, J., {Barnes}, A., {et~al.} 2018, \apj, 853, 171

\bibitem[{{GRAVITY Collaboration} {et~al.}(2022){GRAVITY Collaboration},
  {Abuter}, {Aimar}, {Amorim}, {Ball}, {Baub{\"o}ck}, {Berger}, {Bonnet},
  {Bourdarot}, {Brandner}, {Cardoso}, {Cl{\'e}net}, {Dallilar}, {Davies}, {de
  Zeeuw}, {Dexter}, {Drescher}, {Eisenhauer}, {F{\"o}rster Schreiber},
  {Foschi}, {Garcia}, {Gao}, {Gendron}, {Genzel}, {Gillessen}, {Habibi},
  {Haubois}, {Hei{\ss}el}, {Henning}, {Hippler}, {Horrobin}, {Jochum}, {Jocou},
  {Kaufer}, {Kervella}, {Lacour}, {Lapeyr{\`e}re}, {Le Bouquin}, {L{\'e}na},
  {Lutz}, {Ott}, {Paumard}, {Perraut}, {Perrin}, {Pfuhl}, {Rabien},
  {Shangguan}, {Shimizu}, {Scheithauer}, {Stadler}, {Stephens}, {Straub},
  {Straubmeier}, {Sturm}, {Tacconi}, {Tristram}, {Vincent}, {von Fellenberg},
  {Widmann}, {Wieprecht}, {Wiezorrek}, {Woillez}, {Yazici}, \&
  {Young}}]{Gravity2022}
{GRAVITY Collaboration}, {Abuter}, R., {Aimar}, N., {et~al.} 2022, \aap, 657,
  L12

\bibitem[{{Hatchfield} {et~al.}(2020){Hatchfield}, {Battersby}, {Keto},
  {Walker}, {Barnes}, {Callanan}, {Ginsburg}, {Henshaw}, {Kauffmann},
  {Kruijssen}, {Longmore}, {Lu}, {Mills}, {Pillai}, {Zhang}, {Bally},
  {Butterfield}, {Contreras}, {Ho}, {Ott}, {Patel}, \&
  {Tolls}}]{Hatchfield2020}
{Hatchfield}, H.~P., {Battersby}, C., {Keto}, E., {et~al.} 2020, \apjs, 251, 14

\bibitem[{{Henshaw} {et~al.}(2023){Henshaw}, {Barnes}, {Battersby}, {Ginsburg},
  {Sormani}, \& {Walker}}]{Henshaw2023Review}
{Henshaw}, J.~D., {Barnes}, A.~T., {Battersby}, C., {et~al.} 2023, in
  Astronomical Society of the Pacific Conference Series, Vol. 534, Protostars
  and Planets VII, ed. S.~{Inutsuka}, Y.~{Aikawa}, T.~{Muto}, K.~{Tomida}, \&
  M.~{Tamura}, 83

\bibitem[{{Heywood} {et~al.}(2022){Heywood}, {Rammala}, {Camilo}, {Cotton},
  {Yusef-Zadeh}, {Abbott}, {Adam}, {Adams}, {Aldera}, {Asad}, {Bauermeister},
  {Bennett}, {Bester}, {Bode}, {Botha}, {Botha}, {Brederode}, {Buchner},
  {Burger}, {Cheetham}, {de Villiers}, {Dikgale-Mahlakoana}, {du Toit},
  {Esterhuyse}, {Fanaroff}, {February}, {Fourie}, {Frank}, {Gamatham}, {Geyer},
  {Goedhart}, {Gouws}, {Gumede}, {Hlakola}, {Hokwana}, {Hoosen}, {Horrell},
  {Hugo}, {Isaacson}, {J{\'o}zsa}, {Jonas}, {Joubert}, {Julie}, {Kapp},
  {Kenyon}, {Kotz{\'e}}, {Kriek}, {Kriel}, {Krishnan}, {Lehmensiek},
  {Liebenberg}, {Lord}, {Lunsky}, {Madisa}, {Magnus}, {Mahgoub}, {Makhaba},
  {Makhathini}, {Malan}, {Manley}, {Marais}, {Martens}, {Mauch}, {Merry},
  {Millenaar}, {Mnyandu}, {Mokone}, {Monama}, {Mphego}, {New}, {Ngcebetsha},
  {Ngoasheng}, {Ockards}, {Oozeer}, {Otto}, {Passmoor}, {Patel}, {Peens-Hough},
  {Perkins}, {Ramaila}, {Ramanujam}, {Ramudzuli}, {Ratcliffe}, {Robyntjies},
  {Salie}, {Sambu}, {Schollar}, {Schwardt}, {Schwartz}, {Serylak}, {Siebrits},
  {Sirothia}, {Slabber}, {Smirnov}, {Sofeya}, {Taljaard}, {Tasse}, {Tiplady},
  {Toruvanda}, {Twum}, {van Balla}, {van der Byl}, {van der Merwe}, {Van
  Tonder}, {Van Wyk}, {Venter}, {Venter}, {Wallace}, {Welz}, {Williams}, \&
  {Xaia}}]{Heywood2022}
{Heywood}, I., {Rammala}, I., {Camilo}, F., {et~al.} 2022, \apj, 925, 165

\bibitem[{{Hildebrand}(1983)}]{Hildebrand1983}
{Hildebrand}, R.~H. 1983, \qjras, 24, 267

\bibitem[{{Hirano} {et~al.}(2024){Hirano}, {Sahu}, {Liu}, {Liu}, {Tatematsu},
  {Dutta}, {Li}, {Lee}, {Li}, {Hsu}, {Lin}, {Johnstone}, {Bronfman}, {Chen},
  {Eden}, {Kuan}, {Kwon}, {Lee}, {Liu}, {Rawlings}, {Ristorcelli}, \&
  {Traficante}}]{Hirano2024}
{Hirano}, N., {Sahu}, D., {Liu}, S.-Y., {et~al.} 2024, \apj, 961, 123

\bibitem[{{Hirashita} \& {Omukai}(2009)}]{Hirashita2009}
{Hirashita}, H. \& {Omukai}, K. 2009, \mnras, 399, 1795

\bibitem[{{Ho} {et~al.}(1991){Ho}, {Ho}, {Szczepanski}, {Jackson}, \&
  {Armstrong}}]{Ho1991}
{Ho}, P. T.~P., {Ho}, L.~C., {Szczepanski}, J.~C., {Jackson}, J.~M., \&
  {Armstrong}, J.~T. 1991, \nat, 350, 309

\bibitem[{{Ho} {et~al.}(1985){Ho}, {Jackson}, {Barrett}, \&
  {Armstrong}}]{Ho1985}
{Ho}, P.~T.~P., {Jackson}, J.~M., {Barrett}, A.~H., \& {Armstrong}, J.~T. 1985,
  \apj, 288, 575

\bibitem[{{Hoare} {et~al.}(1991){Hoare}, {Roche}, \& {Glencross}}]{Hoare1991}
{Hoare}, M.~G., {Roche}, P.~F., \& {Glencross}, W.~M. 1991, \mnras, 251, 584

\bibitem[{{Hosek} {et~al.}(2019){Hosek}, {Lu}, {Anderson}, {Najarro}, {Ghez},
  {Morris}, {Clarkson}, \& {Albers}}]{Hosek2019}
{Hosek}, Matthew~W., J., {Lu}, J.~R., {Anderson}, J., {et~al.} 2019, \apj, 870,
  44

\bibitem[{{Hunter} {et~al.}(2023){Hunter}, {Indebetouw}, {Brogan}, {Berry},
  {Chang}, {Francke}, {Geers}, {G{\'o}mez}, {Hibbard}, {Humphreys}, {Kent},
  {Kepley}, {Kunneriath}, {Lipnicky}, {Loomis}, {Mason}, {Masters}, {Maud},
  {Muders}, {Sabater}, {Sugimoto}, {Sz{\H{u}}cs}, {Vasiliev}, {Videla},
  {Villard}, {Williams}, {Xue}, \& {Yoon}}]{Hunter2023}
{Hunter}, T.~R., {Indebetouw}, R., {Brogan}, C.~L., {et~al.} 2023, \pasp, 135,
  074501

\bibitem[{{Immer} {et~al.}(2012){Immer}, {Menten}, {Schuller}, \&
  {Lis}}]{Immer2012}
{Immer}, K., {Menten}, K.~M., {Schuller}, F., \& {Lis}, D.~C. 2012, \aap, 548,
  A120

\bibitem[{{Inoue}(2002)}]{Inoue2002}
{Inoue}, A.~K. 2002, \apj, 570, 688

\bibitem[{{Kauffmann} {et~al.}(2017{\natexlab{a}}){Kauffmann}, {Pillai},
  {Zhang}, {Menten}, {Goldsmith}, {Lu}, \& {Guzm{\'a}n}}]{Kauffmann2017a}
{Kauffmann}, J., {Pillai}, T., {Zhang}, Q., {et~al.} 2017{\natexlab{a}}, \aap,
  603, A89

\bibitem[{{Kauffmann} {et~al.}(2017{\natexlab{b}}){Kauffmann}, {Pillai},
  {Zhang}, {Menten}, {Goldsmith}, {Lu}, {Guzm{\'a}n}, \&
  {Schmiedeke}}]{Kauffmann2017b}
{Kauffmann}, J., {Pillai}, T., {Zhang}, Q., {et~al.} 2017{\natexlab{b}}, \aap,
  603, A90

\bibitem[{{Kendrew} {et~al.}(2013){Kendrew}, {Ginsburg}, {Johnston}, {Beuther},
  {Bally}, {Cyganowski}, \& {Battersby}}]{Kendrew2013}
{Kendrew}, S., {Ginsburg}, A., {Johnston}, K., {et~al.} 2013, \apjl, 775, L50

\bibitem[{{Kruijssen} \& {Longmore}(2013)}]{Kruijssen2013}
{Kruijssen}, J.~M.~D. \& {Longmore}, S.~N. 2013, \mnras, 435, 2598

\bibitem[{{Kruijssen} {et~al.}(2014){Kruijssen}, {Longmore}, {Elmegreen},
  {Murray}, {Bally}, {Testi}, \& {Kennicutt}}]{Kruijssen2014}
{Kruijssen}, J.~M.~D., {Longmore}, S.~N., {Elmegreen}, B.~G., {et~al.} 2014,
  \mnras, 440, 3370

\bibitem[{{Kurtz}(2005)}]{Kurtz2005HCHII}
{Kurtz}, S. 2005, in IAU Symposium, Vol. 227, Massive Star Birth: A Crossroads
  of Astrophysics, ed. R.~{Cesaroni}, M.~{Felli}, E.~{Churchwell}, \&
  M.~{Walmsley}, 111--119

\bibitem[{{Lada} {et~al.}(2010){Lada}, {Lombardi}, \& {Alves}}]{Lada2010}
{Lada}, C.~J., {Lombardi}, M., \& {Alves}, J.~F. 2010, \apj, 724, 687

\bibitem[{{Lis} \& {Carlstrom}(1994)}]{Lis1994}
{Lis}, D.~C. \& {Carlstrom}, J.~E. 1994, \apj, 424, 189

\bibitem[{{Liu}(2019)}]{Liu2019SelfScatter}
{Liu}, H.~B. 2019, \apjl, 877, L22

\bibitem[{{Liu} {et~al.}(2019){Liu}, {Chen}, {Rom{\'a}n-Z{\'u}{\~n}iga},
  {Galv{\'a}n-Madrid}, {Ginsburg}, {Ho}, {Minh}, {Jim{\'e}nez-Serra}, {Testi},
  \& {Zhang}}]{Liu2019G33}
{Liu}, H.~B., {Chen}, H.-R.~V., {Rom{\'a}n-Z{\'u}{\~n}iga}, C.~G., {et~al.}
  2019, \apj, 871, 185

\bibitem[{{Liu} {et~al.}(2014){Liu}, {Galv{\'a}n-Madrid}, {Forbrich},
  {Rodr{\'\i}guez}, {Takami}, {Costigan}, {Manara}, {Yan}, {Karr}, {Chou},
  {Ho}, \& {Zhang}}]{Liu2014}
{Liu}, H.~B., {Galv{\'a}n-Madrid}, R., {Forbrich}, J., {et~al.} 2014, \apj,
  780, 155

\bibitem[{{Liu} {et~al.}(2013){Liu}, {Ho}, {Wright}, {Su}, {Hsieh}, {Sun},
  {Kim}, \& {Minh}}]{Liu2013CMZ}
{Liu}, H.~B., {Ho}, P. T.~P., {Wright}, M. C.~H., {et~al.} 2013, \apj, 770, 44

\bibitem[{{Longmore} {et~al.}(2013){Longmore}, {Bally}, {Testi}, {Purcell},
  {Walsh}, {Bressert}, {Pestalozzi}, {Molinari}, {Ott}, {Cortese}, {Battersby},
  {Murray}, {Lee}, {Kruijssen}, {Schisano}, \& {Elia}}]{Longmore2013}
{Longmore}, S.~N., {Bally}, J., {Testi}, L., {et~al.} 2013, \mnras, 429, 987

\bibitem[{{Lu} {et~al.}(2003){Lu}, {Wang}, \& {Lang}}]{Lu2003NT}
{Lu}, F.~J., {Wang}, Q.~D., \& {Lang}, C.~C. 2003, \aj, 126, 319

\bibitem[{{Lu} {et~al.}(2020){Lu}, {Cheng}, {Ginsburg}, {Longmore},
  {Kruijssen}, {Battersby}, {Zhang}, \& {Walker}}]{Lu2020Jeans}
{Lu}, X., {Cheng}, Y., {Ginsburg}, A., {et~al.} 2020, \apjl, 894, L14

\bibitem[{{Lu} {et~al.}(2021){Lu}, {Li}, {Ginsburg}, {Longmore}, {Kruijssen},
  {Walker}, {Feng}, {Zhang}, {Battersby}, {Pillai}, {Mills}, {Kauffmann},
  {Cheng}, \& {Inutsuka}}]{Lu2021Outflow}
{Lu}, X., {Li}, S., {Ginsburg}, A., {et~al.} 2021, \apj, 909, 177

\bibitem[{{Lu} {et~al.}(2019{\natexlab{a}}){Lu}, {Mills}, {Ginsburg}, {Walker},
  {Barnes}, {Butterfield}, {Henshaw}, {Battersby}, {Kruijssen}, {Longmore},
  {Zhang}, {Bally}, {Kauffmann}, {Ott}, {Rickert}, \& {Wang}}]{Lu2019Census}
{Lu}, X., {Mills}, E. A.~C., {Ginsburg}, A., {et~al.} 2019{\natexlab{a}},
  \apjs, 244, 35

\bibitem[{{Lu} {et~al.}(2019{\natexlab{b}}){Lu}, {Zhang}, {Kauffmann},
  {Pillai}, {Ginsburg}, {Mills}, {Kruijssen}, {Longmore}, {Battersby}, {Liu},
  \& {Gu}}]{Lu2019SFR}
{Lu}, X., {Zhang}, Q., {Kauffmann}, J., {et~al.} 2019{\natexlab{b}}, \apj, 872,
  171

\bibitem[{{Lu} {et~al.}(2015){Lu}, {Zhang}, {Kauffmann}, {Pillai}, {Longmore},
  {Kruijssen}, {Battersby}, \& {Gu}}]{Lu2015}
{Lu}, X., {Zhang}, Q., {Kauffmann}, J., {et~al.} 2015, \apjl, 814, L18

\bibitem[{{Lu} {et~al.}(2017){Lu}, {Zhang}, {Kauffmann}, {Pillai}, {Longmore},
  {Kruijssen}, {Battersby}, {Liu}, {Ginsburg}, {Mills}, {Zhang}, \&
  {Gu}}]{Lu2017}
{Lu}, X., {Zhang}, Q., {Kauffmann}, J., {et~al.} 2017, \apj, 839, 1

\bibitem[{{McKee} {et~al.}(1987){McKee}, {Hollenbach}, {Seab}, \&
  {Tielens}}]{McKee1987GrainDestruction}
{McKee}, C.~F., {Hollenbach}, D.~J., {Seab}, G.~C., \& {Tielens}, A.~G.~G.~M.
  1987, \apj, 318, 674

\bibitem[{{Meng} {et~al.}(2019){Meng}, {S{\'a}nchez-Monge}, {Schilke},
  {Padovani}, {Marcowith}, {Ginsburg}, {Schmiedeke}, {Schw{\"o}rer}, {DePree},
  {Veena}, \& {M{\"o}ller}}]{Meng2019}
{Meng}, F., {S{\'a}nchez-Monge}, {\'A}., {Schilke}, P., {et~al.} 2019, \aap,
  630, A73

\bibitem[{{Men'shchikov}(2021)}]{Men2021getsf}
{Men'shchikov}, A. 2021, \aap, 649, A89

\bibitem[{{Mezger} {et~al.}(1996){Mezger}, {Duschl}, \& {Zylka}}]{Mezger1996}
{Mezger}, P.~G., {Duschl}, W.~J., \& {Zylka}, R. 1996, \aapr, 7, 289

\bibitem[{{Mills}(2017)}]{Mills2017Review}
{Mills}, E.~A.~C. 2017, arXiv e-prints, arXiv:1705.05332

\bibitem[{{Morris} \& {Serabyn}(1996)}]{Morris1996}
{Morris}, M. \& {Serabyn}, E. 1996, \araa, 34, 645

\bibitem[{{Motte} {et~al.}(2022){Motte}, {Bontemps}, {Csengeri}, {Pouteau},
  {Louvet}, {Stutz}, {Cunningham}, {L{\'o}pez-Sepulcre}, {Brouillet},
  {Galv{\'a}n-Madrid}, {Ginsburg}, {Maud}, {Men'shchikov}, {Nakamura}, {Nony},
  {Sanhueza}, {{\'A}lvarez-Guti{\'e}rrez}, {Armante}, {Baug}, {Bonfand},
  {Busquet}, {Chapillon}, {D{\'\i}az-Gonz{\'a}lez}, {Fern{\'a}ndez-L{\'o}pez},
  {Guzm{\'a}n}, {Herpin}, {Liu}, {Olguin}, {Towner}, {Bally}, {Battersby},
  {Braine}, {Bronfman}, {Chen}, {Dell'Ova}, {Di Francesco}, {Gonz{\'a}lez},
  {Gusdorf}, {Hennebelle}, {Izumi}, {Joncour}, {Lee}, {Lefloch}, {Lesaffre},
  {Lu}, {Menten}, {Mignon-Risse}, {Molet}, {Moraux}, {Mundy}, {Nguyen Luong},
  {Reyes}, {Reyes Reyes}, {Robitaille}, {Rosolowsky}, {Sandoval-Garrido},
  {Schuller}, {Svoboda}, {Tatematsu}, {Thomasson}, {Walker}, {Wu}, {Whitworth},
  \& {Wyrowski}}]{Motte2022}
{Motte}, F., {Bontemps}, S., {Csengeri}, T., {et~al.} 2022, \aap, 662, A8

\bibitem[{{Motte} {et~al.}(2018{\natexlab{a}}){Motte}, {Bontemps}, \&
  {Louvet}}]{Motte2018AR}
{Motte}, F., {Bontemps}, S., \& {Louvet}, F. 2018{\natexlab{a}}, \araa, 56, 41

\bibitem[{{Motte} {et~al.}(2018{\natexlab{b}}){Motte}, {Nony}, {Louvet},
  {Marsh}, {Bontemps}, {Whitworth}, {Men'shchikov}, {Nguyen Luong}, {Csengeri},
  {Maury}, {Gusdorf}, {Chapillon}, {K{\"o}nyves}, {Schilke}, {Duarte-Cabral},
  {Didelon}, \& {Gaudel}}]{Motte2018NA}
{Motte}, F., {Nony}, T., {Louvet}, F., {et~al.} 2018{\natexlab{b}}, Nature
  Astronomy, 2, 478

\bibitem[{{Murray} \& {Rahman}(2010)}]{Murray2010}
{Murray}, N. \& {Rahman}, M. 2010, \apj, 709, 424

\bibitem[{{Navarro-Almaida} {et~al.}(2024){Navarro-Almaida}, {Lebreuilly},
  {Hennebelle}, {Fuente}, {Commer{\c{c}}on}, {Le Gal}, {Wakelam}, {Gerin},
  {Rivi{\'e}re-Marichalar}, {Beitia-Antero}, \&
  {Ascasibar}}]{Navarro-Almaida2024}
{Navarro-Almaida}, D., {Lebreuilly}, U., {Hennebelle}, P., {et~al.} 2024, \aap,
  685, A112

\bibitem[{{Nozari} {et~al.}(2025){Nozari}, {Sadavoy}, {Chapillon}, {Mason},
  {Friesen}, {Lowe}, {Stanke}, {Di Francesco}, {Henning}, {Zhang}, \&
  {Stutz}}]{Nozari2025}
{Nozari}, P., {Sadavoy}, S., {Chapillon}, E., {et~al.} 2025, \apj, 979, 142

\bibitem[{{Ossenkopf} \& {Henning}(1994)}]{Ossenkopf1994}
{Ossenkopf}, V. \& {Henning}, T. 1994, \aap, 291, 943

\bibitem[{{Padovani} {et~al.}(2019){Padovani}, {Marcowith},
  {S{\'a}nchez-Monge}, {Meng}, \& {Schilke}}]{Padovani2019}
{Padovani}, M., {Marcowith}, A., {S{\'a}nchez-Monge}, {\'A}., {Meng}, F., \&
  {Schilke}, P. 2019, \aap, 630, A72

\bibitem[{{Pillai} {et~al.}(2019){Pillai}, {Kauffmann}, {Zhang}, {Sanhueza},
  {Leurini}, {Wang}, {Sridharan}, \& {K{\"o}nig}}]{Pillai2019}
{Pillai}, T., {Kauffmann}, J., {Zhang}, Q., {et~al.} 2019, \aap, 622, A54

\bibitem[{{Pouteau} {et~al.}(2022){Pouteau}, {Motte}, {Nony},
  {Galv{\'a}n-Madrid}, {Men'shchikov}, {Bontemps}, {Robitaille}, {Louvet},
  {Ginsburg}, {Herpin}, {L{\'o}pez-Sepulcre}, {Dell'Ova}, {Gusdorf},
  {Sanhueza}, {Stutz}, {Brouillet}, {Thomasson}, {Armante}, {Baug}, {Bonfand},
  {Busquet}, {Csengeri}, {Cunningham}, {Fern{\'a}ndez-L{\'o}pez}, {Liu},
  {Olguin}, {Towner}, {Bally}, {Braine}, {Bronfman}, {Joncour}, {Gonz{\'a}lez},
  {Hennebelle}, {Lu}, {Menten}, {Moraux}, {Tatematsu}, {Walker}, \&
  {Whitworth}}]{Pouteau2022}
{Pouteau}, Y., {Motte}, F., {Nony}, T., {et~al.} 2022, \aap, 664, A26

\bibitem[{{Rathborne} {et~al.}(2015){Rathborne}, {Longmore}, {Jackson},
  {Alves}, {Bally}, {Bastian}, {Contreras}, {Foster}, {Garay}, {Kruijssen},
  {Testi}, \& {Walsh}}]{Rathborne2015}
{Rathborne}, J.~M., {Longmore}, S.~N., {Jackson}, J.~M., {et~al.} 2015, \apj,
  802, 125

\bibitem[{{Schnee} {et~al.}(2014){Schnee}, {Mason}, {Di Francesco}, {Friesen},
  {Li}, {Sadavoy}, \& {Stanke}}]{Schnee2014}
{Schnee}, S., {Mason}, B., {Di Francesco}, J., {et~al.} 2014, \mnras, 444, 2303

\bibitem[{{Silsbee} {et~al.}(2022){Silsbee}, {Akimkin}, {Ivlev}, {Testi},
  {Gong}, \& {Caselli}}]{Silsbee2022}
{Silsbee}, K., {Akimkin}, V., {Ivlev}, A.~V., {et~al.} 2022, \apj, 940, 188

\bibitem[{{Smith} {et~al.}(2009){Smith}, {Longmore}, \& {Bonnell}}]{Smith2009}
{Smith}, R.~J., {Longmore}, S., \& {Bonnell}, I. 2009, \mnras, 400, 1775

\bibitem[{{Stetson}(1987)}]{Stetson1987}
{Stetson}, P.~B. 1987, \pasp, 99, 191

\bibitem[{{Traficante} {et~al.}(2023){Traficante}, {Jones}, {Avison}, {Fuller},
  {Benedettini}, {Elia}, {Molinari}, {Peretto}, {Pezzuto}, {Pillai}, {Rygl},
  {Schisano}, \& {Smith}}]{Traficante2023}
{Traficante}, A., {Jones}, B.~M., {Avison}, A., {et~al.} 2023, \mnras, 520,
  2306

\bibitem[{{Tsuboi} {et~al.}(1999){Tsuboi}, {Handa}, \& {Ukita}}]{Tsuboi1999}
{Tsuboi}, M., {Handa}, T., \& {Ukita}, N. 1999, \apjs, 120, 1

\bibitem[{{Tsukamoto} {et~al.}(2021){Tsukamoto}, {Machida}, \&
  {Inutsuka}}]{Tsukamoto2021}
{Tsukamoto}, Y., {Machida}, M.~N., \& {Inutsuka}, S.-i. 2021, \apjl, 920, L35

\bibitem[{{Tychoniec} {et~al.}(2018){Tychoniec}, {Tobin}, {Karska}, {Chandler},
  {Dunham}, {Harris}, {Kratter}, {Li}, {Looney}, {Melis}, {P{\'e}rez},
  {Sadavoy}, {Segura-Cox}, \& {van Dishoeck}}]{Tychoniec2018}
{Tychoniec}, {\L}., {Tobin}, J.~J., {Karska}, A., {et~al.} 2018, \apjs, 238, 19

\bibitem[{{Walker} {et~al.}(2021){Walker}, {Longmore}, {Bally}, {Ginsburg},
  {Kruijssen}, {Zhang}, {Henshaw}, {Lu}, {Alves}, {Barnes}, {Battersby},
  {Beuther}, {Contreras}, {G{\'o}mez}, {Ho}, {Jackson}, {Kauffmann}, {Mills},
  \& {Pillai}}]{Walker2021}
{Walker}, D.~L., {Longmore}, S.~N., {Bally}, J., {et~al.} 2021, \mnras, 503, 77

\bibitem[{{Walker} {et~al.}(2015){Walker}, {Longmore}, {Bastian}, {Kruijssen},
  {Rathborne}, {Jackson}, {Foster}, \& {Contreras}}]{Walker2015}
{Walker}, D.~L., {Longmore}, S.~N., {Bastian}, N., {et~al.} 2015, \mnras, 449,
  715

\bibitem[{{Wang} {et~al.}(2014){Wang}, {Zhang}, {Testi}, {van der Tak}, {Wu},
  {Zhang}, {Pillai}, {Wyrowski}, {Carey}, {Ragan}, \& {Henning}}]{Wang2014}
{Wang}, K., {Zhang}, Q., {Testi}, L., {et~al.} 2014, \mnras, 439, 3275

\bibitem[{{Wang} {et~al.}(2006){Wang}, {Dong}, \& {Lang}}]{Wang2006}
{Wang}, Q.~D., {Dong}, H., \& {Lang}, C. 2006, \mnras, 371, 38

\bibitem[{{Wong} {et~al.}(2016){Wong}, {Hirashita}, \& {Li}}]{Wong2016Origin}
{Wong}, Y. H.~V., {Hirashita}, H., \& {Li}, Z.-Y. 2016, \pasj, 68, 67

\bibitem[{{Xu} {et~al.}(2024{\natexlab{a}}){Xu}, {Wang}, {Liu}, {Eden}, {Liu},
  {Juvela}, {He}, {Johnstone}, {Goldsmith}, {Garay}, {Wu}, {Soam},
  {Traficante}, {Ristorcelli}, {Falgarone}, {Chen}, {Hirano}, {Doi}, {Kwon},
  {White}, {Whitworth}, {Sanhueza}, {Rawlings}, {Alina}, {Ren}, {Lee},
  {Tatematsu}, {Zhang}, {Zhou}, {Lai}, {Ward-Thompson}, {Liu}, {Gu}, {Chakali},
  {Zhu}, {Mardones}, \& {T{\'o}th}}]{Xu2024Scarcity}
{Xu}, F., {Wang}, K., {Liu}, T., {et~al.} 2024{\natexlab{a}}, \apjl, 963, L9

\bibitem[{{Xu} {et~al.}(2024{\natexlab{b}}){Xu}, {Wang}, {Liu}, {Tang},
  {Evans}, {Palau}, {Morii}, {He}, {Sanhueza}, {Liu}, {Stutz}, {Zhang}, {Chen},
  {Li}, {G{\'o}mez}, {V{\'a}zquez-Semadeni}, {Li}, {Mai}, {Lu}, {Liu}, {Chen},
  {Li}, {Shi}, {Ren}, {Li}, {Garay}, {Bronfman}, {Dewangan}, {Juvela}, {Lee},
  {Zhang}, {Yue}, {Wang}, {Ge}, {Jiao}, {Luo}, {Zhou}, {Tatematsu}, {Chibueze},
  {Su}, {Sun}, {Ristorcelli}, \& {Toth}}]{Xu2024Protocluster}
{Xu}, F., {Wang}, K., {Liu}, T., {et~al.} 2024{\natexlab{b}}, \apjs, 270, 9

\bibitem[{{Xu} {et~al.}(2024{\natexlab{c}}){Xu}, {Wang}, {Liu}, {Zhu}, {Garay},
  {Liu}, {Goldsmith}, {Zhang}, {Sanhueza}, {Qin}, {He}, {Juvela}, {Tej}, {Liu},
  {Li}, {Morii}, {Zhang}, {Zhou}, {Stutz}, {Evans}, {Kim}, {Liu}, {Mardones},
  {Li}, {Bronfman}, {Tatematsu}, {Lee}, {Lu}, {Mai}, {Jiao}, {Chibueze}, {Su},
  \& {T{\'o}th}}]{Xu2024QUARKS}
{Xu}, F., {Wang}, K., {Liu}, T., {et~al.} 2024{\natexlab{c}}, RAA, 24, 065011

\bibitem[{{Xu} {et~al.}(2023){Xu}, {Wang}, {Liu}, {Goldsmith}, {Zhang},
  {Juvela}, {Liu}, {Qin}, {Li}, {Tej}, {Garay}, {Bronfman}, {Li}, {Wu},
  {G{\'o}mez}, {V{\'a}zquez-Semadeni}, {Tatematsu}, {Ren}, {Zhang}, {Toth},
  {Liu}, {Yue}, {Zhang}, {Baug}, {Issac}, {Stutz}, {Liu}, {Fuller}, {Tang},
  {Zhang}, {Dewangan}, {Lee}, {Zhou}, {Xie}, {Jiao}, {Wang}, {Liu}, {Luo},
  {Soam}, \& {Eswaraiah}}]{Xu2023SDC335}
{Xu}, F.-W., {Wang}, K., {Liu}, T., {et~al.} 2023, \mnras, 520, 3259

\bibitem[{{Yusef-Zadeh} {et~al.}(2013){Yusef-Zadeh}, {Hewitt}, {Wardle},
  {Tatischeff}, {Roberts}, {Cotton}, {Uchiyama}, {Nobukawa}, {Tsuru}, {Heinke},
  \& {Royster}}]{Yusef-Zadeh2013}
{Yusef-Zadeh}, F., {Hewitt}, J.~W., {Wardle}, M., {et~al.} 2013, \apj, 762, 33

\bibitem[{{Zhang} {et~al.}(2015){Zhang}, {Wang}, {Lu}, \&
  {Jim{\'e}nez-Serra}}]{Zhang2015}
{Zhang}, Q., {Wang}, K., {Lu}, X., \& {Jim{\'e}nez-Serra}, I. 2015, \apj, 804,
  141

\bibitem[{{Zhang} {et~al.}(2025{\natexlab{a}}){Zhang}, {Lu}, {Ginsburg},
  {Budaiev}, {Cheng}, {Liu}, {Liu}, {Zhang}, {Qiu}, {Feng}, {Pillai}, {Tang},
  {Mills}, {Luo}, {Li}, {Issac}, {Liu}, {Xu}, {Wallace}, {Mai}, {Zhang},
  {Battersby}, {Longmore}, \& {Shen}}]{Zhang2025disk}
{Zhang}, S., {Lu}, X., {Ginsburg}, A., {et~al.} 2025{\natexlab{a}}, arXiv
  e-prints, arXiv:2503.00878

\bibitem[{{Zhang} \& {Tan}(2018)}]{Zhang2018SED}
{Zhang}, Y. \& {Tan}, J.~C. 2018, \apj, 853, 18

\bibitem[{{Zhang} {et~al.}(2025{\natexlab{b}}){Zhang}, {Lu}, {Liu}, {Qin},
  {Ginsburg}, {Cheng}, {Liu}, {Walker}, {Tang}, {Li}, {Zhang}, {Pillai},
  {Kauffmann}, {Battersby}, {Feng}, {Zhang}, {Gu}, {Xu}, {Jiao}, {Liu}, {Chen},
  {Luo}, {Mai}, {Li}, {Yang}, {Shen}, {Liu}, \& {Shen}}]{Zhang2025}
{Zhang}, Z., {Lu}, X., {Liu}, T., {et~al.} 2025{\natexlab{b}}, \apj, 980, 44

\end{thebibliography}

\begin{appendix}

\section{ALMA 3 mm observation supplements} \label{app:b3}

\begin{figure*}[!th]
\centering
\includegraphics[width=\linewidth]{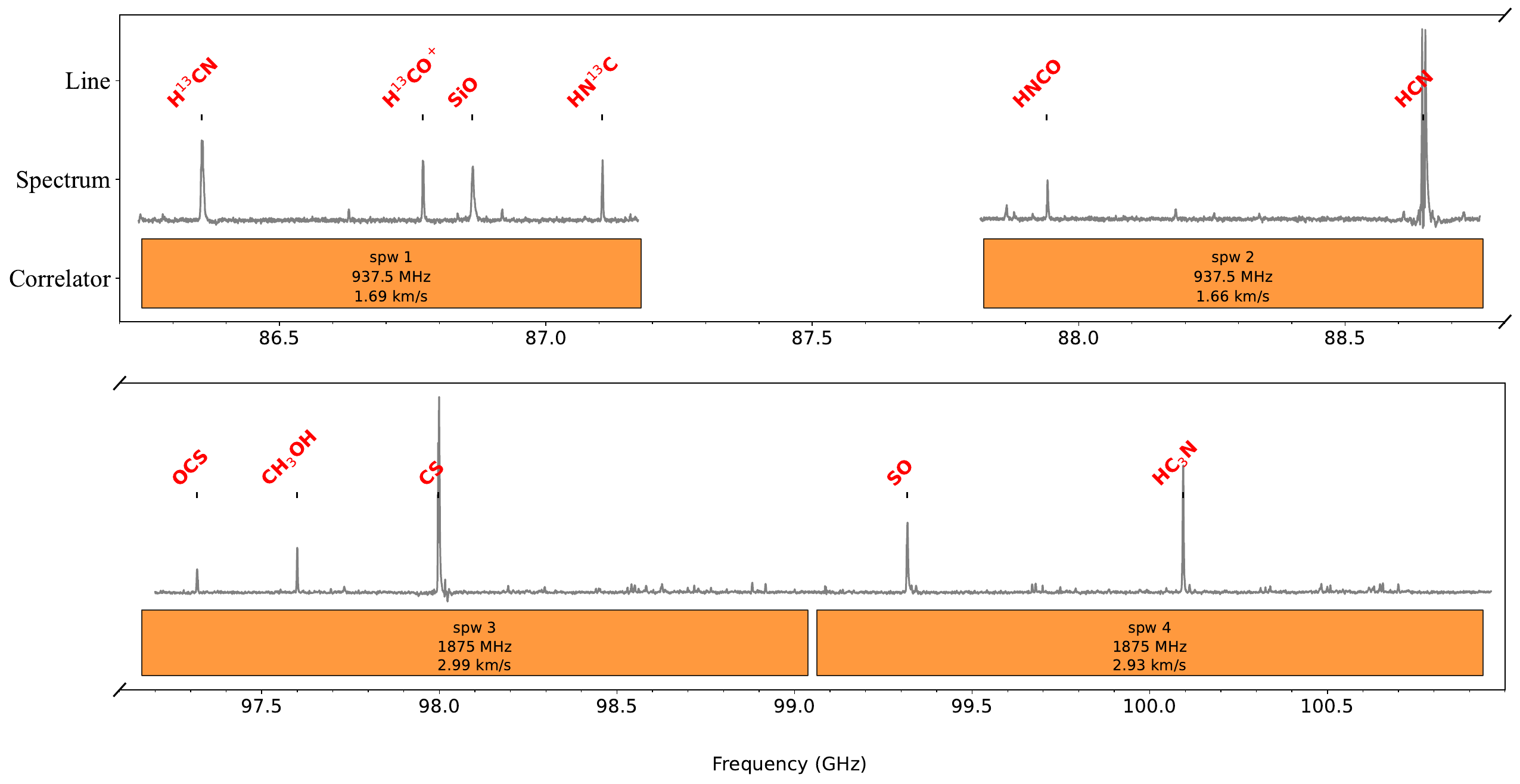}
\caption{ALMA 3~mm spectral window setting, an example spectrum from the Sgr~C protocluster region, and the identification of strong spectral lines. Complex molecular species (COMs) will be identified in forthcoming papers. Four spectral windows are named spw 1--4 from low to high frequency. Two setups of spectral correlators are used: the narrow one has 937.5~MHz bandwidth and 0.488~MHz ($\sim$1.67~\kms) resolution, and the wide one has 1875~MHz bandwidth and 0.976~MHz ($\sim$2.95~\kms) resolution. \label{fig:spw}}
\end{figure*}

We present the log of the ALMA 3~mm observations in Table~\ref{tab:obslogs}, and the spectral window in Fig.~\ref{fig:spw}. The exemplar observed spectra is averaged from the densest part of the Sgr~C cloud \footnote{With 4\arcsec-radius aperture centered at 17h44m40.3s, $-$29d28m14.7s.}.

\begin{table*}[!ht]
\centering
\caption{Summary of the ALMA 3~mm observations \label{tab:obslogs}}
\begin{tabular}{ccccccc}
\hline
\hline \noalign{\smallskip}
Observational & Array & \multicolumn{2}{c}{Baseline range} & Antenna & \multicolumn{2}{c}{Calibrators} \\
\cmidrule(r){3-4} \cmidrule(r){6-7} 
Dates & configuration & (km) & (k$\lambda$) & (number) & Bnadpass/flux & Phase \\
\hline \noalign{\smallskip}
\multicolumn{7}{c}{20 \kms, \sgc} \\
\hline \noalign{\smallskip}
2018-11-27 & C43-4 & 0.015--1.3 & 5--433 & 44 & J1617$-$5848 & J1744$-$3116 \\
2018-11-27 & C43-4 & 0.015--1.3 & 5--433 & 44 & J1617$-$5848 & J1744$-$3116 \\
2018-11-27 & C43-4 & 0.015--1.3 & 5--433 & 45 & J1924$-$2914 & J1744$-$3116 \\
2019-08-28 & C43-7 & 0.038--3.6 & 13--1200 & 48 & J1924$-$2914 & J1744$-$3116 \\
2019-08-28 & C43-7 & 0.038--3.6 & 13--1200 & 48 & J1924$-$2914 & J1752$-$2956 \\
2019-08-29 & C43-7 & 0.038--3.6 & 13--1200 & 47 & J1517$-$2422 & J1744$-$3116 \\
2019-08-29 & C43-7 & 0.038--3.6 & 13--1200 & 46 & J1924$-$2914 & J1744$-$3116 \\
2019-08-30 & C43-7 & 0.038--3.6 & 13--1200 & 47 & J1924$-$2914 & J1744$-$3116 \\
2021-08-02 & C43-7 & 0.047--5.2 & 16--1733 & 37 & J1924$-$2914 & J1744$-$3116 \\
\hline \noalign{\smallskip}
\multicolumn{7}{c}{cloud~e} \\
\hline \noalign{\smallskip}
2018-11-23 & C43-4 & 0.015--1.4 & 5--466 & 47 & J1924$-$2914 & J1744$-$3116 \\
2019-08-27 & C43-7 & 0.038--3.6 & 13--1200 & 48 & J1924$-$2914 & J1744$-$3116 \\
2021-07-30 & C43-7 & 0.047--5.2 & 16--1733 & 41 & J1924$-$2914 & J1744$-$3116 \\
\hline \\
\end{tabular}
\tablefoot{
Two scheduling blocks: \ctw{} and \sgc{} are observed in one and cloud~e is in another one. The minimum and maximum baselines are shown with both spatial distance (km) and wavelength number (k$\lambda$).
}
\end{table*}

\section{Source extraction procedure and catalogs} \label{app:getsf} 

Two scale-related inputs of the getsf algorithm are 1) the minimum size and 2) the maximum size of sources. The minimum size of sources for the images of the two bands are defined by the full width at half maximum of the synthesized beams as listed in column (4) of Table~\ref{tab:dualband_obs}. The maximum size is used to avoid false positive detections of extended structures from flat background emission. We visually estimate it to be 3--4 beam sizes, and then give a more conservative value of 5 beam sizes ($1.2\arcsec$) when extracting sources. The corresponding physical scale, $\sim$0.05 pc or $\sim$10,000 au at the distance of CMZ, is close to the largest identified source size in previous surveys with similar resolution \citep[e.g.,][]{Motte2018NA,Pouteau2022,Xu2024Protocluster}. As is shown in Fig.~\ref{fig:flux}, the maximum source size is $\sim$0\parcsec8, which is much smaller than the adopted maximum size. Therefore, the adopted maximum source size of $1.2\arcsec$ will not lead to incompleteness of the source catalog.

Our getsf procedure is spitted into three rounds. In the first round, single-wavelength (monochromatic) \texttt{separation} of structural components of sources and filaments from their backgrounds is performed individually on the band 3 and band 6 continuum images without primary beam corrections. To avoid false detections at the image boundary, pixels with a primary beam response lower than 0.3 are masked and do not participate in source extraction. This run prepares flattened components of sources and filaments over selected wavebands.

The second round focuses on creating monochromatic core catalogs for both 1.3~mm and 3~mm. In this round, the \texttt{detect} and \texttt{measure} procedures are used to spatially decompose the flattened components, detect the positions of sources, and then measure the sources individually for the two bands. Consequently, two mono-band core catalogs (\texttt{mbcat}), which only contain detections at their individual bands, are obtained. The \texttt{mbcat}s are shown in Table~\ref{tab:mbcat01} and ~\ref{tab:mbcat02}.

In the third round, the \texttt{detect} and \texttt{measure} procedures are restarted to detect cores and remeasure their properties in the two bands. Core identification considers both bands simultaneously, which means that a marginal detection in one band will be excluded if there is no significant detection in the other band. For cores that are detected in one band with high S/Ns, they are included in the combined dual-band catalog and measured in the other band. The dual-band catalog (\texttt{dbcat}) is shown in Table~\ref{tab:dbcat}.

\begin{sidewaystable*}
\centering
\caption{Dual-band source catalog \label{tab:dbcat}}
\begin{tabular}{cccccccccccccc}
\hline
\hline \noalign{\smallskip}
ID & R.A. & Dec. & Sig & \multicolumn{5}{c}{1.3~mm measurement} & \multicolumn{5}{c}{3~mm measurement} \\
\cmidrule(r){5-9} \cmidrule(r){10-14}
 & comb. & $\theta_{\rm maj}\times\theta_{\rm min}$ & PA & Size & $I^{\rm peak}_\nu$ & $S^{\rm tot}_\nu$ & $\theta_{\rm maj}\times\theta_{\rm min}$ & PA & Size & $I^{\rm peak}_\nu$ & $S^{\rm tot}_\nu$ \\
 & (hh:mm:ss) & (dd:mm:ss) & & ($\arcsec\times\arcsec$) & (deg) &(au) & (\mjybeam) & (mJy) & ($\arcsec\times\arcsec$) & (deg) & (au) & (\mjybeam) & (mJy) \\
\hline \noalign{\smallskip}
\multicolumn{14}{c}{20~\kms} \\ 
\hline \noalign{\smallskip}
1 & 17:45:37.47 & -29:03:49.94 & 514.5 & 0.36$\times$0.29 & 67.07 & 2000 & 14.18(0.62) & 47.44(0.92) & 0.41$\times$0.31 & 81.63 & 1800 & 2.34(0.06) & 5.40(0.09) \\
2 & 17:45:37.74 & -29:03:46.83 & 326.2 & 0.30$\times$0.27 & 35.84 & 1500 & 8.68(0.22) & 24.49(0.33) & 0.34$\times$0.30 & 73.73 & 1200 & 1.42(0.04) & 2.57(0.05) \\
3 & 17:45:36.33 & -29:05:49.71 & 307.7 & 0.28$\times$0.23 & 131.1 & 1100 & 8.45(0.12) & 21.18(0.18) & 0.36$\times$0.28 & 110.8 & 1300 & 0.77(0.02) & 1.58(0.02) \\
\hline \noalign{\smallskip}
\multicolumn{14}{c}{Sgr~C} \\ 
\hline \noalign{\smallskip}
1 & 17:44:40.57 & -29:28:15.89 & 913.4 & 0.29$\times$0.26 & 159.8 & 1400 & 39.69(0.71) & 110.75(1.09) & 0.35$\times$0.31 & 84.86 & 1000 & 2.67(0.09) & 5.04(0.09) \\
2 & 17:44:40.15 & -29:28:12.85 & 714.0 & 0.31$\times$0.21 & 157.4 & 1200 & 32.72(0.82) & 89.70(1.29) & 0.35$\times$0.30 & 140.1 & 1200 & 2.84(0.08) & 5.86(0.10) \\
3 & 17:44:41.38 & -29:28:29.81 & 278.1 & 0.28$\times$0.22 & 116.5 & 1000 & 7.98(0.04) & 20.35(0.10) & 0.35$\times$0.25 & 101.0 & 900 & 1.05(0.01) & 1.72(0.01) \\
\hline \noalign{\smallskip}
\multicolumn{14}{c}{cloud~e} \\
\hline \noalign{\smallskip}
1 & 17:46:47.06 & -28:32:07.13 & 628.0 & 0.32$\times$0.21 & 102.6 & 1300 & 19.29(0.24) & 52.29(0.39) & 0.41$\times$0.25 & 92.03 & 1600 & 1.16(0.03) & 2.96(0.04) \\
2 & 17:46:46.21 & -28:31:55.11 & 153.1 & 0.32$\times$0.25 & 159.6 & 1600 & 4.17(0.20) & 10.73(0.24) & 0.34$\times$0.31 & 123.3 & 1500 & 0.53(0.02) & 1.22(0.02) \\
3 & 17:46:48.23 & -28:32:02.06 & 147.7 & 0.28$\times$0.21 & 132.6 & 1000 & 4.13(0.05) & 8.61(0.06) & 0.32$\times$0.24 & 95.26 &  900 & 0.44(0.01) & 0.79(0.01) \\
\hline
\end{tabular}
\parbox{0.95\textwidth}{\textbf{Notes}. The dual-band source catalog. Only the top three rows of each cloud are shown and the complete version can be found in \zenodo. The sources are sorted by the dual-band combined significance level of getsf core extraction. R.A. and Dec.: right ascension and declination of equatorial coordinate system. $\theta_{\rm maj}\times\theta_{\rm min}$: measured major and minor FWHM angular sizes. PA: position angle. Size: the linear FWHM size deconvolved from the synthetic beam. $I^{\rm peak}_{\nu}$: peak intensity. $S^{\rm tot}_{\nu}$: integrated flux over source.}
\end{sidewaystable*}

\section{Source deconvolution from beam} \label{app:deconv}

Astronomical images and therefore the extracted sources are convolved with a point spread function (radio beam in our case). The conventional way of deconvolution is to fit the observed source with the 2D Gaussian model (with FWHM $\theta_{\rm obs}$) and then subtract the contribution of the Gaussian beam (with FWHM $\theta_{\rm beam}$) to obtain the intrinsic source size following $\theta_{\rm source} = \sqrt{\theta_{\rm obs}^2 - \theta_{\rm beam}^2}$. 

For interferometric data, the ellipticity of the beam cannot be ignored. The relative angular position between the observed source and the beam ellipses must be carefully considered, especially for unresolved and marginally resolved sources. As is illustrated in Figure \ref{fig:deconvolve}, there is a clear peak in the position angle distribution in the Sgr~C 1.3~mm \texttt{mbcat}, which aligns with the beam position angle. This indicates that a significant number of sources are marginally resolved, with their measured shapes predominantly determined by the beam size. Proper deconvolution of these sources is crucial. To address this, we utilize covariance matrices for a more accurate solution.

We take both the observed sources and the beam as 2D Gaussians characterized by variance $\sigma_{\rm maj}$, $\sigma_{\rm min}$, and position angle $\phi$.
For the observed elliptical source,
\begin{equation}
\Sigma_{\rm obs} = \mathcal{R} (\phi_{\rm obs})
\begin{pmatrix}
\sigma^2_{\rm maj,obs} & 0 \\
0 & \sigma^2_{\rm min,obs}
\end{pmatrix}
\mathcal{R}^{T} (\phi_{\rm obs}).
\end{equation}
For the synthetic beam,
\begin{equation}
\Sigma_{\rm beam} = \mathcal{R} (\phi_{\rm beam})
\begin{pmatrix}
\sigma^2_{\rm maj,beam} & 0 \\
0 & \sigma^2_{\rm min,beam}
\end{pmatrix}
\mathcal{R}^{T} (\phi_{\rm beam}),
\end{equation}
where $\mathcal{R}(\phi)$ is the rotation matrix,
\begin{equation}
\mathcal{R}(\phi) = 
\begin{pmatrix}
\cos\phi & -\sin\phi \\
\sin\phi & \cos\phi \\
\end{pmatrix}.
\end{equation}
The deconvolution can be written as,
\begin{equation}
    \Sigma_{\rm source} = \Sigma_{\rm obs} - \Sigma_{\rm beam}
\end{equation}
Diagonalization gives the eigenvalues on the basis where the new rotation matrix spans,
\begin{equation}
\Sigma_{\rm source} = \mathcal{R}(\phi_{\rm source})
\begin{pmatrix}
\sigma^2_{\rm maj,source} & 0 \\
0 & \sigma^2_{\rm min,source}
\end{pmatrix}.
\end{equation}
The solid deconvolution solution demands the matrix to be positive definite; that is, the eigenvalues, $\sigma^2_{\rm maj,source}$ and $\sigma^2_{\rm min,source}$, should be positive. As such, the elliptical model of the source deconvolved from the beam is obtained as $\sigma_{\rm maj,source}$, $\sigma_{\rm min,source}$, and $\phi_{\rm source}$. If the matrix is semi-definite (non-negative eigenvalues) or indefinite (negative eigenvalues), the real physical scenario is that a source is completely unresolved. In such a case, the source size is undetermined. 

\section{Spectral index approximation} \label{app:tau_effect}

We discuss several approximation case of spectral index. 
In the optical thin regime, Eq.~(\ref{eq:graybody}) reduces to,
\begin{equation} \label{eq:thindust}
    I_\nu \simeq B(\nu,T)\kappa_\nu\Sigma.
\end{equation}
The Taylor expansion of the exponent term in Eq.~(\ref{eq:blackbody}) can be written as,
\begin{equation} \label{eq:taylor}
    e^{h\nu/kT} - 1 \simeq \frac{h\nu}{kT} + \frac{1}{2} \left( \frac{h\nu}{kT} \right)^{2} + \frac{1}{6} \left( \frac{h\nu}{kT} \right)^{3} \cdots.
\end{equation}
If considering the first two terms in Eq.~(\ref{eq:taylor}), we define the critical temperature ($T_{\rm crit}$) as the temperature beyond which the second term accounts for less than $\epsilon$ of the first order approximation. It can be parameterized as,
\begin{equation} \label{eq:Tcrit}
    T_{\rm crit} \simeq 36 \left(\frac{\epsilon}{10\%}\right)^{-1} \left(\frac{\nu}{150\, \mathrm{GHz}}\right) \mathrm{K}.
\end{equation}
Eq.~(\ref{eq:Tcrit}) gives us a quick idea of how high temperature should be to alleviate deviation from the R-J approximation. It can also be written as
\begin{equation} \label{eq:epsilon}
    \epsilon \simeq 10\% \left(\frac{T}{36\,\mathrm{K}}\right)^{-1} \left(\frac{\nu}{150\, \mathrm{GHz}}\right),
\end{equation}
which gives how the deviation $\epsilon$ changes with temperature and frequency. In the R-J regime ($T>T_{\rm crit}$), the frequency dependence of Eq.~(\ref{eq:blackbody}) should be approximated as $B(\nu,T) \sim \nu^2$. 

\begin{figure*}[!ht]
\centering
\includegraphics[width=0.8\linewidth]{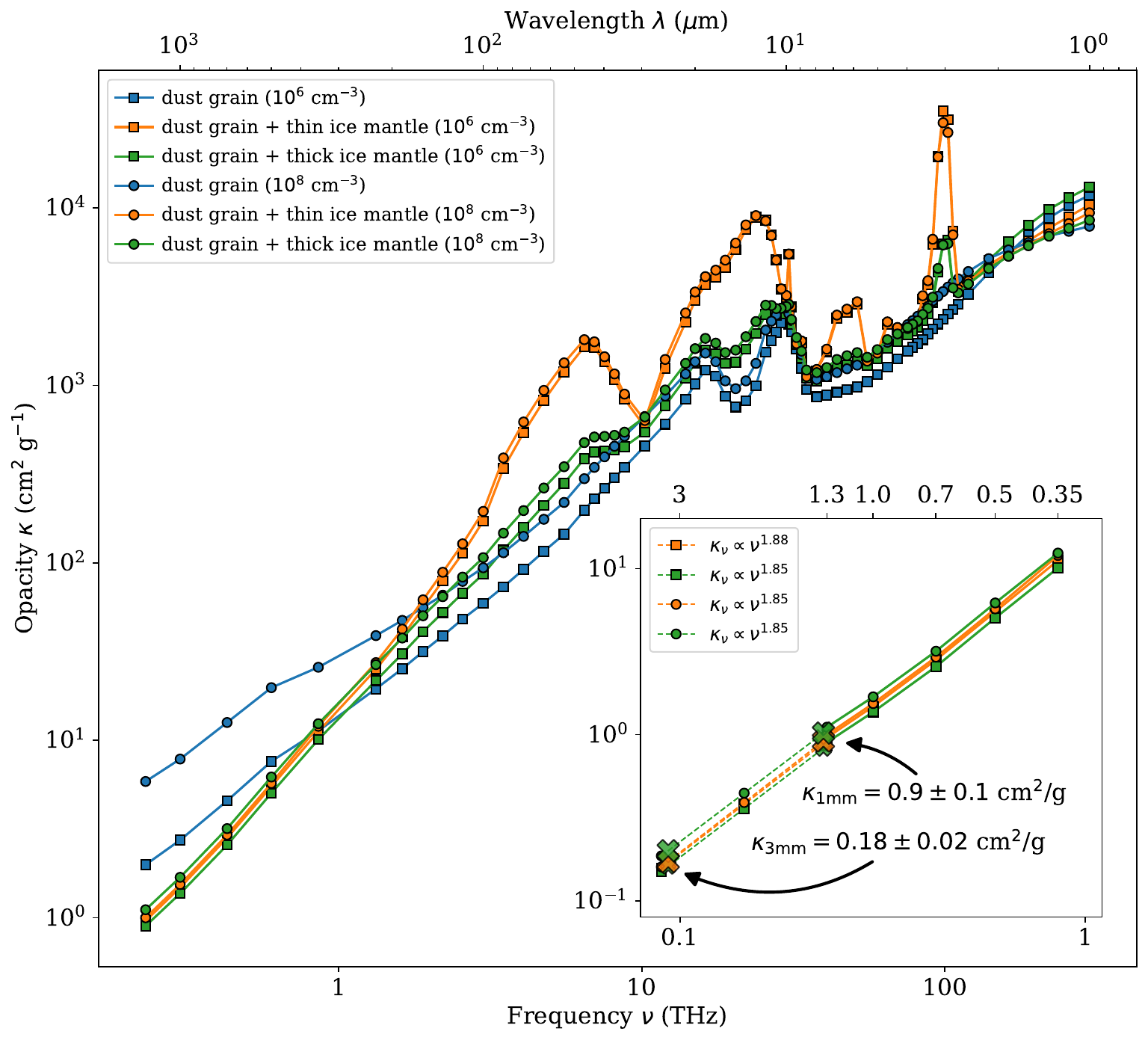}
\caption{Dust opacity to the 1 and 3~mm from the coagulation model in protostellar cores. The data points in the main panel are retrieved from \citet{Ossenkopf1994}. The blue, orange, and green colors represent the dust grain, dust grain with thin ice mantle, and dust grain with thick ice mantle, respectively. The square and circle line markers represent for the case of volume density of $10^6$ and $10^8$ cm$^{-3}$, respectively. The subpanel zooms in the frequency range of $\sim$0.1--1 THz, i.e., 300 $\mu$m to 3~mm. Based on the retrieved data points in a solid line, dust opacity is extrapolated by a power-law form $\kappa_\nu \propto \nu^\beta$ up to the 3~mm band in a dashed line. The extrapolated dust opacity is $\kappa_{\rm 1mm}=0.9 \pm 0.1$ cm$^2$/g and $\kappa_{\rm 3mm}=0.18 \pm 0.02$ cm$^2$/g.}
\label{fig:opacity}
\end{figure*}

At (sub)millimeter and centimeter wavelengths, $\kappa_{\nu}$ is approximately expressed by $\kappa_{\nu}\propto\nu^{\beta}$ (cf.\ \citealt{Hildebrand1983}). To demonstrate it, we adopt the protostellar opacity model from \citep{Ossenkopf1994} and make an extrapolation for the 1.3~mm and 3~mm wavelengths. As is shown in Figure \ref{fig:opacity}, we consider the grain either with thin or thick ice mantle, which are respectively shown in orange and green data points. Since the structures of our interest are mostly dense cores ($\gtrsim10^6$ cm$^{-3}$), the density regime for the model is chosen to be $10^6$ or $10^8$ cm$^{-3}$. These two regimes are marked with square and round line markers. In the zoom-in panel (0.1--1 THz), the four opacity models are linearly extrapolated by the five nearest data points (0.35--1.3 $\mu$m). The extrapolated opacities are $\kappa_{\rm 1.3~mm}=0.9\pm0.1$ cm$^2$ g$^{-1}$ and $\kappa_{\rm 3~mm}=0.18\pm0.02$~cm$^2$~g$^{-1}$. The dust opacity spectral index $\beta$, defined by,
\begin{equation} \label{eq:kappa}
    \kappa_\nu \sim \nu^\beta,
\end{equation}
is measured to be $\sim1.8$.

In this case, $I_{\nu}$ in the optically thin limit (Equation \ref{eq:thindust}) is proportional to $\nu^{2+\beta}$. So one obtains the spectral index as
\begin{equation}\label{eq:alphathin}
    \alpha_{\rm thin,RJ} = 2 + \beta \simeq 3.8.
\end{equation}
In optically thick regime, the high-order term in Eq.~(\ref{eq:taylor}) dominates in the low temperature regime, and the spectral index is always lower than $\alpha_{\rm thin, RJ}$. In the extreme case where $\tau\to\infty$, the spectral index reduces to the case of blackbody,
\begin{equation}\label{eq:alphablack}
    \alpha_{\rm black,RJ} = 2.
\end{equation}

\section{Common $uv$-range at dual bands} \label{app:cbdual}

Our dual-band ALMA observations have different $uv$ ranges at 1.3 and 3~mm, so their original synthesized beams and maximum recoverable scales (MRS) are different. This can bring inconsistency issues of dual-band analyses, especially when absolute flux matters. In our case, the 3~mm continuum images have generally larger MRS and then can recover more extended emission. To this end, we perform another round of continuum image cleaning. We use the \texttt{uvrange} parameter in the CASA task \texttt{tclean} to obtain the same $uv$ ranges for the two bands in each cloud. The shortest $uv$ length is 8.7622--8.8138~$k\lambda$ while the longest $uv$ length is 1718.65--1744.71~$k\lambda$. However, even though we have limited the $uv$ ranges, the resulting beam sizes of the continuum images of the two bands are still slightly different, because the $uv$ sample distribution of the two bands in the 2D $uv$ plain are still different. Therefore, we additionally perform 2D convolution to smooth the dual-band continuum images to a common beam in the image domain. 

We perform the same extraction as described in Appendix~\ref{app:getsf} but on the common-beam images to obtain the common-beam \texttt{dbcat}. We find that the common-beam extraction reduces the numbers of detections both in 1.3~mm \texttt{mbcat} (from 563 to 484 sources) and \texttt{dbcat} (from 450 to 402 sources), but not in 3~mm. It is because the common-beam cleaning drops longer baselines in 1.3~mm observations and shorter baselines in 3~mm observations. This leads to a lower resolution for the 1.3~mm images and reduces the number of detections, but for the 3~mm image, this only leads to a smaller MRS and does not really influence the detection of compact sources. 
The caveat of common-beam \texttt{dbcat} lies in losing the advantage of higher resolution at 1.3~mm, so some blended sources are not resolved. The common-beam catalogs are only used for our analyses of spectral indices. 

\section{Modified blackbody isothermal emission track} \label{app:isothermal}

Taking advantage of the dual-band observations, one can break the degeneracy between temperature and surface density in the optically thick regime. For dual-band observations, MBB emission Eq.~(\ref{eq:graybody}) reads as
\begin{equation} \label{eq:duet_mbb}
\left\{
\begin{aligned}
I_1 &= B_1 (T) \left(1-e^{-\kappa_1\Sigma}\right) \\
I_2 &= B_2 (T) \left(1-e^{-\kappa_2\Sigma}\right) 
\end{aligned}
\right. \\
\Rightarrow \quad
\left\{
\begin{aligned}
e^{-\kappa_1\Sigma} &= 1 - \frac{I_1}{B_1} \\
e^{-\kappa_2\Sigma} &= 1 - \frac{I_2}{B_2} \\
\end{aligned}
\right.,
\end{equation}
where $I_i$, $B_i(T)$ and $\kappa_i$ are the monochromatic values at 1.3~mm ($i=1$) and 3~mm ($i=2$). In principle, the solution of ($T$, $\Sigma$) for each source can be obtained from Eq.~(\ref{eq:duet_mbb}) as long as the other parameters are measured. 

By raising both sides of the second equation of Eq.~(\ref{eq:duet_mbb}) to the power of $\Xi\equiv\kappa_1/\kappa_2$, we obtain
\begin{equation}
\left\{
\begin{aligned}
e^{-\kappa_1\Sigma} &= 1 - \frac{I_1}{B_1} \\
e^{-\kappa_1\Sigma} &= \left[1 - \frac{I_2}{B_2}\right]^{\Xi} \\
\end{aligned}
\right.
\Rightarrow \quad
1 - \frac{I_1}{B_1} = \left[1 - \frac{I_2}{B_2}\right]^{\Xi}.
\end{equation}

Following the definition of spectral index $\alpha$, one can obtain a single-variable function mapping from $x\equiv I_1/B_1$ to $\alpha$,
\begin{equation} \label{eq:inverse}
    \alpha = f(x) = \frac{\log(x\Gamma)-\log\left[1-(1-x)^{1/\Xi}\right]}{\log(\nu_1/\nu_2)},
\end{equation}
where $\Gamma\equiv B_1/B_2$. The inverse function of Eq.~(\ref{eq:inverse}) gives $I_1 = B_1 f^{-1}(\alpha)$. 

For observations of beam-sized or larger objects, the 1.3~mm peak flux within the beam should be
\begin{equation} \label{eq:S1}
I^{\rm peak}_1 = I_1 \Omega_{b,1} = B_1 \Omega_{b,1} f^{-1}(\alpha),
\end{equation}
where $\Omega_{b,1}$ is the solid angle of the beam at 1.3~mm. We can also obtain the 3~mm peak flux in the same way,
\begin{equation}\label{eq:S2}
    I^{\rm peak}_2 = I_2 \Omega_{b,2} = B_1 f^{-1}(\alpha) \left(\frac{\nu_1}{\nu_2}\right)^{-\alpha} \Omega_{b,2}.
\end{equation}
We refer to them as isothermal tracks, meaning that sources with a given temperature but with different surface densities should follow this curve. 

\section{Centimeter counterpart cross-match and free-free emission subtraction} \label{app:ff}

\begin{figure*}
\centering
\includegraphics[width=0.48\linewidth]{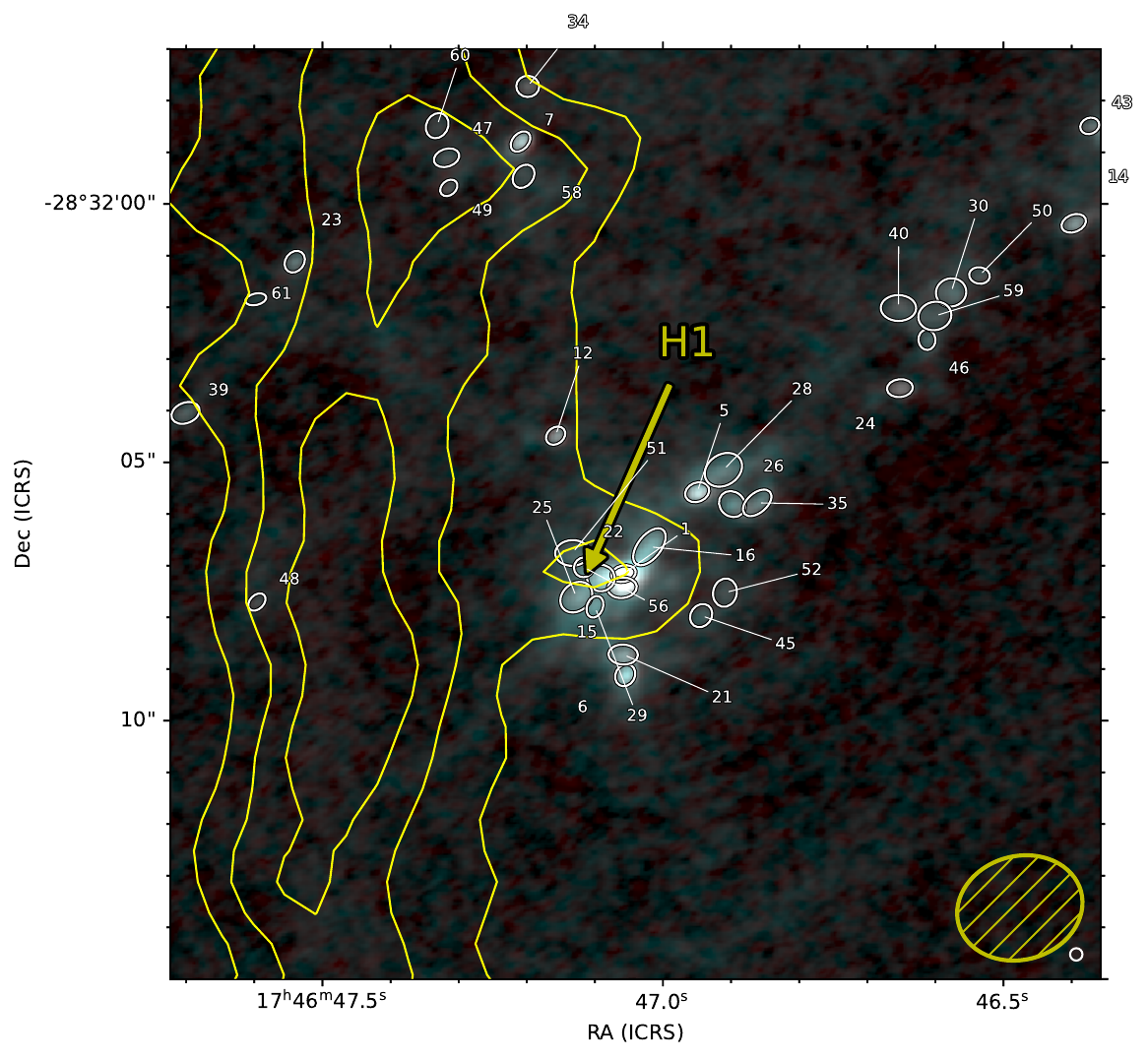}
\includegraphics[width=0.48\linewidth]{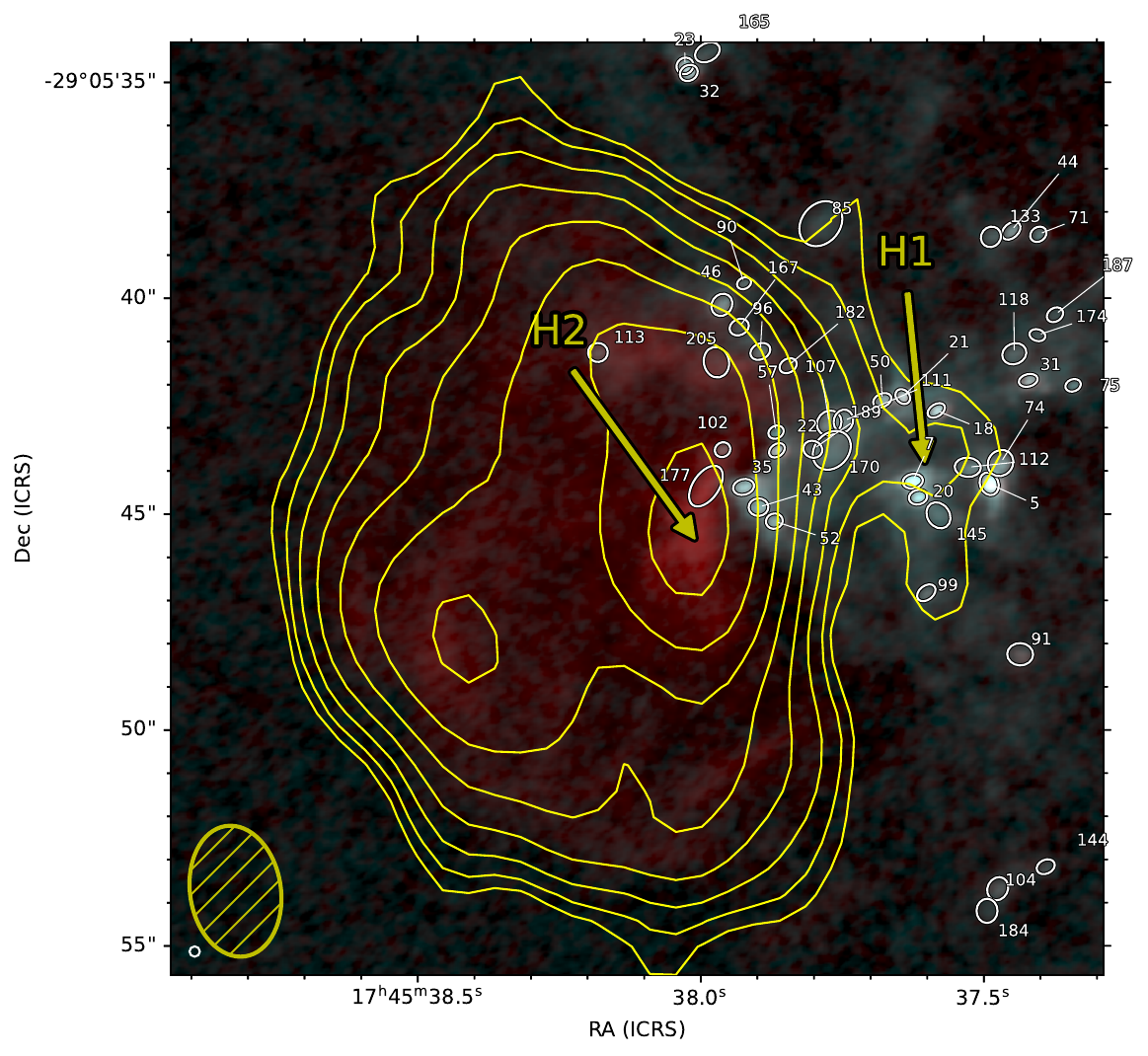}
\includegraphics[width=0.7\linewidth]{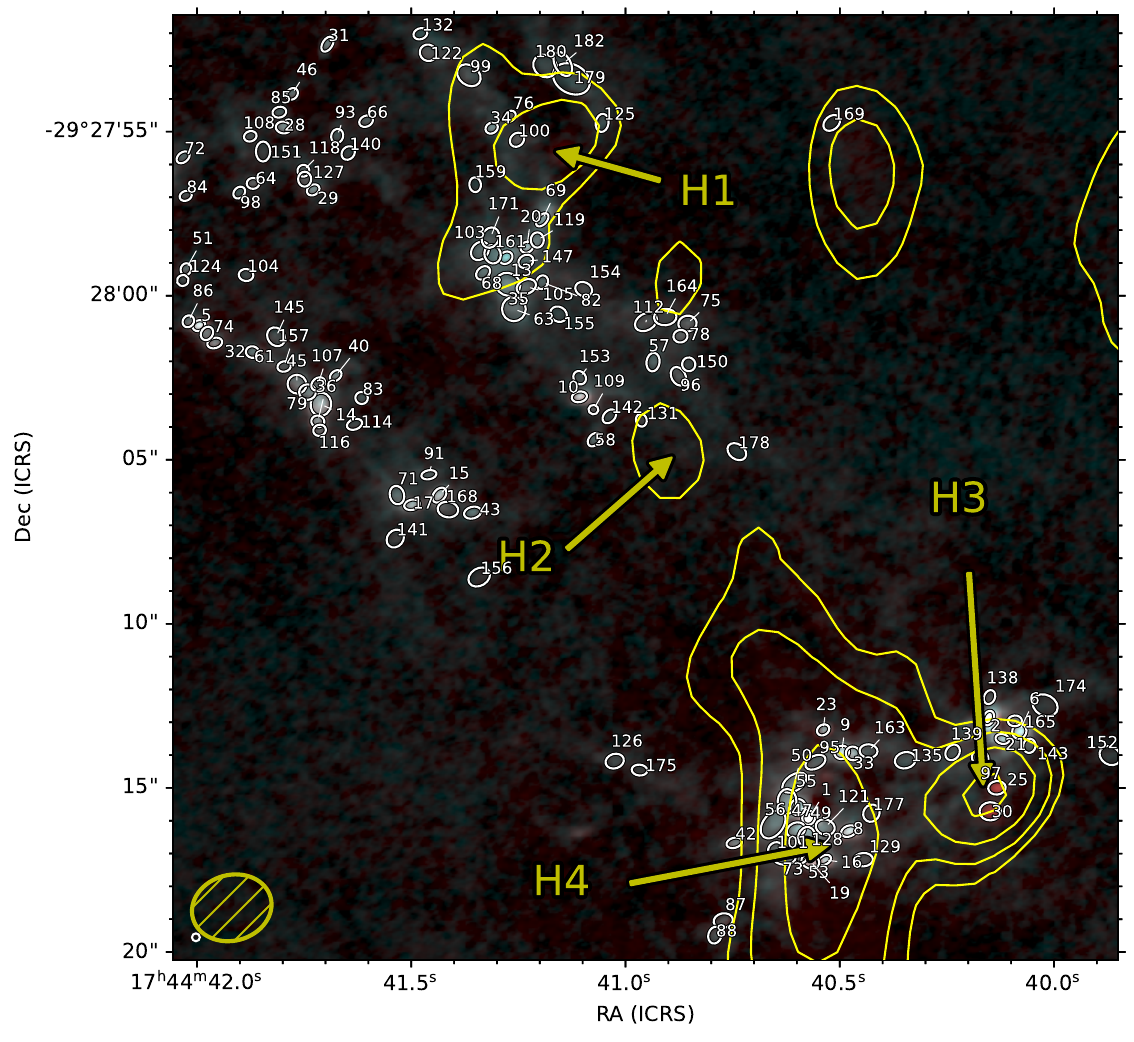}
\caption{Cross-matching JVLA 1.3~cm emission with \texttt{dbcat}. The background color map shows the ALMA 1.3~mm (cyan) and 3~mm (red) continuum emission. The overlaid yellow contours are JVLA 1.3~cm continuum emission. For the \ctw, the contour levels are 0.5, 0.8, 1.3, 2.1, 3.4, 5.5, 8.9, 14.4~mJy\,beam$^{-1}$. For the cloud~e, they are 0.105, 0.210, 0.315~mJy\,beam$^{-1}$. For the Sgr C, they are 0.25, 0.50, 1.0, 2.0~mJy\,beam$^{-1}$. All the identified UCH{\sc ii} region candidates are marked with ``H''s followed by ID numbers.
\label{fig:jvla}}
\end{figure*}

Here we calculate the free-free contamination at the observed bands by extrapolating 1.3~cm flux densities with a spectral index of $-0.1$ for the three clouds. The spatial correspondence is visually inspected in Fig.~\ref{fig:jvla}. 

In Sgr~C, there are four radio compact sources H1--H4. H1 has a cm flux of 2.3~mJy, which can contribute 2~mJy at 3~mm. It is associated with a compact source SgrC-mb\#55 that is only detected in 3~mm with a total flux of 3.6~mJy. We suggest that this source could be a UCH{\sc ii} region (with a size of $\sim$\,2000~au) with more than 50\% of its 3-mm flux from free-free emission, but without detectable dust emission down to 50~$\mu$Jy at 1.3~mm. H2 is not associated with any sources. H3 is associated with two 3~mm bright sources (the red dots) SgrC-db\#25 and SgrC-db\#30. The millimeter spectral index $\alpha_{\rm mm}$ of SgrC-db\#30 is as flat as $-0.1$ which is consistent with free-free emission spectrum, while SgrC-db\#25 has $\alpha_{\rm mm}$ of 0.07, suggesting that there is dust emission. The 3~mm integrated flux of the two sources are 3.1~mJy, while the expected integrated flux at 23~GHz would be 3.6~mJy if the 3~mm flux is from free-free emission, which is lower than the measured cm flux of H3, 5~mJy. However, H4 is more extended along the north-south direction, so we cannot determine the cm flux within the extent of SgrC-db\#1. Therefore, we only subtract the peak flux of $\sim$\,1~mJy to make the correction. The spectral index changes from 3.3 to 3.5 for SgrC-db\#1. 

In cloud~e, only one compact radio source H1 has been detected with an integrated cm flux of 0.5~mJy \citep{Lu2019SFR}. H1 is spatially overlapped with cloude-db\#1. The free-free emission contribution could be as high as 0.43~mJy at 3~mm and 0.40~mJy at 1.3~mm. It is negligible for 1.3~mm but can contaminate the flux by over 10\% at 3~mm. For cloude-db\#1, $\alpha$ would decrease from 3.8 to 3.7.

In \ctw, there are two radio sources H1 and H2 identified. H1 has an integrated cm flux of 1.5~mJy and is associated with a cluster of sources including 20kms-db\#7, 18, 20, 21, and 50. But due to the limited resolution, we cannot tell which one of the DUET sources is associated with H1. Therefore, we evenly distribute the 1.3~cm flux to the five sources, and find that their spectral indices would increase from 2.6--3.5 to 2.9--3.9. The other radio source H2 is an extended H{\sc ii} region, which has a counterpart in our 3~mm continuum image. Two weak detections 20kms-db\#133 and 177 are at the edge of the H{\sc ii} region. However, the contaminated flux cannot be evaluated, because of the resolution difference between 1.3~cm and 3~mm images. So we do not correct their spectral indices.

\end{appendix}

\end{document}